\documentclass[10pt,aps,prb,twocolumn,export,superscriptaddress,bibnotes,reprint,longbibliography]{revtex4-2}

\usepackage[%
colorlinks=true,
linkcolor=blue,
citecolor=blue,
urlcolor=blue,
pdftitle=Braiding games]{hyperref}

\usepackage[utf8]{inputenc}
\usepackage[dvipsnames]{xcolor}
\usepackage{amsthm}
\usepackage{mathtools}
\usepackage{comment}

\usepackage{tikz}
\usetikzlibrary{decorations.markings} 
\usetikzlibrary{decorations.shapes,shapes.geometric} 
\usepackage{tikz-3dplot}

\tikzset{
  c/.style={every coordinate/.try}
}
\tikzset{%
    x=2.4ex,                
    y=2.4ex,                
    baseline={(0,-0.15)}    
}

\usepackage{xpatch}
\makeatletter
\patchcmd{\@ssect@ltx}
    {\addcontentsline{toc}{#1}{\protect\numberline{}#8}}
    {}
    {}
    {}
\makeatother

\usepackage{amsfonts}
\usepackage{amssymb}
\usepackage{braket}
\usepackage{subfigure}
\usepackage[export]{adjustbox} 
\usepackage{dsfont}
\usepackage[inline]{enumitem}  

\usepackage{booktabs}
\usepackage{multirow}
\usepackage{tabularx}
\newcolumntype{Y}{>{\centering\arraybackslash}X}

\usepackage{microtype}
\usepackage{newtxtext}
\usepackage[smallerops]{newtxmath}
\usepackage{anyfontsize}

\newcommand{\id}{\mathds{1}}

\usepackage[normalem]{ulem}

\DeclareMathAlphabet{\mathcal}{OMS}{cmsy}{m}{n}
\DeclareMathAlphabet{\mathbb}{U}{msb}{m}{n}
\usepackage{courierten} 

\newcommand{\expval}[1]{\langle #1 \rangle}

\renewcommand{\vec}[1]{\mathbf{#1}}

\newcommand{\Xe}{X_e}

\newcommand{\Ze}{Z_e}

\DeclareMathOperator{\Tr}{Tr}
\DeclareMathOperator{\im}{im}

\usetikzlibrary{decorations.pathreplacing,decorations.markings,arrows,decorations.pathmorphing}

\usetikzlibrary{arrows.meta,arrows}
\usetikzlibrary{math}
\tikzmath{\c= cos(30); \s = sin(30);} 

\tikzset{snake it/.style={decorate, decoration={snake, segment length=0.1cm, amplitude=0.05ex}}}

\pgfarrowsdeclaredouble{doubled}{doubled}{stealth}{stealth}

\definecolor{xpauli}{HTML}{FF0000}
\definecolor{zpauli}{HTML}{1F77B4}

\newcommand{\vertexop}[2]{%
    \draw [line width=1.25, zpauli, preaction={draw, line width=4, white}, line join=round] ({-1+#1}, {-1+#2}) rectangle ({1+#1}, {1+#2});
}

\newcommand{\faceop}[2]{%
    \draw [line width=1.25, xpauli, preaction={draw, line width=4, white}, line join=round] ({-1+#1}, {-1+#2}) rectangle ({1+#1}, {1+#2});
}

\newcommand{\XL}[2]{%
    \begin{scope}[very thick,decoration={markings,
    mark=at position 0.16667 with {\fill[white] +(-2pt,-2pt) rectangle +(2pt,2pt);},
    mark=at position 0.5 with {\fill[white] +(-2pt,-2pt) rectangle +(2pt,2pt);},
    mark=at position 0.83333 with {\fill[white] +(-2pt,-2pt) rectangle +(2pt,2pt);}}
    ]
        \draw [line width=1.25, xpauli, preaction={decorate}, line cap=round, line join=round] ({1+#1}, {-1+#2}) to ({-1+#1}, {-1+#2}) to ({-1+#1}, {1+#2}) to ({1+#1}, {1+#2});
    \end{scope}
}

\newcommand{\XR}[2]{%
    \begin{scope}[very thick,decoration={markings,
    mark=at position 0.5 with {\fill[white] +(-2pt,-2pt) rectangle +(2pt,2pt);}}
    ]
        \draw [line width=1.25, xpauli, preaction={decorate}, line cap=round, line join=round] ({1+#1}, {-1+#2}) to ({1+#1}, {1+#2});
    \end{scope}
}

\newcommand{\ZL}[2]{%
    \begin{scope}[very thick,decoration={markings,
    mark=at position 0.5 with {\fill[white] +(-2pt,-2pt) rectangle +(2pt,2pt);}}
    ]
        \draw [line width=1.25, zpauli, preaction={decorate}, line cap=round, line join=round] ({-1+#1}, {-1+#2}) to ({-1+#1}, {1+#2});
    \end{scope}
}

\newcommand{\ZR}[2]{%
    \begin{scope}[very thick,decoration={markings,
    mark=at position 0.16667 with {\fill[white] +(-2pt,-2pt) rectangle +(2pt,2pt);},
    mark=at position 0.5 with {\fill[white] +(-2pt,-2pt) rectangle +(2pt,2pt);},
    mark=at position 0.83333 with {\fill[white] +(-2pt,-2pt) rectangle +(2pt,2pt);}}
    ]
        \draw [line width=1.25, zpauli, preaction={decorate}, line cap=round, line join=round] ({-1+#1}, {-1+#2}) to ({1+#1}, {-1+#2}) to ({1+#1}, {1+#2}) to ({-1+#1}, {1+#2});
    \end{scope}
}

\newcommand{\Xbackground}[2]{%
    \draw [line width=1., lightgray, opacity=0.5] ({1+#1}, {-1+#2}) to ({-1+#1}, {-1+#2}) to ({-1+#1}, {1+#2}) to ({1+#1}, {1+#2}) -- cycle;
}

\newcommand{\Xtop}[2]{%
    \begin{scope}[very thick,decoration={markings,
    mark=at position 0.5 with {\fill[white] +(-2pt,-2pt) rectangle +(2pt,2pt);}}
    ]
        \draw [line width=1.25, xpauli, preaction={decorate}, line cap=round] ({-1+#1}, {1+#2}) to ({1+#1}, {1+#2});
    \end{scope}
}

\newcommand{\Xbot}[2]{%
    \begin{scope}[very thick,decoration={markings,
    mark=at position 0.5 with {\fill[white] +(-2pt,-2pt) rectangle +(2pt,2pt);}}
    ]
        \draw [line width=1.25, xpauli, preaction={decorate}, line cap=round] ({-1+#1}, {-1+#2}) to ({1+#1}, {-1+#2});
    \end{scope}
}

\newcommand{\Xsides}[2]{%
    \begin{scope}[very thick,decoration={markings,
    mark=at position 0.5 with {\fill[white] +(-2pt,-2pt) rectangle +(2pt,2pt);}}
    ]
        \draw [line width=1.25, xpauli, preaction={decorate}, line cap=round] ({-1+#1}, {-1+#2}) to ({-1+#1}, {1+#2});
    \end{scope}
    \begin{scope}[very thick,decoration={markings,
    mark=at position 0.5 with {\fill[white] +(-2pt,-2pt) rectangle +(2pt,2pt);}}
    ]
        \draw [line width=1.25, xpauli, preaction={decorate}, line cap=round] ({1+#1}, {-1+#2}) to ({1+#1}, {1+#2});
    \end{scope}
}

\newcommand{\Zbackground}[2]{%
    \draw [line width=1., lightgray, opacity=0.5] ({#1}, {2+#2}) to ({#1}, {-2+#2});
    \draw [line width=1., lightgray, opacity=0.5] ({-1+#1}, {1+#2}) to ({1+#1}, {1+#2});
    \draw [line width=1., lightgray, opacity=0.5] ({-1+#1}, {-1+#2}) to ({1+#1}, {-1+#2});
}

\newcommand{\Ztop}[2]{%
    \begin{scope}[very thick,decoration={markings,
    mark=at position 0.16667 with {\fill[white] +(-2pt,-2pt) rectangle +(2pt,2pt);},
    mark=at position 0.5 with {\fill[white] +(-2pt,-2pt) rectangle +(2pt,2pt);},
    mark=at position 0.83333 with {\fill[white] +(-2pt,-2pt) rectangle +(2pt,2pt);}}
    ]
        \draw [line width=1.25, zpauli, preaction={decorate}, line cap=round, line join=round] ({-1+#1}, {-1+#2}) to ({-1+#1}, {1+#2}) to ({1+#1}, {1+#2}) to ({1+#1}, {-1+#2});
    \end{scope}
}

\newcommand{\Zbot}[2]{%
    \begin{scope}[very thick,decoration={markings,
    mark=at position 0.16667 with {\fill[white] +(-2pt,-2pt) rectangle +(2pt,2pt);},
    mark=at position 0.5 with {\fill[white] +(-2pt,-2pt) rectangle +(2pt,2pt);},
    mark=at position 0.83333 with {\fill[white] +(-2pt,-2pt) rectangle +(2pt,2pt);}}
    ]
        \draw [line width=1.25, zpauli, preaction={decorate}, line cap=round, line join=round] ({-1+#1}, {1+#2}) to ({-1+#1}, {-1+#2}) to ({1+#1}, {-1+#2}) to ({1+#1}, {1+#2});
    \end{scope}
}

\newcommand{\ZH}[2]{%
    \begin{scope}[very thick,decoration={markings,
    mark=at position 0.5 with {\fill[white] +(-2pt,-2pt) rectangle +(2pt,2pt);}}
    ]
        \draw [line width=1.25, zpauli, preaction={decorate}, line cap=round, line join=round] ({-1+#1}, {#2}) to ({1+#1}, {#2});
    \end{scope}
}

\newcommand{\GHZops}{%
\begin{tikzpicture}[scale=0.45]
    \draw [line width=1.25, lightgray, dotted, line cap=round, dash pattern=on 0pt off 2\pgflinewidth] (-4.57, 0)--(4.57, 0);
    \draw [line width=1.25, xpauli] (-2.5, 0)--(2.5, 0);
    \fill [white] (0, 0) circle [radius=5.5pt];
    \fill [zpauli] (0, 0) circle [radius=4pt];
\end{tikzpicture}
}

\newcommand{\GHZstrat}{%
\begin{tikzpicture}[scale=0.45]
    \pgfmathsetmacro{\dx}{0.2}
    \draw [line width=1.25, xpauli] ({-4.5+\dx}, 0)--({-1.5-\dx}, 0);
    \draw [line width=1.25, xpauli] ({-1.5+\dx}, 0)--({1.5-\dx}, 0);
    \draw [line width=1.25, xpauli] ({1.5+\dx}, 0)--({4.5-\dx}, 0);

    \fill [white] (0, 0) circle [radius=5.5pt];
    \fill [zpauli] (0, 0) circle [radius=4pt];

    \fill [white] (-3, 0) circle [radius=5.5pt];
    \fill [zpauli] (-3, 0) circle [radius=4pt];
    
    \fill [white] (3, 0) circle [radius=5.5pt];
    \fill [zpauli] (3, 0) circle [radius=4pt];
\end{tikzpicture}
}

\newcommand{\TCIIops}{%
\begin{tikzpicture}[scale=0.45]
    \draw [line width=1.25, xpauli] (-2.5, -2.5)--(2.5, 2.5);
    \draw [line width=1.25, zpauli] (-2.5, 2.5)--(2.5, -2.5);
    \draw [thick, fill=white] (0, 0) circle [radius=4pt];
\end{tikzpicture}
}

\newcommand{\TCIIstrat}{%
\begin{tikzpicture}[scale=0.45, baseline={(0, -0.15)}]
    \draw [line width=1.25, xpauli] (0, 0) circle [radius=3];
    \draw [line width=1.25, zpauli] (-3, 0) circle [radius=3];
    \draw [line width=1.25, zpauli] (-6, 0) -- (0, 0);
    \draw [thick, fill=white] (-1.5, -2.55) circle [radius=4pt];
    \draw [thick, fill=white] (-1.5, 2.55) circle [radius=4pt];
    \draw [thick, fill=white] (-3, 0) circle [radius=4pt];
\end{tikzpicture}
}

\newcommand{\twoformstrat}{%
\tdplotsetmaincoords{70}{-30}
\begin{tikzpicture}[scale=0.2,tdplot_main_coords,baseline={(0, -0.2)}]

    \begin{scope}[canvas is xz plane at y={0}]
        \draw[xpauli, line width=1.25] (-1.5,{-3*0.866}) arc (-60:0:3);
    \end{scope}

    \begin{scope}[canvas is xy plane at z={0}]
        \fill [zpauli, opacity=0.25] (0, 0) circle [radius=3];
    \end{scope}

    \begin{scope}[canvas is xy plane at z={0}]
        \fill [white, draw=black, line width=0.75] (0, 0) circle [radius=0.35];
    \end{scope}

    \begin{scope}[canvas is xz plane at y={0}]
        \draw[xpauli, line width=1.25] (-1.5,{3*0.866}) arc (60:0:3);
    \end{scope}

    \fill[color = zpauli,
        opacity = 0.25
    ] (0,0,0) circle (3cm);

    \begin{scope}[canvas is xz plane at y={0}]
        \fill [white, draw=black, line width=0.75] (-1.5, {3*0.866}) circle [radius=0.35];
    \end{scope}
    
    \begin{scope}[canvas is xz plane at y={0}]
        \draw[xpauli, line width=1.25, line cap=round] (-1.5,{-3*0.866}) arc (-60:-299:3);
    \end{scope}

    \begin{scope}[canvas is xz plane at y={0}]
        \fill [white, draw=black, line width=0.75, rotate around={57:(-1.5, {-3*0.866})}] (-1.5, {-3*0.866}) ellipse (0.04 and 0.35);
    \end{scope}
\end{tikzpicture}  
}

\newcommand{\oneformops}{%
\tdplotsetmaincoords{70}{-50}
\begin{tikzpicture}[scale=0.2,tdplot_main_coords,baseline={(0, -0.6)}]

\draw [line width=1.25, zpauli] (0, 3, -2) -- (5, 3, -2);

\begin{scope}[canvas is yz plane at x=0]
\fill[red,opacity=0.3] (0,-4) 
--(0,0) 
to [out=20,in=200] (6,0) 
--(6,-4)                         
to [out=220,in=30]  ( 0, -4) --cycle ;
\end{scope}

\begin{scope}[canvas is yz plane at x=0]
\fill [white, draw=black, line width=0.75] (3, -2) circle [radius=0.35];
\end{scope}

\draw [line width=1.25, zpauli, line cap=round] (-0.1, 3, -2) -- (-1, 3, -2);
\draw [line width=1.25, zpauli] (-1, 3, -2) -- (-5, 3, -2);

\end{tikzpicture}
}

\newcommand{\oneformstrat}{%
\tdplotsetmaincoords{70}{-30}
\begin{tikzpicture}[scale=0.2,tdplot_main_coords,baseline={(0, -0.15)}]

    \begin{scope}[transform canvas={rotate=180}, rotate=180]
      \clip [rotate=30] (3cm,0) arc [x radius=3cm, y radius=1.6cm, start angle=0, end angle=180] -- (-3cm,3cm) -- (3cm,3cm) -- (3cm,0);
      \clip [rotate=-20] (3cm,0) arc [x radius=3cm, y radius=0.39cm, start angle=0, end angle=180] -- (-3cm,-3cm) -- (3cm,-3cm) -- (3cm,0);

      \fill [ color = red!50!black, opacity = 0.35] (0,0) circle [radius=3cm];
    \end{scope}

    \begin{scope}
      \clip [rotate={-20+180}] (3cm,0) arc [x radius=3cm, y radius=0.39cm, start angle=0, end angle=-180] -- (-3cm,-3cm) -- (3cm,-3cm) -- (3cm,0);
      \clip [rotate={180}] (3cm,0) arc [x radius=3cm, y radius=1.03cm, start angle=0, end angle=-180] -- (-3cm,-3cm) -- (3cm,-3cm) -- (3cm,0);
      \fill [ color = red!50!black, opacity = 0.175] (0,0) circle [radius=3cm];
    \end{scope}

    \begin{scope}
      \clip [rotate={30+180}] (3cm,0) arc [x radius=3cm, y radius=1.6cm, start angle=0, end angle=-180] -- (-3cm,3cm) -- (3cm,3cm) -- (3cm,0);
      \clip [rotate={0}] (3cm,0) arc [x radius=3cm, y radius=1.03cm, start angle=0, end angle=180] -- (-3cm,-3cm) -- (3cm,-3cm) -- (3cm,0);

      \fill [ color = red!50!black, opacity = 0.05] (0,0) circle [radius=3cm];
    \end{scope}

    \begin{scope}[canvas is xz plane at y={0}]
        \draw[zpauli, line width=1.25] (-1.5,{-3*0.866}) arc (-60:0:3);
    \end{scope}

    \draw[zpauli, line width=1.25] (0, 0, 0) -- (-3, 0, 0);

    \begin{scope}[canvas is xz plane at y={0}]
        \draw[zpauli, line width=1.25] (-1.5,{3*0.866}) arc (60:0:3);
    \end{scope}

    \begin{scope}
      \clip [rotate=-20] (3cm,0) arc [x radius=3cm, y radius=0.39cm, start angle=0, end angle=-180] -- (-3cm,-3cm) -- (3cm,-3cm) -- (3cm,0);
      \clip [rotate=30] (3cm,0) arc [x radius=3cm, y radius=1.6cm, start angle=0, end angle=-180] -- (-3cm,3cm) -- (3cm,3cm) -- (3cm,0);
      \fill [ color = red, opacity = 0.6] (0,0) circle [radius=3cm];
    \end{scope}

    \begin{scope}
        \clip [rotate={-20+180}] (3cm,0) arc [x radius=3cm, y radius=0.39cm, start angle=0, end angle=180] -- (-3cm,-3cm) -- (3cm,-3cm) -- (3cm,0);
        \clip [rotate={180}] (3cm,0) arc [x radius=3cm, y radius=1.03cm, start angle=0, end angle=180] -- (-3cm,-3cm) -- (3cm,-3cm) -- (3cm,0);
        \fill [ color = red, opacity = 0.35] (0,0) circle [radius=3cm];
    \end{scope}

    \begin{scope}
      \clip [rotate={30+180}] (3cm,0) arc [x radius=3cm, y radius=1.6cm, start angle=0, end angle=180] -- (-3cm,3cm) -- (3cm,3cm) -- (3cm,0);
      \clip [rotate={0}] (3cm,0) arc [x radius=3cm, y radius=1.03cm, start angle=0, end angle=-180] -- (-3cm,-3cm) -- (3cm,-3cm) -- (3cm,0);

      \fill [ color = red, opacity = 0.25] (0,0) circle [radius=3cm];
    \end{scope}

    \begin{scope}[canvas is xz plane at y={0}]
        \fill [white, draw=black, line width=0.75] (-1.5, {3*0.866}) circle [radius=0.35];
    \end{scope}

    \begin{scope}[canvas is yz plane at x={-3}]
        \fill [white, draw=black, line width=0.75] (0, 0) circle [radius=0.35];
    \end{scope}

    \draw[zpauli, line width=1.25] (-6, 0, 0) -- (-3, 0, 0);

    \begin{scope}[canvas is xz plane at y={0}]
        \draw[zpauli, line width=1.25, line cap=round] (-1.5,{-3*0.866}) arc (-60:-299:3);
    \end{scope}

    \begin{scope}[canvas is xz plane at y={0}]
        \fill [white, draw=black, line width=0.75, rotate around={57:(-1.5, {-3*0.866})}] (-1.5, {-3*0.866}) ellipse (0.02 and 0.35);
    \end{scope}

\end{tikzpicture}
}

\newcommand{\twoformops}{%
\tdplotsetmaincoords{70}{-50}
\begin{tikzpicture}[scale=0.2,tdplot_main_coords,baseline={(0, -0.6)}]

\draw [line width=1.25, xpauli] (0, 3, -2) -- (5, 3, -2);

\begin{scope}[canvas is yz plane at x=0]
\fill[zpauli,opacity=0.3] (0,-4) 
--(0,0) 
to [out=20,in=200] (6,0) 
--(6,-4)                         
to [out=220,in=30]  ( 0, -4) --cycle ;
\end{scope}

\begin{scope}[canvas is yz plane at x=0]
\fill [white, draw=black, line width=0.75] (3, -2) circle [radius=0.35];
\end{scope}

\draw [line width=1.25, xpauli, line cap=round] (-0.1, 3, -2) -- (-1, 3, -2);
\draw [line width=1.25, xpauli] (-1, 3, -2) -- (-5, 3, -2);

\end{tikzpicture}
}

\newcommand{\XCstrat}{%
\tdplotsetmaincoords{40}{27-90}
\begin{tikzpicture}[scale=0.35,tdplot_main_coords,baseline={(0, 1.1)}]

    \pgfmathsetmacro{\cubex}{2}
    \pgfmathsetmacro{\cubey}{2}
    \pgfmathsetmacro{\cubez}{2}

    \draw[line width=1,zpauli, line join=round, line cap=round] (0, \cubey/2, 0) -- (0,\cubey,0);

    \draw[line width=1,zpauli, line join=round, line cap=round] (-\cubex/2,0,0) -- ++(\cubex/2,0,0);

    \draw[line width=1,zpauli, line join=round, line cap=round] (\cubex/2,0,0) -- ++(\cubex/2,0,0);

    \fill [fill=red,fill opacity=0.25] (-1, -1, -1)--++(0, 2, 0)--++(0, 0, 2)--++(0, -2, 0)--cycle;

    \fill [fill=red,fill opacity=0.25] (-1, -1, -1)--++(2, 0, 0)--++(0, 0, 2)--++(-2, 0, 0)--cycle;
    \fill [fill=red,fill opacity=0.25] (-1, 1, -1)--++(2, 0, 0)--++(0, 0, 2)--++(-2, 0, 0)--cycle;

    \fill [fill=red,fill opacity=0.25] (1, -1, -1)--++(0, 2, 0)--++(0, 0, 2)--++(0, -2, 0)--cycle;

    \begin{scope}[canvas is yz plane at x=-1]
        \draw [line width=0.75, fill=white] (0, 0) circle [radius=0.225];
    \end{scope}

    \begin{scope}[canvas is yz plane at x=1]
        \draw [line width=0.75, fill=white] (0, 0) circle [radius=0.225];
    \end{scope}

    \begin{scope}[canvas is xz plane at y=1]
        \draw [line width=0.75, fill=white] (0, 0) circle [radius=0.225];
    \end{scope}

    \draw[line width=1,zpauli, line join=round, line cap=round] (0,0,0) -- ++(\cubex/2,0,0);

    \draw[line width=1,zpauli, line join=round] (0, \cubey/2, 0) -- (0,0,0) -- ++(0,0,\cubez) -- ++(0,\cubey,0) -- ++(0,0,-\cubez);

    \draw[line width=1,zpauli, line join=round, line cap=round] (0,\cubey,0) -- ++(\cubex,0,0);
    \draw[line width=1,zpauli, line join=round, line cap=round] (0,0,\cubez) -- ++(\cubex,0,0);
    \draw[line width=1,zpauli, line join=round, line cap=round] (0,\cubey,\cubez) -- ++(\cubex,0,0);
    \draw[line width=1,zpauli, line join=round, line cap=round] (\cubex,0,0) -- ++(0,0,\cubez) -- ++(0,\cubey,0) -- ++(0,0,-\cubez) -- cycle;
    
    \draw[line width=1,zpauli, line join=round, line cap=round] (-\cubex/2,0,0) -- ++(-\cubex/2,0,0);
    \draw[line width=1,zpauli, line join=round, line cap=round] (0,\cubey,0) -- ++(-\cubex,0,0);
    \draw[line width=1,zpauli, line join=round, line cap=round] (0,0,\cubez) -- ++(-\cubex,0,0);
    \draw[line width=1,zpauli, line join=round, line cap=round] (0,\cubey,\cubez) -- ++(-\cubex,0,0);
    
    \draw[line width=1,zpauli, line join=round, line cap=round] (-\cubex,0,0) -- ++(0,0,\cubez) -- ++(0,\cubey,0) -- ++(0,0,-\cubez) -- cycle;
    \end{tikzpicture}
}

\newcommand{\XCops}{%
\tdplotsetmaincoords{70}{-60}
\begin{tikzpicture}[scale=0.3,tdplot_main_coords,baseline={(0, 0.55)}]

    \pgfmathsetmacro{\cubex}{2}
    \pgfmathsetmacro{\cubey}{2}
    \pgfmathsetmacro{\cubez}{2}
    
    \draw[line width=1,zpauli, line join=round] ({-\cubex/2},0,0) -- ++({\cubex/2},0,0);
    
    \fill [fill=red,fill opacity=0.25] (-1, -1, -1)--++(0, 2, 0)--++(0, 0, 2)--++(0, -2, 0)--cycle;
    
    \begin{scope}[canvas is yz plane at x=-1]
        \fill [white, draw=black, line width=0.75] (0, 0) circle [radius=0.225];
    \end{scope}

    \draw[line width=1,zpauli, line cap=round] (-\cubex,0,0) -- ++({\cubex/2-0.05},0,0);
    
    \draw[line width=1,zpauli, line join=round] (0,\cubey,0) -- ++(-\cubex,0,0);
    \draw[line width=1,zpauli, line join=round] (0,0,\cubez) -- ++(-\cubex,0,0);
    \draw[line width=1,zpauli, line join=round] (0,\cubey,\cubez) -- ++(-\cubex,0,0);
    
    \draw[line width=1,zpauli, line join=round] (-\cubex,0,0) -- ++(0,0,\cubez) -- ++(0,\cubey,0) -- ++(0,0,-\cubez) -- cycle;
    
\end{tikzpicture}
}

\newcommand{\XCopalt}{%
\tdplotsetmaincoords{70}{-60}
\begin{tikzpicture}[scale=0.3,tdplot_main_coords,baseline={(0, -0.25)}]

    \pgfmathsetmacro{\cubex}{2}
    \pgfmathsetmacro{\cubey}{2}
    \pgfmathsetmacro{\cubez}{2}

    \draw[line width=1,xpauli, line join=round] (0,{-\cubey/2}, 0) -- ++(0,\cubey/2,0) -- ++(0,0,-\cubez) -- ++(0, {-\cubey/2}, 0);

    \fill [fill=zpauli,fill opacity=0.25] (-1, -1, -1)--++(0, 2, 0)--++(0, 0, 2)--++(0, -2, 0)--cycle;

    \fill [fill=zpauli,fill opacity=0.25] (-1, -1, -1)--++(2, 0, 0)--++(0, 0, 2)--++(-2, 0, 0)--cycle;
    \fill [fill=zpauli,fill opacity=0.25] (-1, 1, -1)--++(2, 0, 0)--++(0, 0, 2)--++(-2, 0, 0)--cycle;

    \begin{scope}[canvas is xz plane at y=-1]
        \fill [white, draw=black, line width=0.75] (0, 0) circle [radius=0.225];
    \end{scope}

  \draw[line width=1,xpauli, line join=round, line cap=round] (0,{-\cubey/2-0.05}, 0) -- ++(0,-\cubey/2+0.05,0) -- ++(0,0,-\cubez) -- ++(0, {\cubey/2}, 0);

\end{tikzpicture}

}

\newcommand{\XCstratalt}{%
\tdplotsetmaincoords{40}{27-90}
\begin{tikzpicture}[scale=0.35,tdplot_main_coords,baseline={(0, 0.8)}]

    \pgfmathsetmacro{\cubex}{2}
    \pgfmathsetmacro{\cubey}{2}
    \pgfmathsetmacro{\cubez}{2}

    \draw[line width=1,xpauli, line join=round] (0,{-\cubey/2}, 0) -- ++(0,\cubey/2,0) -- ++(0,0,-\cubez) -- ++(0, {-\cubey/2}, 0);
    \draw[line width=1,xpauli, line join=round] (\cubex,{-\cubey/2}, 0) -- ++(0,\cubey/2,0) -- ++(0,0,-\cubez) -- ++(0, {-\cubey/2}, 0);
    \draw[line width=1,xpauli, line join=round, line cap=round] (\cubex/2,0, 0) -- ++(\cubex/2, 0, 0);

    \draw[line width=1,xpauli, line join=round, line cap=round] (0,0, -\cubez) -- ++(\cubex, 0, 0);

    \fill [fill=zpauli,fill opacity=0.25] (-1, -1, -1)--++(0, 2, 0)--++(0, 0, 2)--++(0, -2, 0)--cycle;

    \fill [fill=zpauli,fill opacity=0.25] (-1, -1, -1)--++(2, 0, 0)--++(0, 0, 2)--++(-2, 0, 0)--cycle;
    \fill [fill=zpauli,fill opacity=0.25] (-1, 1, -1)--++(2, 0, 0)--++(0, 0, 2)--++(-2, 0, 0)--cycle;

    \fill [fill=zpauli,fill opacity=0.25] (1, -1, -1)--++(2, 0, 0)--++(0, 0, 2)--++(-2, 0, 0)--cycle;
    \fill [fill=zpauli,fill opacity=0.25] (1, 1, -1)--++(2, 0, 0)--++(0, 0, 2)--++(-2, 0, 0)--cycle;

    \fill [fill=zpauli,fill opacity=0.25] (1, -1, -1)--++(0, 2, 0)--++(0, 0, 2)--++(0, -2, 0)--cycle;

    \fill [fill=zpauli,fill opacity=0.25] (3, -1, -1)--++(0, 2, 0)--++(0, 0, 2)--++(0, -2, 0)--cycle;

    \begin{scope}[canvas is xz plane at y=-1]
        \fill [white, draw=black, line width=0.75] (0, 0) circle [radius=0.225];
    \end{scope}

    \begin{scope}[canvas is xz plane at y=-1]
        \fill [white, draw=black, line width=0.75] (2, 0) circle [radius=0.225];
    \end{scope}

    \begin{scope}[canvas is yz plane at x=1]
        \fill [white, draw=black, line width=0.75] (0, 0) circle [radius=0.225];
    \end{scope}

  \draw[line width=1,xpauli, line join=round] (0,{-\cubey/2}, 0) -- ++(0,-\cubey/2,0) -- ++(0,0,-\cubez) -- ++(0, {\cubey/2}, 0);

  \draw[line width=1,xpauli, line join=round, line cap=round] (0,{-\cubey}, 0) -- ++(\cubex, 0, 0);
  \draw[line width=1,xpauli, line join=round, line cap=round] (0,0, 0) -- ++(\cubex/2, 0, 0);

  \draw[line width=1,xpauli, line join=round, line cap=round] (0,{-\cubey}, -\cubez) -- ++(\cubex, 0, 0);

  \draw[line width=1,xpauli, line join=round] (\cubex,{-\cubey/2}, 0) -- ++(0,-\cubey/2,0) -- ++(0,0,-\cubez) -- ++(0, {\cubey/2}, 0);

\end{tikzpicture}
}

\DeclarePairedDelimiter\abs{\lvert}{\rvert}%
\DeclarePairedDelimiter\norm{\lVert}{\rVert}%

\makeatletter
\let\oldabs\abs
\def\abs{\@ifstar{\oldabs}{\oldabs*}}
\let\oldnorm\norm
\def\norm{\@ifstar{\oldnorm}{\oldnorm*}}
\makeatother

\usepackage{bbold}

\makeatletter
\@twosidetrue 
\makeatother
\usepackage{fancyhdr}
\begin{document}

\title{%
Braiding for the win: Harnessing braiding statistics in topological states to win quantum games}

\date{\today}

\author{Oliver Hart\,\href{https://orcid.org/0000-0002-5391-7483}{\includegraphics[width=6.5pt]{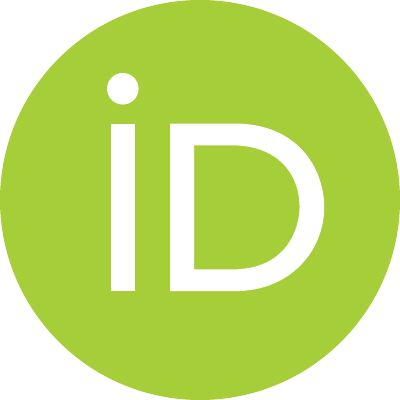}}}
\affiliation{Department of Physics and Center for Theory of Quantum Matter, \href{https://ror.org/02ttsq026}{University of Colorado Boulder}, Boulder, Colorado 80309 USA}

\author{David T. Stephen}
\altaffiliation{Current address: Quantinuum, 303 S Technology Ct, Broomfield, CO 80021, USA}
\affiliation{Department of Physics and Center for Theory of Quantum Matter, \href{https://ror.org/02ttsq026}{University of Colorado Boulder}, Boulder, Colorado 80309 USA}
\affiliation{Department of Physics, \href{https://ror.org/05dxps055}{California Institute of Technology}, Pasadena, California 91125, USA}

\author{Dominic J. Williamson}
\altaffiliation{Current address: IBM Quantum, IBM Almaden Research Center, San Jose, CA 95120, USA}
\affiliation{School of Physics, \href{https://ror.org/0384j8v12}{University of Sydney}, Sydney, New South Wales 2006, Australia}

\author{Rahul Nandkishore}
\affiliation{Department of Physics and Center for Theory of Quantum Matter, \href{https://ror.org/02ttsq026}{University of Colorado Boulder}, Boulder, Colorado 80309 USA}

\begin{abstract}
Nonlocal quantum games provide proof of principle that quantum resources can confer advantage at certain tasks. They also provide a compelling way to explore the computational utility of phases of matter on quantum hardware. In a recent manuscript [Hart \emph{et al}., \href{https://doi.org/10.48550/arXiv.2403.04829}{arXiv:2403.04829}] we demonstrated that a toric code resource state conferred advantage at a certain nonlocal game, which remained robust to small deformations of the resource state. In this manuscript we demonstrate that this robust advantage is a generic property of resource states drawn from topological or fracton ordered phases of quantum matter. To this end, we illustrate how several other states from paradigmatic topological and fracton ordered phases can function as resources for suitably defined nonlocal games, notably the three-dimensional toric-code phase, the X-cube fracton phase, and the double-semion phase. The key in every case is to design a nonlocal game that harnesses the characteristic braiding processes of a quantum phase as a source of contextuality. We unify the strategies that take advantage of mutual statistics by relating the operators to be measured to order and disorder parameters of an underlying generalized symmetry-breaking phase transition. Finally, we massively generalize the family of games that admit perfect strategies when codewords of homological quantum error-correcting codes are used as resources.
\end{abstract}

\maketitle


\tableofcontents

\section{Introduction}

Measurements on disjoint parts of a quantum state can famously be more correlated than any classical hidden-variable model would permit~\cite{Bell}. This result was recast in the language of \emph{nonlocal quantum games} by Mermin~\cite{Mermin,mermin1990quantum}, wherein noncommunicating players can win certain games requiring coordination with higher probability than would be classically possible, if they share an entangled quantum state and condition their strategy on their quantum measurement outcomes. The property of a quantum state that confers advantage at nonlocal games is known as \emph{contextuality}~\cite{kochenSpecker1967}, and the possibility of leveraging contextuality to coordinate without classical communication is known as quantum pseudo-telepathy~\cite{brassard2005pseudotelepathy}.

While the original Mermin games involve a small number of players, in recent years there has been considerable interest in $P$-player nonlocal quantum games in the limit of large $P$. For the quantum-information community, such games provide a valuable proof-of-principle that advantage can be derived from access to (certain) quantum states \cite{DanielStringOrder,Daniel2022Exp, bravyi2020noisyShallow, bravyi2020shallow, jaffali2023new, DallaTorre}. For the many-body theory community, they provide an intriguing perspective on phases of matter on noisy, intermediate-scale quantum (NISQ) hardware~\cite{BBSgame,BulchandaniGames, lin2023quantum}, and, for the quantum complexity theory community, they are a key ingredient in interactive proofs~\cite{natarajan}. 

While theory work on quantum games frequently assumes idealized resource states, it is important to understand to what extent advantage at the game persists if the resource state is imperfect, and how well the advantage scales as we take $P$ to be large. These questions were recently addressed in Ref.~\cite{hart2024playing}, where it was argued that while the advantage afforded by a Greenberger-Horne-Zeilinger (GHZ) resource state considered in the original papers~\cite{Mermin} was not robust to deformations, robust and scalable advantage could be secured by using a (deformed) two-dimensional (2D) toric code state as a resource state. The key idea was to make use of the braiding statistics in the toric code as a source of contextuality. This picture was verified by experiments carried out on one of Quantinuum's trapped-ion quantum processors, connecting also to recent efforts to tune through phase transitions with shallow quantum circuits~\cite{zhu2022, lee2022decoding, chen2023realizing}. This does however prompt the question: what, if anything, is special about the 2D toric code?

In this manuscript we demonstrate that topological and fracton ordered states in general provide a robust and scalable resource for nonlocal quantum games, with the braiding statistics providing the source of contextuality, and that the ``toric code games'' from Refs.~\cite{hart2024playing,BBSgame} may be regarded as specific strategies that take advantage of these nontrivial statistics. These results show that topological (and fracton) phases serve as a generic source of robust and scalable quantum advantage at suitably defined nonlocal quantum games. Equivalently, it paves the way to showing how quantum games may be used to diagnose topological or fracton phases on quantum hardware. 

We begin in Sec.~\ref{sec:games} by reviewing the few-body parity game and its perfect strategy, which requires access to shared quantum resources. We then provide further details on how 2D toric-code ground states may serve as resources for the parity game (expanding on Ref.~\cite{hart2024playing}) in Sec.~\ref{sec:toric-code}, and explain how both the conventional and toric-code strategies can be unified by relating the operators to be measured to order and disorder parameters of an underlying generalized spontaneous-symmetry-breaking (SSB) phase transition. These considerations allow us to harness the mutual statistics of other topological and fracton phases of quantum matter, which we describe explicitly for the 3D toric code and the X-cube models in Secs.~\ref{sec:3D-toric-code} and \ref{sec:X-cube}, respectively. Next, we describe two further generalizations of these ideas. First, in Sec.~\ref{sec:homological-games}, we massively generalize the family of many-body games for which codewords of homological Calderbank-Shor-Steane (CSS) codes enable a perfect quantum strategy. Finally, in Sec.~\ref{sec:self-statistics}, we explore how \emph{self}-statistics can also be harnessed to gain advantage at a different nonlocal quantum game using the double-semion phase as a paradigmatic example. We conclude with a discussion of some interesting open directions in Sec.~\ref{sec:discussion}. 


\section{The parity game}
\label{sec:games}

In this section, we review the construction of a computational task that cannot be completed with certainty with access to only classical resources. The task is framed as a cooperative \emph{nonlocal quantum game} known as the parity game~\cite{ghz1989,GHSZ1990,MerminPolynomials,Brassard2003multiparty,brassard2005recasting,brassard2005pseudotelepathy}, which consists of players that attempt to successfully complete the task (i.e., win the game) without communicating with each other. When the players are given access to shared quantum resources, there exist perfect strategies, which we describe.


\subsection{Three-player game}
\label{sec:parity-game}

The \emph{parity game}~\cite{ghz1989,GHSZ1990,MerminPolynomials,Brassard2003multiparty,brassard2005recasting,brassard2005pseudotelepathy} involves three players, indexed by $i=1, 2, 3$, each of whom is handed one of the three input bits $\{ x_i \}$, which satisfy the constraint $\sum_{i=1}^3 x_i \equiv 0 \mod 2$. Without communicating with one another, their task is to output bits $\{ y_i \}$ such that addition of these output bits mod 2 satisfies
\begin{equation}
    \sum_{i=1}^{3} y_i \equiv \frac12 \sum_{i=1}^3 x_i \mod 2
    \, .
    \label{eqn:parity-criterion}
\end{equation}
If their outputs satisfy~\eqref{eqn:parity-criterion} for a particular set of inputs $\{ x_i \}$ we say that the players \emph{win} the game.
If the players' strategy leads to~\eqref{eqn:parity-criterion} being satisfied for all four choices of input then we refer to their strategy as \emph{perfect}.
Note that the right-hand side can alternatively be written as $x_1 \vee x_2 \vee x_3$, where $\vee$ denotes the nonlinear boolean function `or'.
This connects~\eqref{eqn:parity-criterion} to the computation of nonlinear boolean functions and to measurement-based quantum computation (MBQC)~\cite{Anders2009}.
The performance of an {arbitrary} classical strategy, including probabilistic ones involving shared classical random variables, cannot exceed an optimal deterministic strategy~\cite{Yao1977,brassard2005recasting}. Hence, suppose that the players attempt to win the game by employing a deterministic classical strategy in which the $i$th player outputs $y_i(x_i)$ when handed $x_i$. A perfect deterministic strategy would require that
\begin{subequations}
\begin{align}
    y_1(0) + y_2(1) + y_3(1) \mod 2 &= 1 \label{eqn:deterministic-1} \\
    y_1(1) + y_2(0) + y_3(1) \mod 2 &= 1 \label{eqn:deterministic-2} \\
    y_1(1) + y_2(1) + y_3(0) \mod 2 &= 1 \label{eqn:deterministic-3} \\
    y_1(0) + y_2(0) + y_3(0) \mod 2 &= 0 \, .
\end{align}%
\label{eqn:classical-strat}%
\end{subequations}
This set of equations exhibits a parity inconsistency, which may be revealed by adding the above equations mod 2~\cite{Mermin}, implying that $0 \equiv 1$. This contradiction proves that no such perfect deterministic strategy can exist. Instead, classical strategies must win on \emph{at most} three of the four possible inputs, with the trivial strategy $y_i = 1$, independent of $x_i$, saturating this bound. 

If the players have access to quantum-mechanical resources, the parity game admits a perfect strategy~\cite{Brassard2003multiparty,brassard2005recasting,brassard2005pseudotelepathy}. In particular, if each player has access to one qubit of a three-qubit Greenberger-Horne-Zeilinger (GHZ) state, $\ket{\text{GHZ}} = (\ket{000} + \ket{111})/\sqrt{2}$, they may incorporate local measurements into their strategy. If player $i$ receives $x_i = 0$ ($x_i = 1$), they measure $X_i$ ($Y_i$) on their qubit and, if their measurement outcome reads $(-1)^{s_i}$, they output $y_i = s_i$. Since the operators $\{ X_i, Y_i \}$ satisfy the constraints $X_1Y_2Y_3=Y_1X_2Y_3=Y_1Y_2X_3=-1$ and $X_1X_2X_3=1$ when acting on the GHZ state, while the players' measurement outcomes $s_i$ will be individually random (each player's reduced density matrix is $\rho \propto \id$), when added together mod 2 they will satisfy Eq.~\eqref{eqn:classical-strat} for all choices of inputs. The GHZ state therefore permits a perfect quantum strategy and serves as a resource for the three-player parity game.


\subsection{Increasing the number of players}

The game can also straightforwardly be extended to $P \geq 3$ players~\cite{MerminPolynomials,Brassard2003multiparty,brassard2005recasting}, which further reduces the optimal classical victory probability. The input bits $\{ x_i \}$, where $i \in \{ 1, \dots, P \}$, now satisfy the generalized constraint $\sum_{i=1}^P x_i \equiv 0 \mod 2$ and the players are said to win the game given inputs $\{ x_i \}$ if their outputs $\{ y_i \}$ obey the generalized victory condition
\begin{equation}
    \sum_{i=1}^{P} y_i \equiv \frac12 \sum_{i=1}^P x_i \mod 2
    \, .
    \label{eqn:P-player-victory}
\end{equation}
The constraint on inputs ensures that the right-hand side is always an integer. As in the three-player game, there exists no perfect classical strategy, and the victory probability of any classical strategy $p_\text{cl}$ must satisfy the bound~\cite{brassard2005recasting}
\begin{equation}
    p_\text{cl} \leq \frac12 + \frac{1}{2^{\lceil P/2 \rceil}}
    \, ,
    \label{eqn:parity-game-victory-prob}
\end{equation}
where $\lceil \ \cdot\ \rceil$ denotes the ceiling function. The bound~\eqref{eqn:parity-game-victory-prob} is tight: Simple deterministic classical strategies saturating this bound can be found in Ref.~\cite{brassard2005recasting}. In the limit of large $P$, $p_\text{cl}$ approaches the average victory probability of a classical strategy in which each $y_i$ is drawn from the uniform distribution over $\{0, 1\}$.

If the players share a $P$-qubit GHZ state then the strategy enacted by the individual players described previously wins on all inputs to the $P$-player game~\cite{MerminPolynomials,Brassard2003multiparty,brassard2005recasting}. Namely, if each player measures $X_i$ ($Y_i$) when they are handed $x_i=0$ ($x_i=1$) then, collectively, the players measure GHZ stabilizers up to a phase by virtue of the parity constraint on inputs. Specifically, if the inputs satisfy $\sum_i x_i = 4n$ for $n \in \mathbb{N}$, this phase is equal to $+1$, while if $\sum_i x_i = 4n + 2$ it is equal to $-1$, which is precisely the requirement of the victory condition~\eqref{eqn:P-player-victory}.


\section{Two-dimensional toric code}
\label{sec:toric-code}

We now describe how two-dimensional toric code ground states~\cite{Kitaev2003} are able to function as a resource for the parity game from Sec.~\ref{sec:games}, expanding on the results of Ref.~\cite{hart2024playing} (see also Ref.~\cite{BBSgame}). This construction paves the way to identifying a strategy that can utilize the correlations present in other phases of matter with nontrivial braiding statistics, which is a direction that we explore in Sec.~\ref{sec: top}.


\subsection{The toric code and graphical language}

Place qubits on the edges of an $L \times L$ square lattice in two dimensions.
We use the notation $\Xe$, $\Ze$ for the microscopic Pauli $X$ and $Z$ operators on edge $e$, and denote the basis of $\Ze$ by the states $\ket{0}$, $\ket{1}$, with $\Ze \ket{b}=(-1)^b \ket{b}$. The toric code Hamiltonian~\cite{Kitaev2003} on this lattice is
\begin{equation}
    H = - \sum_{v} A_v -  \sum_p B_p
    \, ,
    \label{eqn:H-toric-code}
\end{equation}
where $A_v = \prod_{\partial e \ni v} \Ze$ and $B_p = \prod_{e \in \partial p} \Xe$ are centered on the vertices $v$ and plaquettes $p$ of the square lattice, respectively. The operators $A_v$ and $B_p$ all mutually commute and square to the identity; eigenstates of~\eqref{eqn:H-toric-code} are therefore also eigenstates of the individual $A_v$ and $B_p$ operators with eigenvalues $\pm 1$. The ground states of the model have all $A_v = B_p = +1$. On the torus, there are four ground states, which are distinguished only by the eigenvalues of operators that correspond to the noncontractible cycles of the torus.

To represent operators, we introduce the following graphical notation. 
Products of Pauli $X$ operators are depicted by shading the corresponding edges of the direct lattice red.
For example, the $B_p$ operator entering~\eqref{eqn:H-toric-code} becomes
\begin{equation}
    B_p = \prod_{e \in \partial p} X_e = \:
    \begin{tikzpicture}[x=2.4ex, y=2.4ex,baseline={(0,-0.2)}]
        \draw [thick, lightgray, opacity=0.75] (-1.5, -1) to (1.5, -1);
        \draw [thick, lightgray, opacity=0.75] (-1.5, 1) to (1.5, 1);
        \draw [thick, lightgray, opacity=0.75] (-1, -1.5) to (-1, 1.5);
        \draw [thick, lightgray, opacity=0.75] (1, -1.5) to (1, 1.5);
        \draw [line width=1.25, xpauli, line join=round] (-1, -1) rectangle (1, 1);
    \end{tikzpicture}
    \, .
    \label{eqn:Bp}
\end{equation}
Products of $B_p$ operators therefore form closed red loops on the direct lattice since $X_e^2=\id$ on all interior edges.
Products of Pauli $Z$ operators are depicted by blue lines on the dual lattice. Hence, the $A_v$ operator in~\eqref{eqn:H-toric-code} becomes
\begin{equation}
    A_v = \prod_{\partial e \ni v} Z_e = \:
    \begin{tikzpicture}[x=2.4ex, y=2.4ex,baseline={(0,-0.2)}]
        \draw [thick, lightgray, opacity=0.75] (-1.5, 0) to (1.5, 0);
        \draw [thick, lightgray, opacity=0.75] (0, -1.5) to (0, 1.5);
        \draw [line width=1.25, zpauli, line join=round] (-1, -1) rectangle (1, 1);
    \end{tikzpicture}
    \, .
\end{equation}
For the sake of simplicity, we often choose not to draw the underlying square lattice (light gray lines). A key property that we use throughout the paper is that, in toric code ground states, any such closed loops assume the definite value $+1$. Open strings of $X$ operators on the direct lattice end on vertices and lead to two excitations ($A_v=-1$) at the endpoints of the string, which we refer to as electric charges ($e$). Similarly, open strings of $Z$ operators on the dual lattice lead to two defective plaquettes ($B_p = -1$) at the endpoints of the dual path, which we refer to as magnetic vortices ($m$).

If two operators are applied to the same edge, we choose to depict the line that corresponds to the operator applied ``most recently'' on top of other lines. That is, the arrow of time in our diagrams flows from below to above the page. Hence, using these conventions, for an adjacent vertex-plaquette pair,
\begin{equation}
    A_v B_p = \:
    \begin{tikzpicture}[x=2.4ex, y=2.4ex,baseline={(0,-0.1)}]
        \faceop{0}{-0.5};
        \vertexop{1}{0.5};
    \end{tikzpicture}
    \, ,
    \qquad 
    B_p A_v = \:
    \begin{tikzpicture}[x=2.4ex, y=2.4ex,baseline={(0,-0.1)}]
        \vertexop{1}{0.5};
        \faceop{0}{-0.5};
    \end{tikzpicture}
    \, .
    \label{eqn:AB-commutation}
\end{equation}
Bringing two differently colored lines through each other on an edge introduces a minus sign due to the anticommutation of $X_e$ and $Z_e$. Since $A_v$ and $B_p$ in~\eqref{eqn:AB-commutation} intersect on \emph{two} edges, the signs cancel and the operators $A_v$ and $B_p$ therefore commute. Conversely, the \emph{twist product}~\cite{haah2016invariant} of these operators is
\begin{equation}
    A_v \infty B_p = \begin{tikzpicture}
    \Xtop{-0.5}{-0.5};\Ztop{0.5}{0.5};\ZH{0.5}{-0.5};\Xsides{-0.5}{-0.5};\Xbot{-0.5}{-0.5};
    \end{tikzpicture}
    ,
    \label{eqn:twist-product}
\end{equation}
which satisfies $\braket{\psi | A_v \infty B_p | \psi} = - 1$ when acting on toric code ground states due to the anticommutation of Pauli $X$ and $Z$. {This twist product is defined by sequentially applying segments of the operators $A_v$ and $B_p$ in the order suggested by the overlapping of the edges in \eqref{eqn:twist-product} \cite{haah2016invariant}. For example, we can apply the bottom edge of $A_v$, then apply $B_p$, then apply the rest of $A_v$. This intertwinement of the operators results in the relative sign of $-1$ compared to the product $A_vB_p$.}


\subsection{As a resource for the parity game}
\label{sec:TC-star}

In Sec.~\ref{sec:games}, we saw that the parity game did not permit a perfect strategy with access to only classical resources. Instead, incorporating measurements of a $P$-particle GHZ state enabled the construction of a perfect quantum strategy. Here, we allow the players to instead share a toric code ground state before playing the game, and construct a strategy that utilizes the correlations therein to function as a resource for the parity game (see also Ref.~\cite{BBSgame}). In what follows, we restrict our attention to single-site Pauli measurements.


\subsubsection{Quantum strategy}

In order to arrive at a perfect quantum mechanical strategy, consider the following loop segments:
\vspace{-15pt}
\begin{equation}
\arraycolsep=6pt\def\arraystretch{4}
\begin{array}{ccc}
    X_1 = \begin{tikzpicture}\Xbackground{0}{0};\Xtop{0}{0};\end{tikzpicture}\, ,
    &
    X_2 = \begin{tikzpicture}\Xbackground{0}{0};\Xsides{0}{0};\end{tikzpicture}\, ,
    &
    X_3 = \begin{tikzpicture}\Xbackground{0}{0};\Xbot{0}{0};\end{tikzpicture}\, , \\
    Z_1 = \begin{tikzpicture}\Zbackground{0}{0};\Ztop{0}{1};\end{tikzpicture}\, ,
    &
    Z_2 = \begin{tikzpicture}\Zbackground{0}{0};\ZH{0}{0};\end{tikzpicture}\, , 
    &
    Z_3 = \begin{tikzpicture}\Zbackground{0}{0};\Zbot{0}{-1};\end{tikzpicture}\, .
    \end{array}
    \label{eqn:TC-state-dependent-loops}
\end{equation}
See~\eqref{eqn:mutual-stats} for the relative position of these operators on the lattice. By construction, $X_i$ and $Z_j$ only intersect exactly once for $i = j$, implying that these operators satisfy the same commutation relations as Pauli matrices. Furthermore, when acting on ideal toric code ground states, we have the constraints
\begin{subequations}
\begin{gather}
    X_1 Y_2 Y_3 = Y_1 X_2 Y_3  = Y_1 Y_2 X_3  = -1 \label{eqn:TC-state-minussigns} \\
    X_1 X_2 X_3 = 1 
    \, ,
\end{gather}%
\label{eqn:measured-loops}%
\end{subequations}
where we defined the Hermitian operator $Y_i \coloneq i X_i Z_i$, which can be decomposed into a product of single-site Pauli $X_e$, $Y_e$, and $Z_e$ operators. Graphically, these constraints may be regarded as following from the diagrams
\begin{subequations}
\begin{align}
    X_1 Y_2 Y_3 = \begin{tikzpicture}\XR{-0.5}{0.5};\ZL{0.5}{-0.5};\XL{-0.5}{0.5};\ZR{0.5}{-0.5};\end{tikzpicture} \quad Y_1 Y_2 X_3 = \begin{tikzpicture}\Xbot{-0.5}{-0.5};\Xsides{-0.5}{-0.5};\ZH{0.5}{-0.5};\Ztop{0.5}{0.5};\Xtop{-0.5}{-0.5};\end{tikzpicture} \\
    Y_1 X_2 Y_3 = \begin{tikzpicture}\Xbot{0}{0};\Zbot{1}{-1};\Xsides{0}{0};\Ztop{1}{1};\Xtop{0}{0};\end{tikzpicture} \quad
    X_1 X_2 X_3 = \begin{tikzpicture}\draw[white](0,0)to(2,0);\faceop{0}{0};\end{tikzpicture}
\end{align}%
\label{eqn:mutual-stats}%
\end{subequations}
Since~\eqref{eqn:mutual-stats} can be interpreted as braiding diagrams,
the minus signs in~\eqref{eqn:TC-state-minussigns} derive from the nontrivial mutual statistics of the $e$ and $m$ excitations of the toric code. Equivalently, we observe that the collective measurements performed by the players lead to the evaluation of the twist product in Eq.~\eqref{eqn:twist-product}. Note that the operators~\eqref{eqn:TC-state-dependent-loops} can be homotopically deformed while maintaining their mutual intersection properties and the constraints~\eqref{eqn:mutual-stats} will still be satisfied. All such smoothly deformed operators can therefore be used in the strategy described below.

Suppose that player $i$ has access only to the qubits in the support of the operators $X_i$ and $Z_i$. A perfect quantum strategy for the parity game using a shared toric code ground state then proceeds as follows. If player $i$ receives $x_i=0$, they measure the individual Pauli operators in the loop segment $X_i$. If they receive $x_i=1$, they measure $Y_e$ at the intersection of the $X$ and $Z$ loop segments, and they measure the individual Pauli operators $X_e$ ($Z_e$) in the remainder of the $X_i$ ($Z_i$) loop segment. Note that since only single-site operators need to be measured, the game can alternatively be phrased in terms of three ``teams'' consisting of players, each of whom has access only to one qubit~\cite{hart2024playing}. Each player then returns the output $y_i = \sum_k s_k \mod 2$, where $(-1)^{s_k}$ is the outcome of the $k$th single-site measurement. That this strategy is perfect follows from the fact that the players effectively measure the operators~\eqref{eqn:mutual-stats}, which assume the definite values required to satisfy the victory condition~\eqref{eqn:parity-criterion} in toric code ground states.

For a generic many-body resource state $\ket{\psi}$, and averaging over all choices of inputs $\{ x_i \}$ with uniform weight, it can be shown that the average victory probability for the strategy just described satisfies 
\begin{equation}
    p_\text{q} (\ket{\psi}) = \frac12 \left[ 1 + \frac14 \braket{\psi | M_3 | \psi}  \right]
    \, ,
    \label{eqn:pq-generic-state}
\end{equation}
where $M_3$ is a third-order Mermin polynomial~\cite{MerminPolynomials} constructed from the partial loop segments in Eq.~\eqref{eqn:TC-state-dependent-loops}, i.e., $M_3 = X_1 X_2 X_3 - X_1 Y_2 Y_3 - Y_1 X_2 Y_3 - Y_1 Y_2 X_3$. The relations presented in Eq.~\eqref{eqn:measured-loops} ensure that $\langle M_3 \rangle = 4$ when toric code ground states are used as a resource. This provides an alternative way to see that the strategy described above represents a perfect quantum strategy, i.e., $p_\text{q} = 1$. The quantum strategy outperforms the optimal classical strategy when $\langle M_3 \rangle > 2$, connecting the players' performance at the task of winning the game to a more conventional Bell-inequality perspective~\cite{ghz1989,GHSZ1990,MerminPolynomials}. We can also view the above strategy as the identification of stringlike operators in the toric code whose measurement statistics reproduce those of single-site Pauli observables in a three-qubit GHZ state; we return to this point in Sec.~\ref{sec:other-states}.

\begin{figure}[t]
    \centering
    \includegraphics[width=\linewidth]{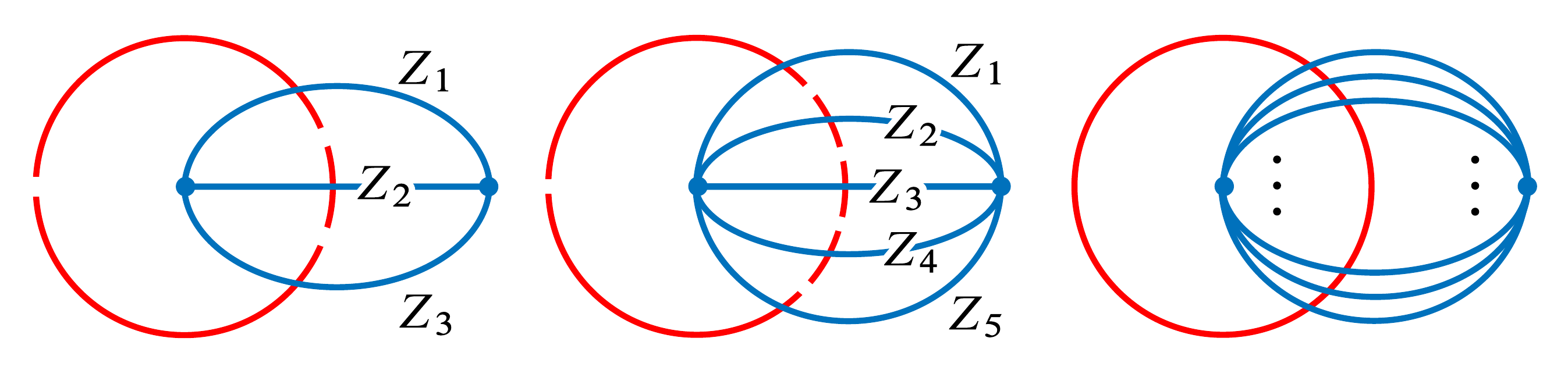}%
    \caption{Lattice-independent illustration of the operator configurations that can be used to win the $P$-player parity game using 2D toric code ground states as a resource for $P=3, 5$, and generic $P$ from left to right. In all cases, the $X$ loop segments are implicitly labeled such that ${X}_i$ only anticommutes with ${Z}_i$.} 
    \label{fig:p_game}
\end{figure}

\begin{table*}[t]
    \centering
    \caption{Schematic overview of the operators $X_i$ (red), $Z_i$ (blue) that enter the perfect quantum strategy for the parity game using the GHZ state, two-dimensional toric code (2DTC), three-dimensional toric code (3DTC), and X-cube (XC) ground states as resources. We also present how these operators are organized relative to one another in the operator arrangement that enables a perfect strategy. In the bottom row, the operators $X_i$ and $Z_i$ satisfy the constraints $X_1 X_2 X_3 = 1 = Z_i Z_j$ required to encode GHZ-like measurement statistics.}
\begin{tabularx}{\textwidth}{@{}XYYYYYY@{}}
\midrule\midrule
\multirow{2}{*}{} & \multicolumn{1}{c}{\multirow{2}{*}{GHZ}} & \multicolumn{1}{c}{\multirow{2}{*}{2DTC}} & \multicolumn{2}{c}{3DTC} & \multicolumn{2}{c}{XC} \\
                  & \multicolumn{1}{c}{}                     & \multicolumn{1}{c}{}                       & 1-form      & 2-form      & prism      & cage      \\ \midrule
Operators {\color{xpauli}$X_i$}, {\color{zpauli}$Z_i$} &        \GHZops          &          \TCIIops         & \oneformops & \twoformops &  \XCops    & \XCopalt  \\
Strategy     &        \GHZstrat        &         \TCIIstrat        &\oneformstrat&\twoformstrat&  \XCstrat  &\XCstratalt\\  \midrule\midrule
\end{tabularx}
\label{tab:summary}
\end{table*}


\subsubsection{Increasing the number of players}

For games involving $P > 3$ players~\cite{MerminPolynomials,Brassard2003multiparty,brassard2005recasting}, the operators and the strategy enacted by the players is analogous. Specifically, the operators are shown schematically in Fig.~\ref{fig:p_game} for $P=3, 5$ and generic $P$ (see also Ref.~\cite{hart2024playing}). Each player has access to the qubits in the support of two loop segments $X_i$ (on the direct lattice) and $Z_i$ (on the dual lattice). These operators are chosen such that they intersect exactly once and furthermore satisfy the constraints $\prod_{i=1}^P X_i = 1$ and $Z_i Z_j=1$ (for all $i, j$) when acting on toric code ground states. Consequently, the measurement statistics of the composite operators $\{ X_i, Y_i \}$ will match those of single-site Pauli operators in a $P$-particle GHZ state. Since the $P$-particle GHZ state acts as a resource for the $P$-player game, it follows that the composite operators illustrated in Fig.~\ref{fig:p_game} will lead to a perfect quantum strategy for all $P$.


\subsection{Order and disorder parameters}
\label{sec:order-disorder}

The strategies used to win the parity game using a GHZ state or a toric code ground state as a resource can both be framed in terms of order and disorder parameters~\cite{fradkin2017disorder} of an underlying generalized symmetry-breaking transition. The GHZ state is the ground state of a 1D transverse-field Ising model (TFIM), which exhibits a phase transition associated with spontaneous symmetry breaking (SSB) of a global (i.e., 0-form) symmetry as a function of transverse field. The 2D toric code can be viewed as the ground space of the local Hamiltonian in Eq.~\eqref{eqn:H-toric-code}. The toric code Hamiltonian with a transverse field $-\Delta\sum_i X_i$ is a 2D $\mathbb{Z}_2$ lattice gauge theory with a 1-form symmetry generated by the algebra of $B_p$ operators. As $\Delta$ decreases, from an arbitrarily large initial value to zero, the toric code with transverse field exhibits a phase transition associated to the SSB of the 1-form symmetry. 

The general strategy to win $P$-player parity games using ordered quantum states is to choose each $X_i$ operator to be a disorder operator, i.e., a truncated symmetry operator, such that $X_1X_2 \cdots X_P$ is a full symmetry operator. At the same time, each $Z_i$ operator is chosen to anticommute with only $X_i$ and hence it is charged under the full symmetry. This can be ensured by choosing $Z_i$ to be supported on a region that only overlaps the support of $X_i$. Furthermore, the $Z_i$ operators are chosen such that $Z_iZ_j$ are symmetric operators with long-range order (LRO). A numerical approach to constructing such pairs of operators is described in Ref.~\cite{Dua2019Sorting}. The strategy to win the parity game with general ordered quantum states is then to have the players measure operators that evaluate to the twist product of the full symmetry operator, e.g., $X_1 X_2 X_3$ for the three-player game, with an operator that exhibits LRO, $Z_i Z_j$. This produces expressions like ``$X_1 X_2 X_3 \infty Z_2 Z_3 = X_1 (X_2 Z_2) (Z_3 X_3)$,'' which evaluate to $-1$ in the appropriate states.

For the 1D TFIM example, the global 0-form symmetry is $S = \prod_i X_i$ and the order parameters $Z_i$ (for any $i$) are charged under the Ising symmetry $S Z_i S = -Z_i$ and detect SSB $\braket{Z} \neq 0$. Products of the order parameters yield symmetric operators with LRO $\lim_{\abs{i-j}\to\infty}\langle Z_i Z_j\rangle \neq 0$. The disorder operators are $\prod_{k=i+1}^{j} X_k$, which create pointlike domain wall excitations at their endpoints. See the first column of Tab.~\ref{tab:summary}.

For the 2D toric code example, the 1-form symmetry is  $\prod_{e \in \partial R} X_e$ for any union of plaquettes $R$. The charged operators under the 1-form symmetry are given by $Z$ string operators $\prod_{e\in \Lambda} Z_e$ where $\Lambda$ is a dual string on the lattice. The corresponding operators that possess LRO are closed dual strings, $\prod_{e \in \partial^\dagger V}Z_e$ for any region $V$ of the dual lattice, where $\partial^\dagger$ denotes the boundary operation on the dual lattice. In this case, LRO is generically indicated by an exponential decay of these operators with length $|\partial^\dagger V|$ (rather than the area $|V|$), i.e., $\langle \prod_{e \in \partial^\dagger V}Z_e\rangle\sim \smash{e^{-c|\partial^\dagger V|}}$ for some $c>0$. The disorder operators correspond to open $X$ strings which create pointlike anyonic excitations at their endpoints. This example extends naturally to higher dimensional toric codes, which can have more general higher-form symmetries~\cite{Nussinov2007, Gaiotto2015,  McGreevy2022}. In a system with a $p$-form symmetry, the symmetry operators have codimension $p$ and the charged operators have dimension $p$. In the presence of explicit symmetry breaking perturbations, $p$-form symmetries with $p>0$ are expected to survive in the sense that the symmetry operators may change but the group structure is approximately preserved in the low-energy subspace. In this setting, finding order and disorder parameters remains possible but is more challenging~\cite{Fredenhagen1983}.

The understanding outlined in this section provides an intuitive explanation for the relative robustness of nonlocal games that exploit topologically ordered resource states. Namely, all strategies involve measuring a twist product of the full symmetry operator with a charged operator. If the symmetry is a {\it global} symmetry, then the full symmetry operator is global, and the requisite twist product is thus subject to an orthogonality catastrophe when the state is perturbed. In contrast, with a topologically ordered resource state, the relevant symmetry is higher form, and there exist symmetry operators that are almost local (viz.~small loops and membranes). Thus, one can construct suitable twist products that are almost local, the measurements of which are not subject to any orthogonality catastrophe upon perturbing the global state.
 

\section{Other topological and fracton orders}
\label{sec: top}

We now explain how mutual statistics of excitations can be harnessed more generally in topological and fracton orders to yield analogous quantum advantage in the parity game. For simplicity we present the analysis for games involving $P=3$ players, but the presented strategies all generalize to arbitrary $P$, as explained in Sec.~\ref{sec:TC-star} (see also Fig.~\ref{fig:p_game}). 


\subsection{3D toric code}
\label{sec:3D-toric-code}

First, we generalize the strategy involving 2D toric code ground states  to the 3D toric code. In 2D, we were able to design a perfect strategy by taking advantage of the mutual statistics between pointlike $e$ and $m$ anyons. In the 3D toric code, we must use the nontrivial braiding statistics between its extended and pointlike excitations.


\subsubsection{1-form symmetry}

We work with a 3D version of Kitaev's toric code~\cite{Kitaev2003,Hamma3DTC} with degrees of freedom defined on the elementary faces $f$ of an $L\times L \times L$ cubic lattice
\begin{equation}
    H_{\text{3TC}} = -\sum_e A_e - \sum_c B_c
    \, ,
    \label{eqn:3D-TC-faces}
\end{equation}
where $A_e = \prod_{\partial f \ni e} Z_f$ and $B_c = \prod_{f \in \partial c} X_f$ are associated to edges $e$ and cubes (or cells) $c$, respectively. All $A_e$ and $B_c$ operators share either two or zero faces and hence commute; the ground states of the model have all $A_e = B_c = 1$. In addition to a number of global constraints, the $A_e$ operators notably satisfy the local constraint $\prod_{\partial e \ni v} A_e = \id$ for each vertex $v$~\cite{Hamma3DTC}. The combination of these constraints gives rise to eight ground states on the 3-torus [see also Eq.~\eqref{eqn:stabilizer-dimensions}].

Graphically, the $A_e$ operators correspond to elementary closed loops on the dual lattice, while the $B_c$ operators are represented by elementary membranes on the direct lattice. These operators may be depicted graphically as
\begin{equation}
    \tdplotsetmaincoords{75}{30}
    A_e =
    \begin{tikzpicture}[scale=0.45,tdplot_main_coords,baseline={(0, 0)}]
        \draw [line width=1.25, lightgray, opacity=0.5] (0, 0, 0) to (0, 2.5, 0);
        \draw [line width=1.25, zpauli] (-1, 0, -1) to (1, 0, -1) to (1, 0, 1) to (-1, 0, 1) --cycle;
        \draw [line width=4., white] (0, -2.5, 0) to (0, -1, 0);
        \draw [line width=1.25, lightgray, opacity=0.5] (0, -2.5, 0) to (0, 0, 0);
    \end{tikzpicture}
    \qquad
    B_c = \:\:
    \begin{tikzpicture}[scale=0.45,tdplot_main_coords,baseline={(0, 1)}]
        \pgfmathsetmacro{\cubex}{2}
        \pgfmathsetmacro{\cubey}{2}
        \pgfmathsetmacro{\cubez}{2}
        \fill[xpauli, opacity=0.4] (0,0,0) -- ++(\cubex,0,0) -- ++(0,\cubey,0) -- ++(-\cubex,0,0) -- cycle; 
        \fill[xpauli, opacity=0.125] (0,0,0) -- ++(0,0,\cubez) -- ++(0,\cubey,0) -- ++(0,0,-\cubez) -- cycle; 
        \fill[xpauli, opacity=0.125] (0,0,0) -- ++(\cubex,0,0) -- ++(0,0,\cubez) -- ++(-\cubex,0,0) -- cycle; 
        \fill[xpauli, opacity=0.05] (\cubex,\cubey,\cubez) -- ++(-\cubex,0,0) -- ++(0,-\cubey,0) -- ++(\cubex,0,0) -- cycle; 
        \fill[xpauli, opacity=0.3] (\cubex,\cubey,\cubez) -- ++(0,0,-\cubez) -- ++(0,-\cubey,0) -- ++(0,0,\cubez) -- cycle; 
        \fill[xpauli, opacity=0.3] (\cubex,\cubey,\cubez) -- ++(-\cubex,0,0) -- ++(0,0,-\cubez) -- ++(\cubex,0,0) -- cycle; 
        \begin{scope}[canvas is xy plane at z=0]
            \fill [red] (1, 1) circle [radius=1.ex];
        \end{scope}
        \begin{scope}[canvas is xy plane at z=\cubez]
            \fill [red] (1, 1) circle [radius=1.ex];
        \end{scope}
        \begin{scope}[canvas is zy plane at x=0]
            \fill [red] (1, 1) circle [radius=1.ex];
        \end{scope}
        \begin{scope}[canvas is zy plane at x=\cubex]
            \fill [red] (1, 1) circle [radius=1.ex];
        \end{scope}
        \begin{scope}[canvas is xz plane at y=0]
            \fill [red] (1, 1) circle [radius=1.ex];
        \end{scope}
        \begin{scope}[canvas is xz plane at y=\cubey]
            \fill [red] (1, 1) circle [radius=1.ex];
        \end{scope}
    \end{tikzpicture}
    \, .
\end{equation}
In the following, we omit the qubits on faces in our graphical language.
The product of $A_e$ operators over some collection of dual faces $R^*$ leads to a closed loop on the boundary: $\prod_{e \in R^*}A_e = \prod_{f \in \partial R^*} Z_f$~\footnote{We will often omit the isomorphism $*$ between a cell complex $X$ and its dual $X^*$ for brevity. Expressions such as $f \in \partial R^*$ should be understood as $f \in *\partial R^*$.}, while the product of $B_c$ operators over elementary cubes $c$ belonging to some volume $V$ leads to the closed membrane operator on its boundary: $\prod_{c \in V}B_c = \prod_{e \in \partial V} X_f$. All such closed loops and membranes are symmetries of 3D toric code ground states, giving rise to 2-form and 1-form symmetries, respectively~\cite{McGreevy2022}.

Open strings of $Z_f$ operators on the dual lattice end on cube centers and lead to pointlike excitations $B_c = -1$ at the endpoints of the string. On the other hand, an open membrane $M$ of $X_f$ operators on the primary lattice excites a closed, one-dimensional loop of operators $A_e = -1$ at its boundary $e \in \partial M$. These excitations exhibit nontrivial mutual statistics in the sense that the wave function acquires a phase of $\pi$ if a string of $Z_f$ operators penetrates an open $X$ membrane.

To proceed, we first make use of the model's $X$-like 1-form symmetry, $\prod_{e \in \partial V} X_f = 1$. Consider some volume $V$ whose boundary $\partial V$ is simply connected. This boundary can then be split into three disjoint surfaces $S_i$ satisfying $\cup_i S_i = \partial V$, which can be used to define the operators $X_1$, $X_2$, and $X_3$, i.e., $X_i = \prod_{f \in S_i} X_f$. By virtue of the model's 1-form symmetry, these operators satisfy the constraint $X_1X_2X_3 = 1$ when acting on 3D toric code ground states. We must then identify operators $Z_1$, $Z_2$, and $Z_3$ charged under the symmetry $X_1 X_2 X_3$. This may be achieved by considering open paths $\Gamma_i^*$ on the dual lattice, where $\Gamma_i^*$ intersects (only) $S_i$ once, and defining the stringlike operators $Z_i = \prod_{f \in \Gamma_i^*} Z_f$. Furthermore, if the endpoints of the paths $\{ \Gamma_i^* \}$ coincide, it follows that each $Z_i Z_j$ forms a closed loop on the dual lattice. Thus, $Z_i Z_j = 1$ for all pairs $(i, j)$ since $Z_i Z_j$ reduces to a product of $A_p$ operators.

The smallest set of such operators on the cubic lattice involves $V$ equal to an elementary cube, while $Z_1Z_2$ and $Z_2 Z_3$ are supported on the faces that border a single edge:
\vspace{-10pt}
\begin{equation}
    \tdplotsetmaincoords{70}{30}
    \arraycolsep=10pt\def\arraystretch{4}
    \begin{array}{ccc}
    {X}_1 = \:
    \begin{tikzpicture}[scale=0.35,tdplot_main_coords,every path/.style={line width=1,line cap=round, line join=round}]
        \coordinate (O) at (-1, -1, -1);
        \coordinate (X) at (1, -1, -1);
        \coordinate (Y) at (-1, 1, -1);
        \coordinate (Z) at (-1, -1, 1);
        \coordinate (XY) at (1, 1, -1);
        \coordinate (YZ) at (-1, 1, 1);
        \coordinate (ZX) at (1, -1, 1);
        \coordinate (XYZ) at (1, 1, 1);
        \begin{scope}[every coordinate/.style={shift={(0,0,0)}}]
            \fill [xpauli, opacity=0.15] ([c]O)--([c]X)--([c]ZX)--([c]Z)--cycle;
            \fill [xpauli, opacity=0.15] ([c]Z)--([c]YZ)--([c]Y)--([c]O)--cycle;
            \draw [xpauli!50, snake it] ([c]O)--([c]X)--([c]ZX)--([c]Z)--([c]YZ)--([c]Y)--([c]O);
        \end{scope}
    \end{tikzpicture}
    &
    {X}_2 = \:
    \begin{tikzpicture}[scale=0.35,tdplot_main_coords,every path/.style={line width=1,line cap=round, line join=round}]
        \coordinate (O) at (-1, -1, -1);
        \coordinate (X) at (1, -1, -1);
        \coordinate (Y) at (-1, 1, -1);
        \coordinate (Z) at (-1, -1, 1);
        \coordinate (XY) at (1, 1, -1);
        \coordinate (YZ) at (-1, 1, 1);
        \coordinate (ZX) at (1, -1, 1);
        \coordinate (XYZ) at (1, 1, 1);
        \begin{scope}[every coordinate/.style={shift={(0, 0, 0)}}]
            \fill [xpauli, opacity=0.15] ([c]X)--([c]ZX)--([c]Z)--([c]YZ)--([c]XYZ)--([c]XY)--cycle;
            \draw [xpauli!50, style={decorate, decoration={snake, segment length=0.115cm, amplitude=0.05ex}}] ([c]X)--([c]ZX)--([c]Z)--([c]YZ)--([c]XYZ)--([c]XY)--cycle;
        \end{scope}
    \end{tikzpicture}
    &
    {X}_3 = \:
    \begin{tikzpicture}[scale=0.35,tdplot_main_coords,every path/.style={line width=1,line cap=round, line join=round}]
        \coordinate (O) at (-1, -1, -1);
        \coordinate (X) at (1, -1, -1);
        \coordinate (Y) at (-1, 1, -1);
        \coordinate (Z) at (-1, -1, 1);
        \coordinate (XY) at (1, 1, -1);
        \coordinate (YZ) at (-1, 1, 1);
        \coordinate (ZX) at (1, -1, 1);
        \coordinate (XYZ) at (1, 1, 1);
        \begin{scope}[every coordinate/.style={shift={(0,0,0)}}]
            \fill [xpauli, opacity=0.15] ([c]O)--([c]Y)--([c]YZ)--([c]XYZ)--([c]XY)--([c]X)--cycle;
            \draw [xpauli!50, style={decorate, decoration={snake, segment length=0.11cm, amplitude=0.05ex}}] ([c]O)--([c]Y)--([c]YZ)--([c]XYZ)--([c]XY)--([c]X)--cycle;
        \end{scope}
    \end{tikzpicture} \\
    {Z}_1 = \:
    \begin{tikzpicture}[scale=0.35,tdplot_main_coords,every path/.style={line width=1,line cap=round, line join=round}]
        \draw [zpauli] (0, 0, 0) to (0, -2, 0) to (2, -2, 0) to (2, 0, 0);
    \end{tikzpicture}
    &
    {Z}_2 = \:
    \begin{tikzpicture}[scale=0.35,tdplot_main_coords,every path/.style={line width=1,line cap=round, line join=round}]
        \draw [zpauli] (0, 0, 0) to (2, 0, 0);
    \end{tikzpicture}
    &
    {Z}_3 = \:
    \begin{tikzpicture}[scale=0.35,tdplot_main_coords,every path/.style={line width=1,line cap=round, line join=round}]
        \draw [zpauli] (0, 0, 0) to (0, 2, 0) to (2, 2, 0) to (2, 0, 0);
    \end{tikzpicture}
    \end{array}
    \label{eqn:3D-TC-1form-ops}
\end{equation}
By construction, the product $X_1 X_2 X_3 = 1$ and $Z_i Z_j=1$ for all pairs $(i, j)$ when acting on 3D toric code ground states. Furthermore, $Z_i$ intersects only $X_i$ and does so exactly once, ensuring that the operators in~\eqref{eqn:3D-TC-1form-ops} furnish a representation of the Pauli algebra on three qubits [see also Tab.~\ref{tab:summary}]. Measurement of the operators $\{ X_i, Y_i \}$ with $Y_i \coloneq i X_i Z_i$ may therefore be used to construct a perfect strategy for the three-player parity game.


\subsubsection{2-form symmetry}

An entirely analogous construction exists that uses the 2-form symmetry of the model. To make conventions consistent with the previous examples, it is convenient to consider a modified version of~\eqref{eqn:3D-TC-faces} in which degrees of freedom are situated on the edges of an $L\times L \times L$ cubic lattice. That is,
\begin{equation}
    H_{\text{3TC}^*} = - \sum_v A_v - \sum_f B_f
\end{equation}
where $A_v = \prod_{\partial e \ni v} Z_e$ and $B_f = \prod_{e \in \partial f} X_e$ are associated to vertices $v$ and faces $f$, respectively. This is equivalent to interchanging Pauli $X$ and $Z$ and labeling operators according to the dual cubic lattice with respect to~\eqref{eqn:3D-TC-faces}. Hence, the $X$-like symmetry is now a 2-form symmetry. A perfect strategy for the parity game is constructed by choosing $X_1 X_2 X_3$ to form a closed loop on the direct lattice with each $X_i$ supported on a disjoint interval. The $Z_i$ are chosen such that they are charged under the symmetry $X_1X_2X_3$, which is achieved by choosing dual surfaces $S_i^*$ that intersect $X_i$. The condition $Z_i Z_j=1$ is ensured by working with $S_i^*$ that share a common boundary, i.e., $\partial S_i^* = \partial S_j^*$ for all $i$, $j$. 

The minimal such configuration on the cubic lattice involves $X_1 X_2 X_3$ being supported around an elementary face $f$:
\vspace{-20pt}
\begin{equation}
\tdplotsetmaincoords{70}{30}
\arraycolsep=10pt\def\arraystretch{4}
\begin{array}{ccc}
    {X}_1 = \:
    \begin{tikzpicture}[scale=0.35,tdplot_main_coords,every path/.style={line width=1,line cap=round, line join=round}]
        \draw [lightgray, opacity=0.5] (-1, 1, 0) to (-1, -1, 0) to (1, -1, 0) to (1, 1, 0)-- cycle;
        \draw [xpauli] (-1, -1, 0) to (1, -1, 0);
    \end{tikzpicture}
    &
    {X}_2 = \:
    \begin{tikzpicture}[scale=0.35,tdplot_main_coords,every path/.style={line width=1,line cap=round, line join=round}]
        \draw [lightgray, opacity=0.5] (-1, 1, 0) to (-1, -1, 0) to (1, -1, 0) to (1, 1, 0)-- cycle;
        \draw [xpauli] (1, -1, 0) to (1, 1, 0);
        \draw [xpauli] (-1, -1, 0) to (-1, 1, 0);
    \end{tikzpicture}
    &
    {X}_3 = \:
    \begin{tikzpicture}[scale=0.35,tdplot_main_coords,every path/.style={line width=1,line cap=round, line join=round}]
        \draw [lightgray, opacity=0.5] (-1, 1, 0) to (-1, -1, 0) to (1, -1, 0) to (1, 1, 0)-- cycle;
        \draw [xpauli] (-1, 1, 0) to (1, 1, 0);
    \end{tikzpicture} \\
    {Z}_1 = \:
    \begin{tikzpicture}[scale=0.35,tdplot_main_coords,every path/.style={line width=1,line cap=round, line join=round}]
        \coordinate (O) at (-1, -1, -1);
        \coordinate (X) at (1, -1, -1);
        \coordinate (Y) at (-1, 1, -1);
        \coordinate (Z) at (-1, -1, 1);
        \coordinate (XY) at (1, 1, -1);
        \coordinate (YZ) at (-1, 1, 1);
        \coordinate (ZX) at (1, -1, 1);
        \coordinate (XYZ) at (1, 1, 1);
        \begin{scope}[every coordinate/.style={shift={(0,0,0)}}]
            \fill [zpauli, opacity=0.35] ([c]O)--([c]X)--([c]ZX)--([c]Z)--cycle;
            \fill [zpauli, opacity=0.15] ([c]Z)--([c]YZ)--([c]Y)--([c]O)--cycle;
            \fill [zpauli, opacity=0.25] ([c]X)--([c]ZX)--([c]Z)--([c]YZ)--([c]XYZ)--([c]XY)--cycle;
            \fill [zpauli, opacity=0.15] ([c]O)--([c]Y)--([c]XY)--([c]X)--cycle;
            \draw [zpauli!65, style={decorate, decoration={snake, segment length=0.1cm, amplitude=0.05ex}}] ([c]Y)--([c]XY)--([c]XYZ)--([c]YZ)--cycle;
        \end{scope}
    \end{tikzpicture}
    &
    {Z}_2 = \:
    \begin{tikzpicture}[scale=0.35,tdplot_main_coords,every path/.style={line width=1,line cap=round, line join=round}]
        \coordinate (O) at (-1, -1, -1);
        \coordinate (X) at (1, -1, -1);
        \coordinate (Y) at (-1, 1, -1);
        \coordinate (Z) at (-1, -1, 1);
        \coordinate (XY) at (1, 1, -1);
        \coordinate (YZ) at (-1, 1, 1);
        \coordinate (ZX) at (1, -1, 1);
        \coordinate (XYZ) at (1, 1, 1);
        \begin{scope}[every coordinate/.style={shift={(0,0,0)}}]
            \fill [zpauli, opacity=0.15] ([c]Y)--([c]XY)--([c]XYZ)--([c]YZ)--cycle;
            \draw [zpauli!65, style={decorate, decoration={snake, segment length=0.1cm, amplitude=0.05ex}}] ([c]Y)--([c]XY)--([c]XYZ)--([c]YZ)--cycle;
        \end{scope}
    \end{tikzpicture}
    &
    {Z}_3 = \:
    \begin{tikzpicture}[scale=0.35,tdplot_main_coords,every path/.style={line width=1,line cap=round, line join=round}]
        \coordinate (O) at (-1, -1, -1);
        \coordinate (X) at (1, -1, -1);
        \coordinate (Y) at (-1, 1, -1);
        \coordinate (Z) at (-1, -1, 1);
        \coordinate (XY) at (1, 1, -1);
        \coordinate (YZ) at (-1, 1, 1);
        \coordinate (ZX) at (1, -1, 1);
        \coordinate (XYZ) at (1, 1, 1);
        \begin{scope}[every coordinate/.style={shift={(0,0,0)}}]
            \fill [zpauli, opacity=0.25] ([c]X)--([c]ZX)--([c]Z)--([c]YZ)--([c]XYZ)--([c]XY)--cycle;
            \fill [zpauli, opacity=0.15] ([c]O)--([c]Y)--([c]YZ)--([c]XYZ)--([c]XY)--([c]X)--cycle;
            \fill [zpauli, opacity=0.15] ([c]O)--([c]Y)--([c]YZ)--([c]Z)--cycle;
            \draw [zpauli!65, style={decorate, decoration={snake, segment length=0.1cm, amplitude=0.05ex}}] ([c]O)--([c]X)--([c]ZX)--([c]Z)--cycle;
        \end{scope}
    \end{tikzpicture}
\end{array}
\label{eqn:3D-TC-2form-ops}
\end{equation}
See also Tab.~\ref{tab:summary} for a lattice-independent perspective on these operators. In the operator arrangements~\eqref{eqn:3D-TC-1form-ops} and \eqref{eqn:3D-TC-2form-ops} the averaged victory probability of the corresponding strategy in given some generic resource state $\ket{\psi}$ is still given by Eq.~\eqref{eqn:pq-generic-state} using the appropriate composite operators $\{ X_i, Y_i \}$. The operator product $X_1 Y_2 Y_3$ and permutations thereof can again be interpreted as the twist products of a symmetry operator $X_1 X_2 X_3$ and the order parameter $Z_2 Z_3$. Diagrammatically, we may interpret the the sign $X_1 Y_2 Y_3 = -1$ as arising from the braiding statistics between the pointlike and linelike excitations of the 3D toric code (whether the pointlike excitations are $X$-like or $Z$-like depends on whether we are considering $H_{\text{3TC}}$ or the dual Hamiltonian $H_{\text{3TC}^*}$).

\begin{figure*}[t]
    \centering%
    \subfigure[\label{subfig:fracton-creation}]{%
    \includegraphics[scale=0.75, valign=b]{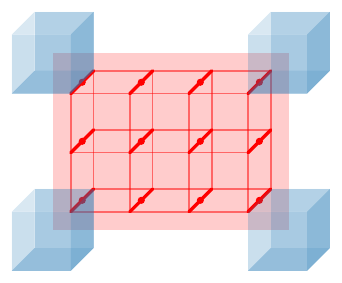}}%
    \hspace{0.075\linewidth}%
    \subfigure[\label{subfig:fracton-lineon-braid}]{%
    \includegraphics[scale=0.75, valign=b]{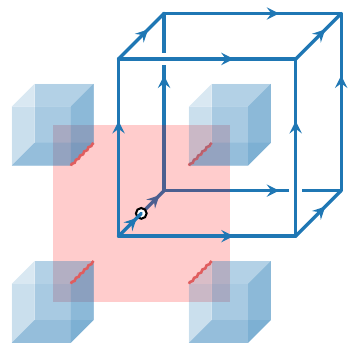}}%
    \hspace{0.075\linewidth}%
    \subfigure[\label{subfig:X-cube-strat}]{%
    \raisebox{0.5cm}{\includegraphics[scale=1.2, valign=b]{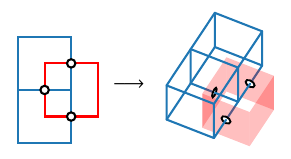}}}%
    \caption{(a) A membrane operator that creates four well-separated fractons (depicted by the shaded blue cubes) at its corners. Bold red edges host an $X_e$ operator, which we depict as a membrane on the dual lattice. (b) The nontrivial ``braiding'' process, which leads to a statistical phase of $\pi$: Three lineons are nucleated at the bottom left corner (say), propagated around the edges of the cube, and eventually annihilate one another at the top right. This process is sensitive to the number of fractons mod 2 contained within the cage. (c) Arrangement of the operators used to win the parity game using X-cube ground states as a resource. These operators may be regarded as a generalization of the operators used to win the 2D toric code game. Each $Z_iZ_j$ with $i \neq j$ forms a cage, while $X_1 X_2 X_3$ forms a $z$-oriented rectangular prism.}
    \label{fig:X-cube}
\end{figure*}


\subsection{X-cube model}
\label{sec:X-cube}

As a second example we show how systems with \emph{fractonic} topological order~\cite{Nandkishore2019fractons} can be utilized as a resource for the parity game, using the X-cube model~\cite{CastelnovoXcube,Vijay2016} as an example. The X-cube model involves qubits on the edges of a three-dimensional $L\times L\times L$ cubic lattice, which interact according to the Hamiltonian
\begin{equation}
    H_\text{XC} = - \sum_{c} A_c - \sum_{v, \mu} B_{v}^{\mu} 
    \, ,
\end{equation}
where $A_c$ is equal to the product of $Z_e$ for all $e$ that bound the cube $c$, and $B_v^{\mu}$ with $\mu \in \{ x, y, z \}$ is equal to the product of $X_e$ over the edges belonging to the dual boundary of the vertex $v$ in the plane perpendicular to $\mu$, i.e., $\{ \partial e \ni v \,|\, e \perp \mu \}$. As in previous sections, we denote the application of the operator $Z_e$ by shading the edge $e$ blue. Meanwhile, the operator $X_e$ is denoted by shading red the dual face that intersects $e$:
\begin{equation}
    Z_e = 
    \tdplotsetmaincoords{75}{30}
    \begin{tikzpicture}[scale=0.5,tdplot_main_coords,baseline={(0, 0.1)}]
        \draw [line width=1., lightgray, opacity=0.5] (-0.75, 0, 0) to (0.75, 0, 0);
        \draw [line width=1., lightgray, opacity=0.5] (0, 0, -0.75) to (0, 0, 0.75);
        \draw [line width=1., lightgray, opacity=0.5] (-0.75, 2, 0) to (0.75, 2, 0);
        \draw [line width=1., lightgray, opacity=0.5] (0, 2, -0.75) to (0, 2, 0.75);
        \draw [line width=1.25, zpauli] (0, 0, 0) to (0, 2, 0);
    \end{tikzpicture}
    \qquad 
    X_e = 
    \begin{tikzpicture}[scale=0.5,tdplot_main_coords,baseline={(0, 0.1)}]
        \draw [line width=1.25, red] (0, 0, 0) to (0, 2, 0);
        \fill [red, fill opacity=0.25] (-1, 1, -1)--++(2, 0, 0)--++(0, 0, 2)--++(-2, 0, 0);
        \begin{scope}[canvas is zx plane at y=1]
            \fill [red] (0, 0) circle [radius=0.75ex];
        \end{scope}
    \end{tikzpicture}
\end{equation}
Using this graphical notation, the stabilizers are expressed as
\begin{equation}
    A_c = \:\:
    \tdplotsetmaincoords{75}{30}
    \begin{tikzpicture}[scale=0.35,tdplot_main_coords,baseline={(0, 1)}]
        \pgfmathsetmacro{\cubex}{2}
        \pgfmathsetmacro{\cubey}{2}
        \pgfmathsetmacro{\cubez}{2}
        \draw[line width=1,zpauli,line cap=round,line join=round] (0,0,0) -- ++(\cubex,0,0) -- ++(0,\cubey,0) -- ++(-\cubex,0,0) -- cycle;
        \draw[line width=1,zpauli,line cap=round,line join=round] (0,0,0) -- ++(0,0,\cubez) -- ++(0,\cubey,0) -- ++(0,0,-\cubez) -- cycle;
        \draw[line width=1,zpauli,line cap=round,line join=round] (0,0,0) -- ++(\cubex,0,0) -- ++(0,0,\cubez) -- ++(-\cubex,0,0) -- cycle;
        \draw[line width=1,zpauli,line cap=round,line join=round] (\cubex,\cubey,\cubez) -- ++(-\cubex,0,0) -- ++(0,-\cubey,0) -- ++(\cubex,0,0) -- cycle;
        \draw[line width=1,zpauli,line cap=round,line join=round] (\cubex,\cubey,\cubez) -- ++(0,0,-\cubez) -- ++(0,-\cubey,0) -- ++(0,0,\cubez) -- cycle;
        \draw[line width=1,zpauli,line cap=round,line join=round] (\cubex,\cubey,\cubez) -- ++(-\cubex,0,0) -- ++(0,0,-\cubez) -- ++(\cubex,0,0) -- cycle;
    \end{tikzpicture}\qquad
    B_v^x = \:\:
    \tdplotsetmaincoords{75}{30}
    \begin{tikzpicture}[scale=0.35,tdplot_main_coords]
        \draw [line width=1.25, red] (-2, 0, 0) to (2, 0, 0);
        \draw [line width=1.25, red] (0, 0, -2) to (0, 0, 2);
        \fill [red, fill opacity=0.25] (-1, -1, -1)--++(0, 2, 0)--++(2, 0, 0)--++(0, -2, 0);
        \fill [red, fill opacity=0.25] (-1, -1, 1)--++(0, 2, 0)--++(2, 0, 0)--++(0, -2, 0);
        \fill [red, fill opacity=0.25] (-1, -1, 1)--++(0, 2, 0)--++(0, 0, -2)--++(0, -2, 0);
        \fill [red, fill opacity=0.25] (1, -1, 1)--++(0, 2, 0)--++(0, 0, -2)--++(0, -2, 0);
        \begin{scope}[canvas is xy plane at z=-1]
            \fill [red] (0, 0) circle [radius=1ex];
        \end{scope}
        \begin{scope}[canvas is xy plane at z=+1]
            \fill [red] (0, 0) circle [radius=1ex];
        \end{scope}
        \begin{scope}[canvas is zy plane at x=+1]
            \fill [red] (0, 0) circle [radius=1ex];
        \end{scope}
        \begin{scope}[canvas is zy plane at x=-1]
            \fill [red] (0, 0) circle [radius=1ex];
        \end{scope}
    \end{tikzpicture} 
\end{equation}
Hence, products of adjacent $A_c$ operators form a cage that bounds an orthogonal polyhedron, and products of adjacent $B_v^\mu$ (with the same orientation $\mu$) form rectilinear polygonal prisms.
As in the toric code, all operators $A_c$ and $B_v$ mutually commute and square to the identity, implying that the ground states of the model satisfy $A_c = B_v = +1$. On an $L\times L \times L$ lattice with periodic boundaries the model has an exponentially large ground-state degeneracy $D$ that scales as $\log_2 D = 6L-3$~\cite{Vijay2016}.

The model has multiple types of excitations~\cite{Vijay2016}. The ones we are interested in are the \emph{fractons}, which live on cubes satisfying $A_c=-1$, and the \emph{lineons}, which correspond to, e.g., $B_v^{x} = B_v^{y} = -1$ and $B_v^{z}=+1$. A single $X_e$ operator creates four adjacent fractons on the cubes touching the edge $e$, and four well-separated fractons can be created at the corners of a rectangular membrane of $X_e$ operators on the dual lattice, as illustrated in Fig.~\ref{subfig:fracton-creation}. Isolated fractons are immobile since they cannot move without creating additional fractons. A single $Z_e$ operator on a $z$-oriented edge creates two neighboring lineons, which may be separated along the $z$ axis by acting with $Z_e$ along a $z$-oriented string. Turning a corner requires the creation of additional lineons so lineon excitations reside at the end of rigid strings.

We now utilize the nontrivial braiding properties of these excitations~\cite{Ma2017Coupled} to construct a perfect strategy for the parity game. A nontrivial statistical phase is picked up when braiding lineons around an isolated fracton: Nucleating three lineons and propagating them around the edges of a cube (nucleating or absorbing lineons at the corners) leads to a statistical phase factor $(-1)^{n_f}$, where $n_f$ is the number of enclosed fractons, as depicted in Fig.~\ref{subfig:fracton-lineon-braid}. Following the 2D and 3D toric code examples, operators that give rise to a perfect strategy may be identified by partitioning the $X$-like symmetry into three disjoint regions. Specifically, we consider a rectangular prism and partition its faces into three disjoint sets, which ensures that $X_1 X_2 X_3 = 1$. Second, we construct operators $Z_i$ such that $Z_1Z_2$ and $Z_2Z_3$ (and their product) correspond to cages, and $Z_1$, $Z_2$, and $Z_3$ are all charged under the symmetry $X_1X_2X_3$. Operators satisfying these conditions are, e.g.,
\vspace{-15pt}
\begin{equation}
        \arraycolsep=5pt\def\arraystretch{4}
        \begin{array}{ccc}
            Z_1 = 
        \tdplotsetmaincoords{75}{30-90}
        \begin{tikzpicture}[scale=0.45,tdplot_main_coords,baseline={(0, 1)}]
        \pgfmathsetmacro{\cubex}{1.5}
        \pgfmathsetmacro{\cubey}{1.5}
        \pgfmathsetmacro{\cubez}{1.5}
        \draw[line width=1,zpauli,line cap=round,line join=round] (0,0,0) -- ++(0,0,\cubez) -- ++(0,\cubey,0) -- ++(0,0,-\cubez) -- cycle;
        \draw[line width=1,zpauli,line cap=round] (0,0,0) -- ++(\cubex,0,0);
        \draw[line width=1,zpauli,line cap=round] (0,\cubey,0) -- ++(\cubex,0,0);
        \draw[line width=1,zpauli,line cap=round] (0,0,\cubez) -- ++(\cubex,0,0);
        \draw[line width=1,zpauli,line cap=round] (0,\cubey,\cubez) -- ++(\cubex,0,0);
    \end{tikzpicture}
    &
    Z_2 = 
    \tdplotsetmaincoords{75}{30-90}
    \begin{tikzpicture}[scale=0.45,tdplot_main_coords,baseline={(0, 1)}]
        \pgfmathsetmacro{\cubex}{1.5}
        \pgfmathsetmacro{\cubey}{1.5}
        \pgfmathsetmacro{\cubez}{1.5}
        \draw[line width=1,zpauli] (0,0,0) -- ++(0,0,\cubez) -- ++(0,\cubey,0) -- ++(0,0,-\cubez) -- cycle;
    \end{tikzpicture}
    &
    Z_3 = 
    \tdplotsetmaincoords{75}{30-90}
    \begin{tikzpicture}[scale=0.45,tdplot_main_coords,baseline={(0, 1)}]
        \pgfmathsetmacro{\cubex}{1.5}
        \pgfmathsetmacro{\cubey}{1.5}
        \pgfmathsetmacro{\cubez}{1.5}
        \draw[line width=1,zpauli,line cap=round,line join=round] (\cubex,\cubey,\cubez) -- ++(0,0,-\cubez) -- ++(0,-\cubey,0) -- ++(0,0,\cubez) -- cycle;
        \draw[line width=1,zpauli,line cap=round] (0,0,0) -- ++(\cubex,0,0);
        \draw[line width=1,zpauli,line cap=round] (0,\cubey,0) -- ++(\cubex,0,0);
        \draw[line width=1,zpauli,line cap=round] (0,0,\cubez) -- ++(\cubex,0,0);
        \draw[line width=1,zpauli,line cap=round] (0,\cubey,\cubez) -- ++(\cubex,0,0);
    \end{tikzpicture}\\
    X_1=
    \tdplotsetmaincoords{75}{30-90}
    \begin{tikzpicture}[scale=0.45,tdplot_main_coords,baseline={(0, 0)}]
        \fill [fill=red,fill opacity=0.25] (-1, -1, -1)--++(0, 2, 0)--++(0, 0, 2)--++(0, -2, 0)--cycle;
        \foreach \z in {0, 1, ..., 4} {
            \draw [thick, lightgray, opacity=0.5] (-1, -1, {-1+0.5*\z})--++(2, 0, 0)--++(0, 2, 0)--++(-2, 0, 0)--cycle;
        }
    \end{tikzpicture} 
    &
    X_2=
    \tdplotsetmaincoords{75}{30-90}
    \begin{tikzpicture}[scale=0.45,tdplot_main_coords,baseline={(0, 0)}]
        \fill [fill=red,fill opacity=0.25] (-1, -1, -1)--++(2, 0, 0)--++(0, 0, 2)--++(-2, 0, 0)--cycle;
        \fill [fill=red,fill opacity=0.25] (-1, 1, -1)--++(2, 0, 0)--++(0, 0, 2)--++(-2, 0, 0)--cycle;
        \foreach \z in {0, 1, ..., 4} {
            \draw [thick, lightgray, opacity=0.5] (-1, -1, {-1+0.5*\z})--++(2, 0, 0)--++(0, 2, 0)--++(-2, 0, 0)--cycle;
        }
    \end{tikzpicture} 
    &
    X_3=
    \tdplotsetmaincoords{75}{30-90}
    \begin{tikzpicture}[scale=0.45,tdplot_main_coords,baseline={(0, 0)}]
        \fill [fill=red,fill opacity=0.25] (1, -1, -1)--++(0, 2, 0)--++(0, 0, 2)--++(0, -2, 0)--cycle;
        \foreach \z in {0, 1, ..., 4} {
            \draw [thick, lightgray, opacity=0.5] (-1, -1, {-1+0.5*\z})--++(2, 0, 0)--++(0, 2, 0)--++(-2, 0, 0)--cycle;
        }
    \end{tikzpicture} 
    \end{array}
\end{equation}
See Fig.~\ref{subfig:X-cube-strat} for the relative positioning of these operators. Since these operators satisfy the constraints~\eqref{eqn:measured-loops}, they provide a perfect strategy for the parity game. Note that any of the three prism orientations could have been utilized to construct a perfect strategy. Second, we could consider a dual X-cube model in which the $X$-like symmetry operators are cages (see Tab.~\ref{tab:summary} for the corresponding strategy). The interpretation in all cases is identical: Operators such as $X_1 Y_2 Y_3$ correspond to the twist product of stabilizer operators $A_c$ and $B_v^\mu$. The nontrivial phase picked up by $\expval{A_c \infty B_v^\mu}$ can then be interpreted in terms of the lineon-fracton braiding process.


\section{Beyond the parity game}
\label{sec:homological-games}

We have seen that the parity game can be won using a variety of topologically ordered states as resources by making use of the braiding processes of their excitations.
In all cases, the perfect quantum strategy for the $P$-player game involved the identification of a set of composite operators (order and disorder parameters, as discussed in Sec.~\ref{sec:order-disorder}) whose measurement statistics are identical to those of single-site Pauli operators in a $P$-qubit GHZ state. 
One might reasonably wonder what other families of states can naturally be encoded (in the sense above) into a given topologically ordered state, such as the toric code, and hence which nonlocal quantum games they function as resource states for.

In this section, we first answer the question above in the context of the 2D toric code by explicitly constructing the family of $N$-qubit stabilizer states whose Pauli measurement statistics are encoded in composite linelike operators. Then, we further generalize to homological codes in $d$ spatial dimensions and identify generalizations of the parity game for which the codewords of such quantum codes act as perfect resources.


\subsection{Embedding other states in two dimensions}
\label{sec:other-states}

Consider an $N$-qubit \emph{stabilizer state} $\ket{\psi}$. Such a state is the simultaneous eigenstate of all operators in the stabilizer group $\mathcal{S} = \langle \{ S_k \} \rangle \subset \mathcal{P}$, a subgroup of the Pauli group on $N$ qubits.
The subgroup is generated by Pauli strings $S_k$, which satisfy $S_k \ket{\psi} = \ket{\psi}$, and we have $\abs{\mathcal{S}} = 2^N$.
In the following, we ask what family of such stabilizer states can be embedded in toric code ground states on composite, linelike degrees of freedom.

We seek $N$ pairs of operators, $\{ X_i, Z_i \}_{i=1}^N$, satisfying the following criteria:
\begin{enumerate*}[label=(\alph*)]
    \item the composite operators furnish a representation of the Pauli algebra on $N$ qubits,
    \item \label{item:closed-cycles} every $X$ ($Z$) loop segment must be part of a \emph{cycle} of $X$ ($Z$) operators, 
    \item \label{item:unique-state} the total number of \emph{independent} cycles of $X$ or $Z$ type must equal $N$.
\end{enumerate*}
Each independent cycle gives rise to an independent stabilizer constraint such as $X_{i_1} \cdots X_{i_k} = 1$. Together, these conditions then ensure that the measurement statistics of the composite operators $\{ X_i, Z_i \}$ will be identical to single-site Pauli operators in the uniquely specified $N$-qubit state stabilized by $\mathcal{S}$.

\begin{figure}[t]
    \subfigure[\label{subfig:generic-graph}]{%
    \includegraphics[width=0.495\linewidth]{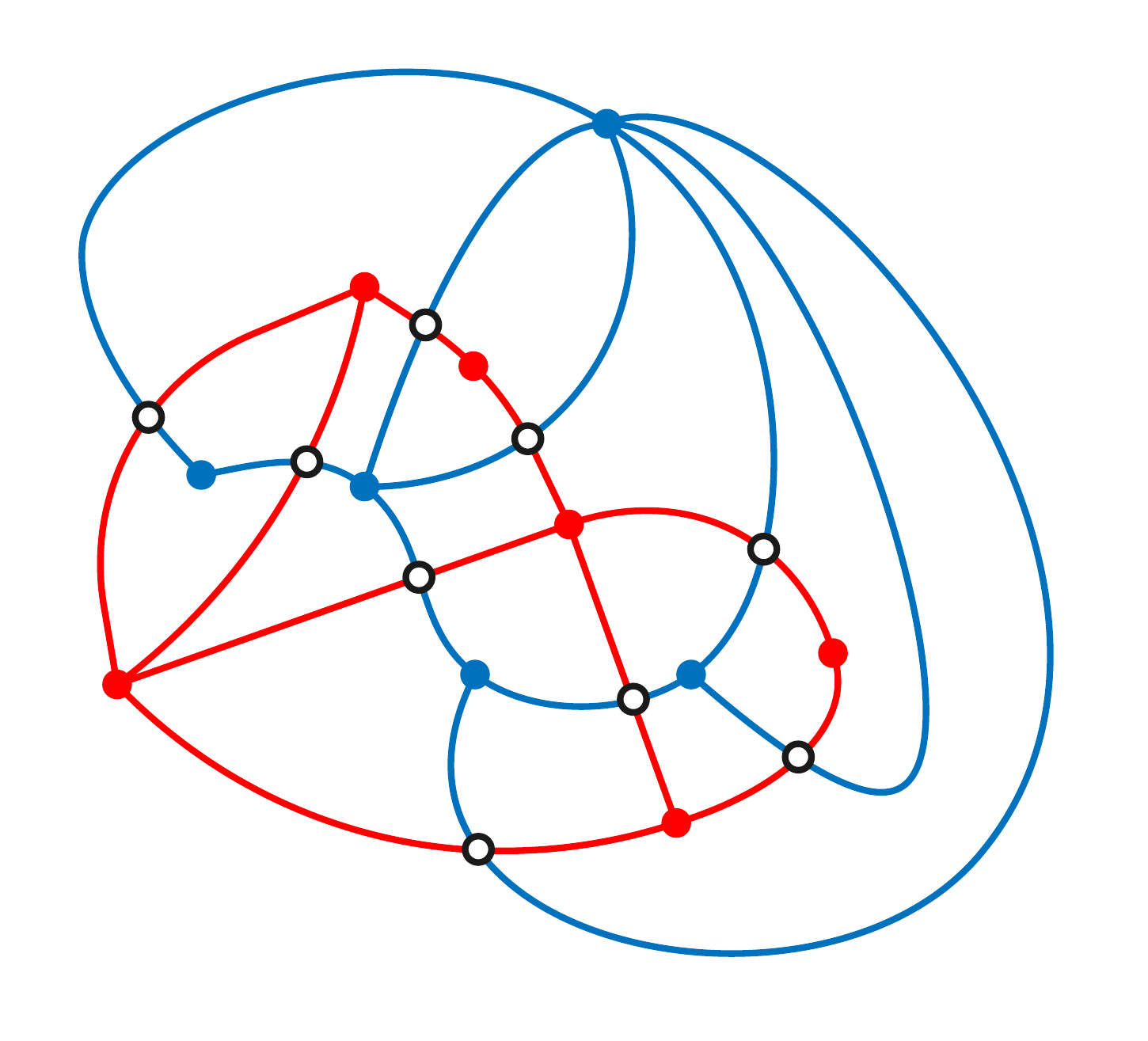}}%
    \subfigure[\label{subfig:wheel-graph}]{%
    \includegraphics[width=0.495\linewidth]{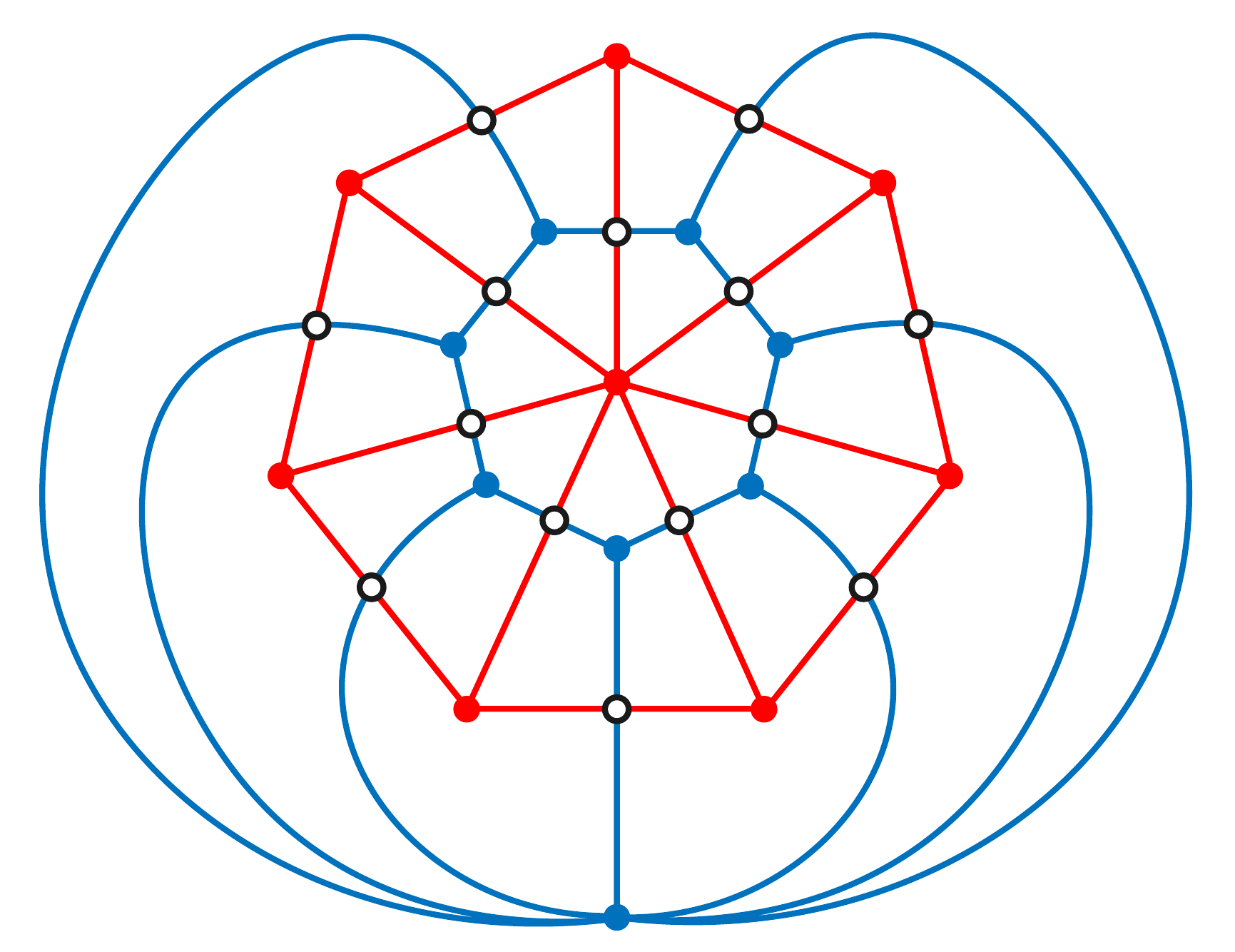}}
    \caption{(a) An effective nine-qubit state defined by a graph. The blue and red graphs are both planar graphs and are dual to each other. The effective $X_i$ and $Z_i$ operators live on the edges of the primary and dual lattices, respectively, and hence intersect exactly once. (b) The self-dual \emph{wheel graph}. A wheel graph built from a regular polygon with $k$ edges corresponds to an embedding of a $2k$-qubit 1D cluster state with periodic boundaries, up to local unitary transformations.}
    \label{fig:graph-state}
\end{figure}

These constraints are satisfied by first constructing a plane graph $G$ (i.e., a planar embedding of a planar graph) for the $X$ loop segments, say, where the edges of $G$ are identified with the collective operators $X_i$~\footnote{We consider graphs $G$ where $G$ and its geometric dual $G^*$ are \emph{loopless}, i.e., they contain no self-loops. A self-loop corresponds to a trivial, disentangled qubit, since either $X_i$ or $Z_i$ belongs to the stabilizer group for some $i$ corresponding to the self-loop.}. If the graph $G$ has $(V, E, F)$ vertices, edges, and faces, respectively, then the $X$ graph encodes $F-1$ $X$-like stabilizers from the graph's bounded faces. The $Z$ loop segments are given by the edges of the geometric dual graph $G^*$. By definition, each edge in $G$ intersects the corresponding dual edge in $G^*$ only, implying that $\{ X_i , Z_i\}$ obey the required Pauli commutation relations. Furthermore, by Euler's formula, we have $F+F^*-2 = N$, where $F^*=V$ is the number of dual faces, implying that the number of independent constraints on the operators $\{ X_i, Z_i\}$ is equal to $N$, such that the embedded state is uniquely specified by the constraints. An example generic arrangement of $X$ and $Z$ loop segments in shown in Fig.~\ref{subfig:generic-graph}. 

The strategy used to win the parity game introduced in Sec.~\ref{sec:TC-star} fits into the above construction by embedding the measurement statistics of single-site Pauli operators in a GHZ state into composite operators with respect to a 2D toric code ground state. From the graph-theoretic perspective, this follows from the duality of cycle and dipole graphs (see Fig.~\ref{fig:p_game}). Similarly, the measurement statistics of Pauli operators in one-dimensional cluster states with periodic boundary conditions correspond to the self-dual wheel graph [see Fig.~\ref{subfig:wheel-graph}]. The measurement statistics encoded therein can be used to win the cubic boolean function games from Ref.~\cite{Daniel2022Exp} (up to a linear transformation of the computed boolean function). In the next subsection we construct systematically a family of games that codewords can be used to win by making single-site Pauli measurements, which includes the cubic boolean function game as an example. More generally, the effective states that can be embedded on composite degrees of freedom may be regarded as the ground state of a toric-code Hamiltonian on a generic cellulation of the 2-sphere $S_2$, since every plane graph is embeddable on $S_2$ via stereographic projection. Because $S_2$ is topologically trivial, i.e., $\pi_1(S_2)=0$, the ground state of this Hamiltonian is unique~\cite{Dennis2002}, as required to exactly reproduce the measurement statistics of single-site Pauli operators in a specified state of $N$ particles.

For reasons of generality we have focused on a lattice-independent perspective. The results above apply irrespective of the underlying microscopic lattice (crystalline or not) on which the toric code Hamiltonian is defined. To implement the above operators on a given microscopic lattice, however, it may be the case that the continuum graph contains vertices of higher valency than the microscopic lattice (or dual lattice). In this case, it is necessary to ``block'' vertices to obtain larger valencies. Physically, if this is necessary, it means that the composite operators $\{ X_i \}$ or $\{ Z_i \}$ must necessarily have overlapping support in the microscopic lattice implementation.

The construction above examines which states can be embedded in \emph{arbitrary} ground states of the toric code. If we work instead with a specific ground state, we can also make use of the noncontractible cycles of the underlying manifold to obtain operator constraints. For instance, on the torus, $X_1 X_2 X_3$ may form a noncontractible loop, while each $Z_i$ winds around the torus in the other direction. In the appropriate ground state(s), the operators $\{ X_i, Y_i \}$ will obey three-qubit-GHZ-like measurement statistics. Thus, the above construction reduces to the ``local'' strategy of Ref.~\cite{hart2024playing} when elementary cycles are used to define the operators $\{ X_i, Y_i \}$, and provides an alternative strategy to the one presented in Ref.~\cite{BBSgame} when noncontractible cycles are used~\footnote{Unlike Ref.~\cite{BBSgame}, the strategy presented herein does not require the application of a nonlocal unitary operator, but gives rise to the same $p_\text{q}(\ket{\psi})$ given some generic resource state $\ket{\psi}$.}.


\subsection{Homological codes and cellulation games}

In the preceding subsection we discussed the set of states that could be embedded into 2D toric code ground states. We now generalize in two directions, firstly 
\begin{enumerate*}[label=(\roman*)]
    \item by extending from two to $d$ spatial dimensions, and
    \item extending to quantum games beyond the parity game. 
\end{enumerate*}
The objective of this section is to clarify which quantum games permit perfect strategies using single-site Pauli measurements on a resource state that is the codeword of a $d$-dimensional homological code~\cite{freedman2001projective,Dennis2002,Kitaev2003,Bombin2006}.

\subsubsection{More spatial dimensions}

Moving from two to $d$ spatial dimensions, we now consider a \emph{cellulation} of $\mathbb{R}^d$ (see, e.g., Ref.~\cite{BreuckmannThesis}). Specifically, we consider the cellulation $\mathcal{X}$, which consists of $0$-cells (vertices), $1$-cells (edges), $2$-cells (faces), and $3$-cells (volumes), etc., and its dual $\mathcal{X}^*$, where dual $i$-cells are in one-to-one correspondence with primary $(d-i)$-cells. Next, introduce the $\mathbb{Z}_2$ vector spaces $C_i$ spanned by, e.g., sets of $i$-cells that contain a single $i$-cell, and the boundary maps between them $\partial_i : C_i \to C_{i-1}$, satisfying $\partial_{i}\partial_{i+1}=0$. For an $i$-cell $c_i \in C_i$, the boundary operator $\partial_i$ returns all $(i-1)$-cells incident to $c_i$. Similarly, we introduce the coboundary operator $\delta_i : C_{i} \to C_{i+1}$. The coboundary operator is related to the boundary maps via $\delta_{i} = \partial_{i+1}^T$, and, for $c_i \in C_i$, returns the $(i+1)$-cells incident to $c_i$.

We fix a dimension $p \in \{1, \dots, d-1 \}$, which corresponds to working with states possessing a $(d-p)$-form $X$ symmetry. We imagine that, microscopically, we are working with a homological code such that $X$ stabilizers correspond to the boundaries of $(p+1)$-cells, while $Z$ stabilizers correspond to the coboundaries of $(p-1)$-cells [equivalently, to the boundaries of dual $(d-p+1)$-cells]. Correspondingly, $X$ operators associated to the boundary space $B_p = \im \partial_{p+1}$ and $Z$ operators belonging to the coboundary space $B^p = \im \delta_{p-1}$ evaluate to unity with respect to codewords (i.e., states that satisfy all microscopic $X$ and $Z$ stabilizers). Next, we associate a composite operator $X_i$ to each $p$-cell in $C_p$, and the composite operator $Z_i$ to the corresponding dual $(d-p)$-cell in $C_{\smash d-p}^*$. This ensures that $X_i$ and $Z_i$, which are constructed by gluing together microscopic $p$-cells and dual $(d-p)$-cells, respectively, intersect exactly once and hence anticommute, as required. We now show that the total number of \emph{independent} $X$ stabilizers and $Z$ stabilizers equals $N$. Specifically, using the rank-nullity theorem,
\begin{subequations}
\begin{align}
    \dim B_p &= \sum_{i=1}^{d-p} {(-1)}^{1+i} \left( \dim C_{p+i} - \dim H_{i+p} \right) \\
    \dim B^p &= \sum_{i=1}^{p} {(-1)}^{1+i} \left( \dim C_{p-i} - \dim H^{p-i} \right)
\end{align}%
\label{eqn:stabilizer-dimensions}%
\end{subequations}
where $H_i = \ker \partial_i /\im \partial_{i+1}$ ($H^i = \ker \delta_i /\im \delta_{i-1}$) are the homology (cohomology) groups, with $\partial_{d+1}$ defined to be the zero map. In the context of the 3D toric code, say, Eq.~\eqref{eqn:stabilizer-dimensions} captures the local constraints obeyed by stabilizer generators, such as $\prod_{\partial e \ni v} A_e = \id$ in Eq.~\eqref{eqn:3D-TC-faces}, in addition to the homology of the manifold. It thus follows that the total number of independent stabilizers satisfies $\dim B_p + \dim B^p = N - \dim H_p$ since $\sum_{i=0}^d (-1)^i \dim C_i = \sum_{i=0}^d (-1)^i \dim H_i = \chi(M)$, the Euler characteristic of the manifold $M$ being cellulated.

Since we consider a cellulation of $\mathbb{R}^d$, the number of constraints on the composite operators $\{ X_i, Z_i \}$ is equal to $N$, and the measurement statistics of the composite operators will be identical to single-site Pauli operators in a \emph{uniquely specified} $N$-qubit state. Note that we assume that the cell complex includes an unbounded $d$-cell such that the cell complex is homotopy equivalent to the $d$-sphere, $S_d$, with Euler characteristic $\chi(S_d) = 1+ (-1)^d$. In summary, given a codeword of some underlying homological code in $d$ spatial dimensions, each cellulation $\mathcal{X}$ of $\mathbb{R}^d$ leads to the construction of $2N$ composite operators $X_i$ and $Z_i$ whose measurement statistics with respect to a codeword are identical to Pauli $X$ and $Z$ in an $N$-qubit state uniquely determined by the chosen cellulation $\mathcal{X}$. The stabilizer group of the embedded state is generated by the boundary and coboundary spaces of $\mathcal{X}$. Minimal examples that correspond to an embedding of the three-qubit GHZ state in the 3D toric code are given in~\eqref{eqn:3D-TC-1form-ops} and \eqref{eqn:3D-TC-2form-ops} (see also Tab.~\ref{tab:summary}). 


\subsubsection{Cellulation games}

Given such a cellulation $\mathcal{X}$, we introduce a game for which codewords of the underlying homological code provide a perfect quantum strategy. Associate players to $p$-cells of $\mathcal{X}$, and classical bits $\vec{x} \in \{ 0, 1 \}^N$ to a fixed basis for $B_p$ and $B^p$~\footnote{The requirement that the bits are associated to a linearly independent set of stabilizer generators can straightforwardly be relaxed. If additional stabilizer generators are included, the map from $\vec{x}$ to bits $\vec{a}$ and $\vec{b}$ is surjective but not injective.}. For example, consisting of boundaries of $(p+1)$-cells and coboundaries of $(p-1)$-cells, taking into account their linear dependence~\eqref{eqn:stabilizer-dimensions}. Each player is then handed two bits
\begin{equation}
    a_{c_p} = \bigoplus_{\delta c_{p-1} \ni c_p} x_{c_{p-1}},  \quad 
    b_{c_p} = \bigoplus_{\partial c_{p+1} \ni c_p} x_{c_{p+1}}
    \, ,
    \label{eqn:homological-game-input}
\end{equation}
namely the sum mod 2 of the bits associated to stabilizers that are incident to the $p$-cell $c_p$. For notational convenience we include all stabilizer generators in~\eqref{eqn:homological-game-input} and assign $x=0$ to those that do not belong to the chosen basis. Denoting the set of $p$-cells by $\{ c_p \}$, the players output a bit $y_{c}$ and are said to win the game if their outputs satisfy
\begin{equation}
    \sum_{c \in \{ c_p \}} y_{c}(a_{c}, b_{c}) \equiv \frac{1}{2} \sum_{c \in \{ c_p \}} a_{c} b_{c} \mod 2
    \, .
\end{equation}
That $\sum_{c \in \{c_p\}} a_{c} b_{c}$ is even follows from the commutation of the stabilizers. The quantum-mechanical strategy in which the player associated to $c \in \{ c_p \}$ measures the operator $P_{c}$ then wins for all choices of inputs to the game, where $P_{c}(a_{c}, b_{c})$ corresponds to the composite Hermitian operator
\begin{equation}
    P_{c}(a, b) \coloneq i^{a b} X_{c}^{a} Z^{b}_{c}
    \, .
    \label{eqn:generalized-Pauli-meas}
\end{equation}
Such a strategy is perfect since the players are effectively measuring the stabilizers of the underlying codeword up to a phase $\pm 1$. This phase encodes precisely the information needed to win the game since
\begin{equation}
     \prod_{c \in \{ c_p \}} P_{c}(a_{c}, b_{c}) = \exp\left(\frac{i\pi}{2} \sum_{c \in \{ c_p \}} a_{c} b_{c} \right) \prod_k S_k^{x_k}
     \, ,
\end{equation}
where $k$ runs over the chosen basis for $B_p$ and $B^p$.

Given a generic resource state $\ket{\psi}$, the strategy in which the player associated to $c_p$ measures the operator~\eqref{eqn:generalized-Pauli-meas} will win with probability given by
\begin{equation}
    p_\text{q} = \frac12 + \frac{1}{2^{N+1}}
    \sum_{\vec{x}} \exp\left(\frac{i\pi}{2} \sum_{c} a_{c} b_{c} \right) \braket{ \psi | \prod_{c} P_{c} | \psi }
    \label{eqn:pq-cellulation}
\end{equation}
averaged over all inputs to the game $\vec{x}$ with equal weight, where we suppressed the dependence of $P_{c}$ on the bits $a_c$, $b_c$~\eqref{eqn:homological-game-input} handed handed to each player (which themselves are functions of the inputs $\vec{x}$). The sum over $\vec{x}$ gives rise to the expectation value of a generalized Mermin polynomial~\cite{MerminPolynomials} of the measured operators. Note that we recover the parity game by choosing the cellulation $\mathcal{X}$ as in Tab.~\ref{tab:summary} \emph{and} restricting the inputs to the game such that $a_{c} = 1$. Furthermore, if the cellulation $\mathcal{X}$ is chosen to be equal to that of the underlying microscopic lattice then~\eqref{eqn:pq-cellulation} evaluates to $( 1 + \Tr[ \rho \Pi_0 ] )/2$, where $\rho = \ket{\psi}\!\bra{\psi}$ and $\Pi_0$ is the projector onto the codespace.

This discussion clarifies the scope of games that can be won using single-site Pauli measurements on a $d$-dimensional homological code. In all cases, the operator $\prod_c P_c$ appearing in Eq.~\eqref{eqn:pq-cellulation} may be interpreted as a braiding diagram between the model's excitations [as in Eq.~\eqref{eqn:mutual-stats}] that gives rise to the required sign and results in a perfect quantum strategy, $p_\text{q}=1$.


\section{Beyond mutual statistics: double semion model}
\label{sec:self-statistics}

Finally, we construct a game for which the double-semion (DS) model~\cite{LevinWen2005} ground states act as a resource. This procedure illustrates how anyons with nontrivial \emph{self}-statistics can also be used to gain quantum advantage. Specifically, we show that the braiding statistics of the DS phase can most naturally be used to win a slight generalization of another prototypical nonlocal quantum game known as the \emph{magic-square game}~\cite{aravind2002bell,aravind2003simple,Mermin1990simple}.

\begin{figure}[t]
    \centering
    \includegraphics[width=0.8\linewidth]{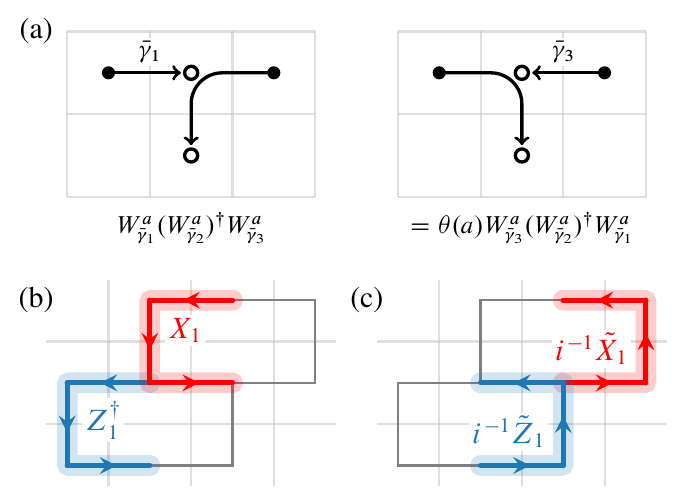}
    \caption{(a) Microscopic process used to extract the exchange statistics of two $a$ anyons. The statistical phase $\theta(a)$ is defined by the relative phase of the two processes. (b) Microscopic definition of the operators used to win the generalized magic square game. Each dual loop segment is corresponds to the application of shift and phase operators according to the definitions in Eq.~\eqref{eqn:W_s}.} 
    \label{fig:DS-operators}
\end{figure}

A stabilizer Hamiltonian for the double-semion phase~\cite{Ellison2022} is defined on the square lattice with four-dimensional qudits on edges. We use the notation $X_e$, $Z_e$ for the shift and phase operators on edge $e$, which satisfy the $Z_eX_e = iX_eZ_e$, and label basis states $\ket{q}$ with $q \in \{ 0, \dots, 3 \}$ such that $Z_e\ket{q} = i^q \ket{q}$. The Hamiltonian is
\begin{equation}
    H_\text{DS} = -\sum_v A_v - \sum_p B_p - \sum_e C_e + \text{H.c.}
    \, ,
    \label{eqn:H_DS}
\end{equation}
where the unitary operators $A_v$, $B_p$, and $C_e$ are defined on vertices $v$, plaquettes $p$, and edges $e$ of the lattice, respectively. Graphically, they assume the form 
\begin{subequations}
\begin{gather}
    A_v = 
    \begin{tikzpicture}[x=3ex,y=3ex,every node/.style={inner sep=0.9,outer sep=0, fill=white, font=\small}]
        \draw [line width=1., lightgray, opacity=0.5] (-2, 0) to (2, 0) to (2, 2) to (0, 2) to (0, -2);
        \node at (-1, 0) {$\color{xpauli}X$};
        \node at (1, 0) {${\color{xpauli}X^\dagger} {\color{zpauli}Z}$};
        \node at (0, -1) {$\color{xpauli}X$};
        \node at (0, 1) {$ {\color{zpauli}Z^\dagger}{\color{xpauli}X^\dagger}$};
        \node at (1, 2) {$\color{zpauli}Z^\dagger$};
        \node at (2, 1) {$\color{zpauli}Z$};
    \end{tikzpicture}
    \qquad 
    B_p = 
    \begin{tikzpicture}[x=3ex,y=3ex,every node/.style={inner sep=0.9,outer sep=0, fill=white, font=\small}]
        \draw [line width=1., lightgray, opacity=0.5] (-1, -1) rectangle (1, 1);
        \node at (-1, 0) {$\color{zpauli}Z^2$};
        \node at (1, 0) {$\color{zpauli}Z^2$};
        \node at (0, 1) {$\color{zpauli}Z^2$};
        \node at (0, -1) {$\color{zpauli}Z^2$};
    \end{tikzpicture}
    \\
    C_e = 
    \begin{tikzpicture}[x=3ex,y=3ex,every node/.style={inner sep=0.9,outer sep=0, fill=white, font=\small}]
        \draw [line width=1., lightgray, opacity=0.5] (-2, -1) to (-2, 1) to (0, 1);
        \node at (-1, 1) {$\color{xpauli}X_e^2$};
        \node at (-2, 0) {$\color{zpauli}Z^2$};
    \end{tikzpicture}
    \:\:\, , \:\:
    \begin{tikzpicture}[x=3ex,y=3ex,every node/.style={inner sep=0.9,outer sep=0, fill=white, font=\small}]
        \draw [line width=1., lightgray, opacity=0.5] (-2, -1) to (0, -1) to (0, 1);
        \node at (0, 0) {$\color{xpauli}X_e^2$};
        \node at (-1, -1) {$\color{zpauli}Z^2$};
    \end{tikzpicture}
\end{gather}%
\label{eqn:DS-stabilizers}%
\end{subequations}
where the light gray lines denote the direct lattice. Note that the definition of $C_e$ depends on the orientation of the edge $e$ (the location of $X_e^2$ defines $e$).
The stabilizers $A_v$, $B_p$, and $C_e$ mutually commute and the ground states of the model~\eqref{eqn:H_DS} belong to the mutual $+1$ eigenspace of these operators. On a torus, the ground state has degeneracy four~\cite{Ellison2022}.
Just like the $\mathbb{Z}_2$ toric code phase, the anyons of the DS phase, which we label as $\{1, s, \bar{s}, s\bar{s} \}$, form a $\mathbb{Z}_2 \times \mathbb{Z}_2$ group under fusion. Their exchange statistics are
\begin{equation}
    \theta(1) = 1, \quad \theta(s) = i, \quad
    \theta(\bar{s}) = -i, \quad \theta(s\bar{s}) = 1
    \, ,
\end{equation}
where $\theta(a)$ is the phase accumulated when two identical $a$ anyons are exchanged.
The anyons are created by open string operators that fail to commute with the operators~\eqref{eqn:DS-stabilizers} at their endpoints. These operators can be decomposed over edges as
\begin{equation}
    W^{s}_{\gamma^*} = \prod_{e^* \in \gamma^*} W^{s}_{e^*} , \:\:
    W^{\bar{s}}_{\gamma^*} = \prod_{e^* \in \gamma^*} W^{\bar{s}}_{e^*}, \:\:
    W^{s\bar{s}}_{\gamma} = \prod_{e \in \gamma} W^{s\bar{s}}_e
    \, ,
\end{equation}
where $\gamma$ ($\gamma^*$) denotes a directed path on the direct (dual) lattice. The ordering of the edges along the path determines the overall phase of the operator; if the path $\gamma^* = (e_1^*, e_2^*, \ldots, e_k^*)$ then $W_{\smash{\gamma^*}}^{\smash{s}} =
W_{\smash{e_k^*}}^{\smash{s}} \cdots W_{\smash{e_2^*}}^{\smash{s}} W_{\smash{e_1^*}}^{\smash{s}}$. Note that $ W^{\smash{s}}_{\smash{\delta v}} \propto A_v $ or $A_v^\dagger$, with $\delta v$ the dual boundary of $v$, depending on whether the vertex $v$ is traversed counterclockwise or clockwise, respectively.
Explicitly, the string operator that creates $s$ anyons is built from
\begin{equation}
    W_{e^*}^s \equiv 
    \begin{tikzpicture}[x=3ex,y=3ex,every node/.style={inner sep=0.9,outer sep=0, fill=white, font=\small}]
        \draw [line width=1., lightgray, opacity=0.5] (-2, -1) to (-2, 1) to (0, 1);
        \draw [line width=1.] (-3, 0) to (-1, 0);
        \draw [->, >=Stealth, line width=0.6] (-2, 0) to (-0.8, 0);
        \node at (-1, 1) {$\color{zpauli}Z$};
        \node at (-2, 0) {$\color{xpauli}X_e$};
    \end{tikzpicture},
    \:\:
    \begin{tikzpicture}[x=3ex,y=3ex,every node/.style={inner sep=0.9,outer sep=0, fill=white, font=\small}]
        \draw [line width=1., lightgray, opacity=0.5] (-2, -1) to (-2, 1) to (0, 1);
        \draw [line width=1.] (-1, 0) to (-3, 0);
        \draw [->, >=Stealth, line width=0.6] (-2, 0) to (-3.2, 0);
        \node at (-1, 1) {$\color{zpauli}Z^\dagger$};
        \node at (-2, 0) {$\color{xpauli}X_e^\dagger$};
    \end{tikzpicture},
    \:\:
    \begin{tikzpicture}[x=3ex,y=3ex,every node/.style={inner sep=0.9,outer sep=0, fill=white, font=\small},baseline={(0,-1.45)}]
        \draw [line width=1., lightgray, opacity=0.5] (-2, -1) to (0, -1) to (0, 1);
        \draw [line width=1.] (-1, 0) to (-1, -2);
        \draw [->, >=Stealth, line width=0.6] (-1, -1) to (-1, -2.2);
        \node at (0, 0) {$\color{zpauli}Z^\dagger$};
        \node at (-1, -1) {$\color{xpauli}X_e$};
    \end{tikzpicture},
    \:\:
    \begin{tikzpicture}[x=3ex,y=3ex,every node/.style={inner sep=0.9,outer sep=0, fill=white, font=\small},baseline={(0,-1.45)}]
        \draw [line width=1., lightgray, opacity=0.5] (-2, -1) to (0, -1) to (0, 1);
        \draw [line width=1.] (-1, -2) to (-1, 0);
        \draw [->, >=Stealth, line width=0.6] (-1, -1) to (-1, 0.2);
        \node at (0, 0) {$\color{zpauli}Z$};
        \node at (-1, -1) {$\color{xpauli}X_e^\dagger$};
    \end{tikzpicture}.
    \label{eqn:W_s}
\end{equation}
where the black arrow denotes the orientation of the dual edge $e^*$.
The expressions for $W_{\smash{e^*}}^{\smash{\bar{s}}}$ are given by interchanging $X_e \leftrightarrow X_e^{\dagger}$ in Eq.~\eqref{eqn:W_s}. The self-statistics of $s$ anyons can be extracted using relative phase of the two processes illustrated in Fig.~\ref{fig:DS-operators}~\cite{Kawagoe2020microscopic}. Microscopically, the nontrivial self-statistics arise from the noncommutation of adjacent $W_{\smash{e^*}}^{\smash{s}}$ operators.

To construct a game for which DS ground states can act as a resource, consider the operators $X_1$, $\tilde{X}_1$ and $Z_1$, $\tilde{Z}_1$ defined in Fig.~\ref{fig:DS-operators}, where the Pauli operators that contribute to each dual edge are determined by~\eqref{eqn:W_s}. This leads to two sets of operators that satisfy the generalized commutation relations $Z_i X_j = i^{\delta_{ij}}X_jZ_i $, equivalent to those of two $d=4$ qudits. 
Note that the factor of $i$ has the same origins as the nontrivial self-statistics of the $s$ anyon, as illustrated in Fig.~\ref{fig:DS-operators}. Furthermore, the operators satisfy the constraints $X_1 \tilde{X}_1=1$ and $Z_1\tilde{Z}_1^\dagger = 1$ when acting on fixed-point DS ground states since they evaluate to closed loops of $W_e^s$ operators (up to a phase), which may be decomposed into a product of $A_v$ operators (with no phase). Note that, if the operators $X_1$, $\tilde{X}_1$ and $Z_1$, $\tilde{Z}_1$ were interpreted as single-site shift and phase operators on two qubits, they would stabilize the state $\ket{\Omega} \propto \sum_{q \in \mathbb{Z}_4} \ket{qq}$, i.e., the qudit Bell state. It therefore stands to reason that DS ground states can be used to win nonlocal quantum games for which $\ket{\Omega}$ is a resource.

While a number of such games are known~\cite{brassard2005pseudotelepathy}, we opt to use DS ground states as a resource for a straightforward generalization of the \emph{magic-square game} for simplicity of presentation. Before designing a perfect strategy for the DS model, we first discuss the generalization of the magic-square game to qudits of even dimension. The game involves two players, $A$ and $B$. Without communicating, the players must fill out the squares of a $3\times 3$ grid with integers mod $d$. Player $A$ ($B$) receives an integer $x_{A(B)} \in \{01, 10, 11 \}$ and fills out the corresponding row (column) of their grid. The game's winning criteria are:
\begin{enumerate*}[label=(\roman*)]
    \item \label{itm:player-A-parity} the integers mod $d$ filled out by player $A$ must sum to 0 mod $d$, 
    \item \label{itm:player-B-parity} the integers mod $d$ filled out by player $B$ must sum to $d/2$ mod $d$,
    \item \label{itm:agree} the players' integers must agree on the square on which the row and column intersect.
\end{enumerate*}
It is straightforward to show that the optimal classical strategy wins on $8/9$ of the inputs since there exists a global inconsistency in the constraints \ref{itm:player-A-parity}--\ref{itm:agree}. However, there exists a perfect quantum strategy using a generalization of Mermin's square~\cite{Mermin} to qudits of even dimension~\cite{LaversanneFinot_2017}. Consider the operators
\begin{equation}
\arraycolsep=5pt\def\arraystretch{1.2}
\begin{array}{cccc}
   & 01                      & 10                      & 11      \\
01 & U_1^\dagger \otimes \id & \id \otimes U_1^\dagger & U_1 \otimes U_1 \\
10 & \id \otimes U_2^\dagger & U_2^\dagger \otimes \id & U_2 \otimes U_2 \\
11 & -U_1 \otimes U_2        & -U_2 \otimes U_1        & U_3 \otimes U_3
\end{array}%
\label{eqn:unitary-square-ops}
\end{equation}
If the unitary operators $U_i$ satisfy the commutation relations $[U_i, U_j] = 2i\epsilon_{ijk}U_k^\dagger$ and $\{ U_i, U_j \} = 2\delta_{ij} U_i^2$, then it may be verified that all operators contained within a row or column of~\eqref{eqn:unitary-square-ops} commute, and the product of all operators belonging to a row (column) evaluates to $\id$ ($-\id$). Suppose that players $A$ and $B$ have access to independent degrees of freedom on which the operators~\eqref{eqn:unitary-square-ops} can be defined. Then, the players are able to satisfy the constraints \ref{itm:player-A-parity} and \ref{itm:player-B-parity} if player $A$ ($B$) measures the operators belonging to the appropriate row (column).

We therefore require two copies of the operators depicted in Fig.~\ref{fig:DS-operators}. Hence, we define $X_2, \tilde{X}_2$ and $Z_2, \tilde{Z}_2$ elsewhere on the lattice. Player $A$ then has access to operators $X_i$, $Z_i$, while player $B$ has access to $\tilde{X}_i$, $\tilde{Z}_i$. Using these operators, the $U_i$ entering the table~\eqref{eqn:unitary-square-ops} can be taken to be
\begin{gather}
    U_1 = Z^\dagger, \quad U_2 = X^2 , \quad U_3 = XZX 
    \, ,
    \label{eqn:U-def}
\end{gather}
and equivalently for $\tilde{U}_i$ using $\tilde{X}$, $\tilde{Z}$. These operators are unitary and have the desired commutation relations. The property $A_v = 1$ immediately implies that these operators satisfy the constraints $U_1 \tilde{U}_1^\dagger = U_2 \tilde{U}_2^\dagger = 1$ and $U_3 \tilde{U}_3^\dagger = -1$ with respect to DS ground states, which ensure that the players' outcomes agree on the square filled out by both players, satisfying \ref{itm:agree}. Note that the minus sign for $U_3$ cancels since the operators $U_3$ and $U_3^\dagger$ appear only in the constraint $U_3 \tilde{U}_3^\dagger \otimes U_3 \tilde{U}_3^\dagger = 1$. Hence,~\eqref{eqn:unitary-square-ops} and \eqref{eqn:U-def} taken together define a perfect strategy for the generalized magic-square game in which players measure the operators belonging to the row or column of~\eqref{eqn:unitary-square-ops} specified by their input bit~\footnote{Measurement of the unitary operators $U_i = \sum_j \omega^{s_j} \ket{j}\bra{j}$ with $s_j \in \mathbb{Z}_d$ may be thought of as measuring the Hermitian operator $\sum_j s_j \ket{j}\bra{j}$. Alternatively, phase estimation can be performed, which is exact in the present setting~\cite{okay2017topological}.}


\section{Discussion}
\label{sec:discussion}

We have demonstrated that the robust advantage discussed in Ref.~\cite{hart2024playing} is not particular to 2D toric code ground states. Indeed, we have shown that analogous advantage at suitably defined quantum games can be obtained for states drawn from, e.g., the 3D toric code, X-cube, or double-semion phases. The key ingredient in every case is to design a game that employs the mutual or self statistics of the phase to construct a set of operators whose measurement statistics provide a perfect strategy for the quantum game.

We have also presented a new perspective that unifies these ``topological games'' with previously discussed games using symmetry-breaking or symmetry protected topological (SPT) resource states. All these games involve measuring operators that correspond to the twist product of a symmetry operator with an operator charged under the symmetry (and exhibiting long-range order). If the symmetry is global, then the symmetry operators are global and hence are subject to an orthogonality catastrophe upon perturbation of the state, and the game is not robust. In contrast, for topological orders, the underlying symmetry is higher form, and there exist symmetry operators that are almost local (small loops or membranes), such that one can construct suitable twist products that are not subject to an orthogonality catastrophe upon weakly perturbing the state. This neatly explains the relative robustness of quantum games exploiting topologically ordered resource states. Note that for a game to be ``successful'' we demand $O(1)$ quantum advantage. If perturbation to a resource state reduces the quantum advantage from something $O(1)$ to something exponentially small in system size \cite{BulchandaniGames} then we declare this game to be ``not robust.'' A game is robust (in our sense) if and only if it preserves $O(1)$ quantum advantage in the presence of perturbations.

A number of questions immediately present themselves. Firstly, while our results strongly suggest that {\it any} nontrivial topological or fracton ordered phase should have a corresponding nonlocal game at which it provides robust advantage, we have not proven this. A generalization of the toric code games to a family of type-II fracton phases can be achieved by applying the fractalization procedure introduced in Ref.~\cite{Devakul2020b} to the operators involved in the games. Can one prove that, in fact, for every topological or fracton phase, there is a corresponding nonlocal game at which it provides robust advantage? Additionally, we have only considered specific types of braiding processes. For example, the X-cube game we considered employed lineon-fracton braiding, but there are also fracton-fracton self-statistics extracted via windmill processes~\cite{SongFractonSelf} -- what games can these be used to win? Can one uniquely identify a topological- or fracton-ordered phase by the set of games at which it provides advantage? Is there a systematic algorithmic way to construct the necessary games, if so? And can one actually carry out this task on quantum hardware? 

The quantum advantage provided by the above games is ``robust'' in the sense that $O(1)$ better-than-classical performance persists over some broad swathe of the topologically (or fracton) ordered phase. However, a \emph{perfect} quantum strategy (with certainty of victory) only exists given a ``fixed point'' resource state. If given a deformed (i.e., non-fixed-point) state, can one judiciously employ quantum error correction to gain access to perfect (or perhaps, arbitrarily good) quantum strategies? Equivalently, can one use error correction to expand the regime of quantum advantage to cover the \emph{entire} topologically ordered phase? We will address these questions elsewhere.

Furthermore, in the presence of faulty measurements in the state-preparation process, to what extent does a quantum advantage persist and is it scalable? What we have in mind is a protocol where the resource state (e.g., toric code) is produced by a protocol that incorporates measurements and feedback, and where faulty measurements may produce, e.g., a toric code with some number of excitations. If the players did not know that excitations were present, they would fail to win the game. Whether this failure mode can be evaded corresponds to finding an implementation of our procedure that is fault-tolerant to measurement errors in the resource state preparation process. For example, with 2D locality $O(L)$ rounds of stabilizer measurement are expected to be required to fault-tolerantly prepare a surface code state, after the measurement results are decoded appropriately. It is unclear how the players in the game can then implement their individual measurements fault tolerantly. We remark that our goal is distinct from simply implementing a contextuality game on logical qubits, as the above quantum error correction procedure can be performed on, e.g., a surface code with boundary conditions that specify a unique ground state and hence no encoded qubits. Such a game can be viewed as a modified stability experiment \cite{Gidney2022stability}. We expect that by using a 3D resource state, the game can be played fault-tolerantly (in the sense above) in a constant amount of time. It would be interesting to explore what minimum spacetime volume is required for fault tolerance.

Finally, we introduced a family of ``cellulation games'' in Sec.~\ref{sec:homological-games}, which significantly broadens the scope of games for which codewords of homological codes in $d$ spatial dimensions act as a resource. Does there exist an analogous family of games for other classes of quantum error-correcting codes? Separately, it is known that the contextuality that underlies success at nonlocal quantum games can be understood as having topological origins~\cite{okay2017topological}. Can the real-space picture of braiding presented herein be understood in this abstract language? A potential path to answering this question is to connect the cohomology theories that are relevant to contextuality~\cite{okay2017topological}, and anyon braiding~\cite{delcamp2019_2form}, respectively. Most broadly, this work sets the stage for an improved understanding of contextuality in a many-body setting, and the tasks for which it may be utilized to complete effectively.

\begin{acknowledgments}
OH would like to thank Charles Stahl and Evan Wickenden for numerous useful discussions. This work was primarily supported by the U.S.~Department of Energy, Office of Science, Basic Energy Sciences, under Award DE-SC0021346. DTS is supported by the Simons Collaboration on Ultra-Quantum Matter, which is a grant from the Simons Foundation (651440). DJW was supported in part by the Australian Research Council Discovery Early Career Research Award (DE220100625). 
\end{acknowledgments}


\bibliography{biblio.bib}

\begin{thebibliography}{60}%
\makeatletter
\providecommand \@ifxundefined [1]{%
 \@ifx{#1\undefined}
}%
\providecommand \@ifnum [1]{%
 \ifnum #1\expandafter \@firstoftwo
 \else \expandafter \@secondoftwo
 \fi
}%
\providecommand \@ifx [1]{%
 \ifx #1\expandafter \@firstoftwo
 \else \expandafter \@secondoftwo
 \fi
}%
\providecommand \natexlab [1]{#1}%
\providecommand \enquote  [1]{``#1''}%
\providecommand \bibnamefont  [1]{#1}%
\providecommand \bibfnamefont [1]{#1}%
\providecommand \citenamefont [1]{#1}%
\providecommand \href@noop [0]{\@secondoftwo}%
\providecommand \href [0]{\begingroup \@sanitize@url \@href}%
\providecommand \@href[1]{\@@startlink{#1}\@@href}%
\providecommand \@@href[1]{\endgroup#1\@@endlink}%
\providecommand \@sanitize@url [0]{\catcode `\\12\catcode `\$12\catcode
  `\&12\catcode `\#12\catcode `\^12\catcode `\_12\catcode `\%12\relax}%
\providecommand \@@startlink[1]{}%
\providecommand \@@endlink[0]{}%
\providecommand \url  [0]{\begingroup\@sanitize@url \@url }%
\providecommand \@url [1]{\endgroup\@href {#1}{\urlprefix }}%
\providecommand \urlprefix  [0]{URL }%
\providecommand \Eprint [0]{\href }%
\providecommand \doibase [0]{https://doi.org/}%
\providecommand \selectlanguage [0]{\@gobble}%
\providecommand \bibinfo  [0]{\@secondoftwo}%
\providecommand \bibfield  [0]{\@secondoftwo}%
\providecommand \translation [1]{[#1]}%
\providecommand \BibitemOpen [0]{}%
\providecommand \bibitemStop [0]{}%
\providecommand \bibitemNoStop [0]{.\EOS\space}%
\providecommand \EOS [0]{\spacefactor3000\relax}%
\providecommand \BibitemShut  [1]{\csname bibitem#1\endcsname}%
\let\auto@bib@innerbib\@empty
\bibitem [{\citenamefont {Bell}(1966)}]{Bell}%
  \BibitemOpen
  \bibfield  {author} {\bibinfo {author} {\bibfnamefont {J.~S.}\ \bibnamefont
  {Bell}},\ }\bibfield  {title} {\bibinfo {title} {On the problem of hidden
  variables in quantum mechanics},\ }\href
  {https://doi.org/10.1103/RevModPhys.38.447} {\bibfield  {journal} {\bibinfo
  {journal} {Rev. Mod. Phys.}\ }\textbf {\bibinfo {volume} {38}},\ \bibinfo
  {pages} {447} (\bibinfo {year} {1966})}\BibitemShut {NoStop}%
\bibitem [{\citenamefont {Mermin}(1993)}]{Mermin}%
  \BibitemOpen
  \bibfield  {author} {\bibinfo {author} {\bibfnamefont {N.~D.}\ \bibnamefont
  {Mermin}},\ }\bibfield  {title} {\bibinfo {title} {Hidden variables and the
  two theorems of john bell},\ }\href
  {https://doi.org/10.1103/RevModPhys.65.803} {\bibfield  {journal} {\bibinfo
  {journal} {Rev. Mod. Phys.}\ }\textbf {\bibinfo {volume} {65}},\ \bibinfo
  {pages} {803} (\bibinfo {year} {1993})}\BibitemShut {NoStop}%
\bibitem [{\citenamefont {Mermin}(1990{\natexlab{a}})}]{mermin1990quantum}%
  \BibitemOpen
  \bibfield  {author} {\bibinfo {author} {\bibfnamefont {N.~D.}\ \bibnamefont
  {Mermin}},\ }\bibfield  {title} {\bibinfo {title} {Quantum mysteries
  revisited},\ }\href {https://doi.org/10.1119/1.16503} {\bibfield  {journal}
  {\bibinfo  {journal} {Am. J. Phys}\ }\textbf {\bibinfo {volume} {58}},\
  \bibinfo {pages} {731} (\bibinfo {year} {1990}{\natexlab{a}})}\BibitemShut
  {NoStop}%
\bibitem [{\citenamefont {Kochen}\ and\ \citenamefont
  {Specker}(1967)}]{kochenSpecker1967}%
  \BibitemOpen
  \bibfield  {author} {\bibinfo {author} {\bibfnamefont {S.}~\bibnamefont
  {Kochen}}\ and\ \bibinfo {author} {\bibfnamefont {E.}~\bibnamefont
  {Specker}},\ }\bibfield  {title} {\bibinfo {title} {The problem of hidden
  variables in quantum mechanics},\ }\href@noop {} {\bibfield  {journal}
  {\bibinfo  {journal} {Journal of Mathematics and Mechanics}\ }\textbf
  {\bibinfo {volume} {17}} (\bibinfo {year} {1967})}\BibitemShut {NoStop}%
\bibitem [{\citenamefont {Brassard}\ \emph
  {et~al.}(2005{\natexlab{a}})\citenamefont {Brassard}, \citenamefont
  {Broadbent},\ and\ \citenamefont {Tapp}}]{brassard2005pseudotelepathy}%
  \BibitemOpen
  \bibfield  {author} {\bibinfo {author} {\bibfnamefont {G.}~\bibnamefont
  {Brassard}}, \bibinfo {author} {\bibfnamefont {A.}~\bibnamefont
  {Broadbent}},\ and\ \bibinfo {author} {\bibfnamefont {A.}~\bibnamefont
  {Tapp}},\ }\bibfield  {title} {\bibinfo {title} {Quantum pseudo-telepathy},\
  }\href {https://doi.org/0.1007/s10701-005-7353-4} {\bibfield  {journal}
  {\bibinfo  {journal} {Foundations of Physics}\ }\textbf {\bibinfo {volume}
  {35}},\ \bibinfo {pages} {1877} (\bibinfo {year}
  {2005}{\natexlab{a}})}\BibitemShut {NoStop}%
\bibitem [{\citenamefont {Daniel}\ and\ \citenamefont
  {Miyake}(2021)}]{DanielStringOrder}%
  \BibitemOpen
  \bibfield  {author} {\bibinfo {author} {\bibfnamefont {A.~K.}\ \bibnamefont
  {Daniel}}\ and\ \bibinfo {author} {\bibfnamefont {A.}~\bibnamefont
  {Miyake}},\ }\bibfield  {title} {\bibinfo {title} {Quantum computational
  advantage with string order parameters of one-dimensional symmetry-protected
  topological order},\ }\href {https://doi.org/10.1103/PhysRevLett.126.090505}
  {\bibfield  {journal} {\bibinfo  {journal} {Phys. Rev. Lett.}\ }\textbf
  {\bibinfo {volume} {126}},\ \bibinfo {pages} {090505} (\bibinfo {year}
  {2021})}\BibitemShut {NoStop}%
\bibitem [{\citenamefont {Daniel}\ \emph {et~al.}(2022)\citenamefont {Daniel},
  \citenamefont {Zhu}, \citenamefont {Alderete}, \citenamefont {Buchemmavari},
  \citenamefont {Green}, \citenamefont {Nguyen}, \citenamefont {Thurtell},
  \citenamefont {Zhao}, \citenamefont {Linke},\ and\ \citenamefont
  {Miyake}}]{Daniel2022Exp}%
  \BibitemOpen
  \bibfield  {author} {\bibinfo {author} {\bibfnamefont {A.~K.}\ \bibnamefont
  {Daniel}}, \bibinfo {author} {\bibfnamefont {Y.}~\bibnamefont {Zhu}},
  \bibinfo {author} {\bibfnamefont {C.~H.}\ \bibnamefont {Alderete}}, \bibinfo
  {author} {\bibfnamefont {V.}~\bibnamefont {Buchemmavari}}, \bibinfo {author}
  {\bibfnamefont {A.~M.}\ \bibnamefont {Green}}, \bibinfo {author}
  {\bibfnamefont {N.~H.}\ \bibnamefont {Nguyen}}, \bibinfo {author}
  {\bibfnamefont {T.~G.}\ \bibnamefont {Thurtell}}, \bibinfo {author}
  {\bibfnamefont {A.}~\bibnamefont {Zhao}}, \bibinfo {author} {\bibfnamefont
  {N.~M.}\ \bibnamefont {Linke}},\ and\ \bibinfo {author} {\bibfnamefont
  {A.}~\bibnamefont {Miyake}},\ }\bibfield  {title} {\bibinfo {title} {Quantum
  computational advantage attested by nonlocal games with the cyclic cluster
  state},\ }\href {https://doi.org/10.1103/PhysRevResearch.4.033068} {\bibfield
   {journal} {\bibinfo  {journal} {Phys. Rev. Res.}\ }\textbf {\bibinfo
  {volume} {4}},\ \bibinfo {pages} {033068} (\bibinfo {year}
  {2022})}\BibitemShut {NoStop}%
\bibitem [{\citenamefont {Bravyi}\ \emph {et~al.}(2020)\citenamefont {Bravyi},
  \citenamefont {Gosset}, \citenamefont {Koenig},\ and\ \citenamefont
  {Tomamichel}}]{bravyi2020noisyShallow}%
  \BibitemOpen
  \bibfield  {author} {\bibinfo {author} {\bibfnamefont {S.}~\bibnamefont
  {Bravyi}}, \bibinfo {author} {\bibfnamefont {D.}~\bibnamefont {Gosset}},
  \bibinfo {author} {\bibfnamefont {R.}~\bibnamefont {Koenig}},\ and\ \bibinfo
  {author} {\bibfnamefont {M.}~\bibnamefont {Tomamichel}},\ }\bibfield  {title}
  {\bibinfo {title} {Quantum advantage with noisy shallow circuits},\ }\href
  {https://doi.org/10.1038/s41567-020-0948-z} {\bibfield  {journal} {\bibinfo
  {journal} {Nature Physics}\ }\textbf {\bibinfo {volume} {16}},\ \bibinfo
  {pages} {1040} (\bibinfo {year} {2020})}\BibitemShut {NoStop}%
\bibitem [{\citenamefont {Bravyi}\ \emph {et~al.}(2018)\citenamefont {Bravyi},
  \citenamefont {Gosset},\ and\ \citenamefont {König}}]{bravyi2020shallow}%
  \BibitemOpen
  \bibfield  {author} {\bibinfo {author} {\bibfnamefont {S.}~\bibnamefont
  {Bravyi}}, \bibinfo {author} {\bibfnamefont {D.}~\bibnamefont {Gosset}},\
  and\ \bibinfo {author} {\bibfnamefont {R.}~\bibnamefont {König}},\
  }\bibfield  {title} {\bibinfo {title} {Quantum advantage with shallow
  circuits},\ }\href {https://doi.org/10.1126/science.aar3106} {\bibfield
  {journal} {\bibinfo  {journal} {Science}\ }\textbf {\bibinfo {volume}
  {362}},\ \bibinfo {pages} {308} (\bibinfo {year} {2018})}\BibitemShut
  {NoStop}%
\bibitem [{\citenamefont {Jaffali}\ and\ \citenamefont
  {Holweck}(2024)}]{jaffali2023new}%
  \BibitemOpen
  \bibfield  {author} {\bibinfo {author} {\bibfnamefont {H.}~\bibnamefont
  {Jaffali}}\ and\ \bibinfo {author} {\bibfnamefont {F.}~\bibnamefont
  {Holweck}},\ }\bibfield  {title} {\bibinfo {title} {{Two new non-equivalent
  three-qubit CHSH games}},\ }\href {https://doi.org/10.26421/QIC24.5-6-4}
  {\bibfield  {journal} {\bibinfo  {journal} {Quant. Inf. Comput.}\ }\textbf
  {\bibinfo {volume} {24}},\ \bibinfo {pages} {0438} (\bibinfo {year}
  {2024})}\BibitemShut {NoStop}%
\bibitem [{\citenamefont {Sheffer}\ \emph {et~al.}(2022)\citenamefont
  {Sheffer}, \citenamefont {Azses},\ and\ \citenamefont
  {Dalla~Torre}}]{DallaTorre}%
  \BibitemOpen
  \bibfield  {author} {\bibinfo {author} {\bibfnamefont {M.}~\bibnamefont
  {Sheffer}}, \bibinfo {author} {\bibfnamefont {D.}~\bibnamefont {Azses}},\
  and\ \bibinfo {author} {\bibfnamefont {E.~G.}\ \bibnamefont {Dalla~Torre}},\
  }\bibfield  {title} {\bibinfo {title} {Playing quantum nonlocal games with
  six noisy qubits on the cloud},\ }\href
  {https://doi.org/https://doi.org/10.1002/qute.202100081} {\bibfield
  {journal} {\bibinfo  {journal} {Advanced Quantum Technologies}\ }\textbf
  {\bibinfo {volume} {5}},\ \bibinfo {pages} {2100081} (\bibinfo {year}
  {2022})}\BibitemShut {NoStop}%
\bibitem [{\citenamefont {Bulchandani}\ \emph
  {et~al.}(2023{\natexlab{a}})\citenamefont {Bulchandani}, \citenamefont
  {Burnell},\ and\ \citenamefont {Sondhi}}]{BBSgame}%
  \BibitemOpen
  \bibfield  {author} {\bibinfo {author} {\bibfnamefont {V.~B.}\ \bibnamefont
  {Bulchandani}}, \bibinfo {author} {\bibfnamefont {F.~J.}\ \bibnamefont
  {Burnell}},\ and\ \bibinfo {author} {\bibfnamefont {S.~L.}\ \bibnamefont
  {Sondhi}},\ }\bibfield  {title} {\bibinfo {title} {A multiplayer multiteam
  nonlocal game for the toric code},\ }\href
  {https://doi.org/10.1103/PhysRevB.107.035409} {\bibfield  {journal} {\bibinfo
   {journal} {Phys. Rev. B}\ }\textbf {\bibinfo {volume} {107}},\ \bibinfo
  {pages} {035409} (\bibinfo {year} {2023}{\natexlab{a}})}\BibitemShut
  {NoStop}%
\bibitem [{\citenamefont {Bulchandani}\ \emph
  {et~al.}(2023{\natexlab{b}})\citenamefont {Bulchandani}, \citenamefont
  {Burnell},\ and\ \citenamefont {Sondhi}}]{BulchandaniGames}%
  \BibitemOpen
  \bibfield  {author} {\bibinfo {author} {\bibfnamefont {V.~B.}\ \bibnamefont
  {Bulchandani}}, \bibinfo {author} {\bibfnamefont {F.~J.}\ \bibnamefont
  {Burnell}},\ and\ \bibinfo {author} {\bibfnamefont {S.~L.}\ \bibnamefont
  {Sondhi}},\ }\bibfield  {title} {\bibinfo {title} {Playing nonlocal games
  with phases of quantum matter},\ }\href
  {https://doi.org/10.1103/PhysRevB.107.045412} {\bibfield  {journal} {\bibinfo
   {journal} {Phys. Rev. B}\ }\textbf {\bibinfo {volume} {107}},\ \bibinfo
  {pages} {045412} (\bibinfo {year} {2023}{\natexlab{b}})}\BibitemShut
  {NoStop}%
\bibitem [{\citenamefont {Lin}\ \emph {et~al.}(2023)\citenamefont {Lin},
  \citenamefont {Bulchandani},\ and\ \citenamefont {Sondhi}}]{lin2023quantum}%
  \BibitemOpen
  \bibfield  {author} {\bibinfo {author} {\bibfnamefont {C.}~\bibnamefont
  {Lin}}, \bibinfo {author} {\bibfnamefont {V.~B.}\ \bibnamefont
  {Bulchandani}},\ and\ \bibinfo {author} {\bibfnamefont {S.~L.}\ \bibnamefont
  {Sondhi}},\ }\href@noop {} {\bibinfo {title} {Quantum tasks assisted by
  quantum noise}} (\bibinfo {year} {2023}),\ \Eprint
  {https://arxiv.org/abs/2308.10969} {arXiv:2308.10969 [quant-ph]} \BibitemShut
  {NoStop}%
\bibitem [{\citenamefont {Natarajan}\ and\ \citenamefont
  {Nirkhe}(2024)}]{natarajan}%
  \BibitemOpen
  \bibfield  {author} {\bibinfo {author} {\bibfnamefont {A.}~\bibnamefont
  {Natarajan}}\ and\ \bibinfo {author} {\bibfnamefont {C.}~\bibnamefont
  {Nirkhe}},\ }\href@noop {} {\bibinfo {title} {The status of the quantum pcp
  conjecture (games version)}} (\bibinfo {year} {2024}),\ \Eprint
  {https://arxiv.org/abs/2403.13084} {arXiv:2403.13084 [quant-ph]} \BibitemShut
  {NoStop}%
\bibitem [{\citenamefont {Hart}\ \emph {et~al.}(2024)\citenamefont {Hart},
  \citenamefont {Stephen}, \citenamefont {Williamson}, \citenamefont
  {Foss-Feig},\ and\ \citenamefont {Nandkishore}}]{hart2024playing}%
  \BibitemOpen
  \bibfield  {author} {\bibinfo {author} {\bibfnamefont {O.}~\bibnamefont
  {Hart}}, \bibinfo {author} {\bibfnamefont {D.~T.}\ \bibnamefont {Stephen}},
  \bibinfo {author} {\bibfnamefont {D.~J.}\ \bibnamefont {Williamson}},
  \bibinfo {author} {\bibfnamefont {M.}~\bibnamefont {Foss-Feig}},\ and\
  \bibinfo {author} {\bibfnamefont {R.}~\bibnamefont {Nandkishore}},\
  }\href@noop {} {\bibinfo {title} {Playing nonlocal games across a topological
  phase transition on a quantum computer}} (\bibinfo {year} {2024}),\ \Eprint
  {https://arxiv.org/abs/2403.04829} {arXiv:2403.04829 [quant-ph]} \BibitemShut
  {NoStop}%
\bibitem [{\citenamefont {Zhu}\ \emph {et~al.}(2023)\citenamefont {Zhu},
  \citenamefont {Tantivasadakarn}, \citenamefont {Vishwanath}, \citenamefont
  {Trebst},\ and\ \citenamefont {Verresen}}]{zhu2022}%
  \BibitemOpen
  \bibfield  {author} {\bibinfo {author} {\bibfnamefont {G.-Y.}\ \bibnamefont
  {Zhu}}, \bibinfo {author} {\bibfnamefont {N.}~\bibnamefont
  {Tantivasadakarn}}, \bibinfo {author} {\bibfnamefont {A.}~\bibnamefont
  {Vishwanath}}, \bibinfo {author} {\bibfnamefont {S.}~\bibnamefont {Trebst}},\
  and\ \bibinfo {author} {\bibfnamefont {R.}~\bibnamefont {Verresen}},\
  }\bibfield  {title} {\bibinfo {title} {Nishimori's cat: Stable long-range
  entanglement from finite-depth unitaries and weak measurements},\ }\href
  {https://doi.org/10.1103/PhysRevLett.131.200201} {\bibfield  {journal}
  {\bibinfo  {journal} {Phys. Rev. Lett.}\ }\textbf {\bibinfo {volume} {131}},\
  \bibinfo {pages} {200201} (\bibinfo {year} {2023})}\BibitemShut {NoStop}%
\bibitem [{\citenamefont {Lee}\ \emph {et~al.}(2022)\citenamefont {Lee},
  \citenamefont {Ji}, \citenamefont {Bi},\ and\ \citenamefont
  {Fisher}}]{lee2022decoding}%
  \BibitemOpen
  \bibfield  {author} {\bibinfo {author} {\bibfnamefont {J.~Y.}\ \bibnamefont
  {Lee}}, \bibinfo {author} {\bibfnamefont {W.}~\bibnamefont {Ji}}, \bibinfo
  {author} {\bibfnamefont {Z.}~\bibnamefont {Bi}},\ and\ \bibinfo {author}
  {\bibfnamefont {M.~P.~A.}\ \bibnamefont {Fisher}},\ }\href@noop {} {\bibinfo
  {title} {Decoding measurement-prepared quantum phases and transitions: from
  ising model to gauge theory, and beyond}} (\bibinfo {year} {2022}),\ \Eprint
  {https://arxiv.org/abs/2208.11699} {arXiv:2208.11699 [cond-mat.str-el]}
  \BibitemShut {NoStop}%
\bibitem [{\citenamefont {Chen}\ \emph {et~al.}(2024)\citenamefont {Chen},
  \citenamefont {Zhu}, \citenamefont {Verresen} \emph
  {et~al.}}]{chen2023realizing}%
  \BibitemOpen
  \bibfield  {author} {\bibinfo {author} {\bibfnamefont {E.}~\bibnamefont
  {Chen}}, \bibinfo {author} {\bibfnamefont {G.}~\bibnamefont {Zhu}}, \bibinfo
  {author} {\bibfnamefont {R.}~\bibnamefont {Verresen}}, \emph {et~al.},\
  }\bibfield  {title} {\bibinfo {title} {Nishimori transition across the error
  threshold for constant-depth quantum circuits},\ }\bibfield  {journal}
  {\bibinfo  {journal} {Nature Physics}\ }\href
  {https://doi.org/10.1038/s41567-024-02696-6} {10.1038/s41567-024-02696-6}
  (\bibinfo {year} {2024})\BibitemShut {NoStop}%
\bibitem [{\citenamefont {Greenberger}\ \emph {et~al.}(1989)\citenamefont
  {Greenberger}, \citenamefont {Horne},\ and\ \citenamefont
  {Zeilinger}}]{ghz1989}%
  \BibitemOpen
  \bibfield  {author} {\bibinfo {author} {\bibfnamefont {D.~M.}\ \bibnamefont
  {Greenberger}}, \bibinfo {author} {\bibfnamefont {M.~A.}\ \bibnamefont
  {Horne}},\ and\ \bibinfo {author} {\bibfnamefont {A.}~\bibnamefont
  {Zeilinger}},\ }\bibfield  {title} {\bibinfo {title} {Going beyond bell’s
  theorem},\ }in\ \href {https://doi.org/10.1007/978-94-017-0849-4_10} {\emph
  {\bibinfo {booktitle} {Bell’s theorem, quantum theory and conceptions of
  the universe}}}\ (\bibinfo  {publisher} {Springer},\ \bibinfo {year} {1989})\
  pp.\ \bibinfo {pages} {69--72}\BibitemShut {NoStop}%
\bibitem [{\citenamefont {Greenberger}\ \emph {et~al.}(1990)\citenamefont
  {Greenberger}, \citenamefont {Horne}, \citenamefont {Shimony},\ and\
  \citenamefont {Zeilinger}}]{GHSZ1990}%
  \BibitemOpen
  \bibfield  {author} {\bibinfo {author} {\bibfnamefont {D.~M.}\ \bibnamefont
  {Greenberger}}, \bibinfo {author} {\bibfnamefont {M.~A.}\ \bibnamefont
  {Horne}}, \bibinfo {author} {\bibfnamefont {A.}~\bibnamefont {Shimony}},\
  and\ \bibinfo {author} {\bibfnamefont {A.}~\bibnamefont {Zeilinger}},\
  }\bibfield  {title} {\bibinfo {title} {{Bell’s theorem without
  inequalities}},\ }\href {https://doi.org/10.1119/1.16243} {\bibfield
  {journal} {\bibinfo  {journal} {American Journal of Physics}\ }\textbf
  {\bibinfo {volume} {58}},\ \bibinfo {pages} {1131} (\bibinfo {year}
  {1990})}\BibitemShut {NoStop}%
\bibitem [{\citenamefont {Mermin}(1990{\natexlab{b}})}]{MerminPolynomials}%
  \BibitemOpen
  \bibfield  {author} {\bibinfo {author} {\bibfnamefont {N.~D.}\ \bibnamefont
  {Mermin}},\ }\bibfield  {title} {\bibinfo {title} {Extreme quantum
  entanglement in a superposition of macroscopically distinct states},\ }\href
  {https://doi.org/10.1103/PhysRevLett.65.1838} {\bibfield  {journal} {\bibinfo
   {journal} {Phys. Rev. Lett.}\ }\textbf {\bibinfo {volume} {65}},\ \bibinfo
  {pages} {1838} (\bibinfo {year} {1990}{\natexlab{b}})}\BibitemShut {NoStop}%
\bibitem [{\citenamefont {Brassard}\ \emph {et~al.}(2003)\citenamefont
  {Brassard}, \citenamefont {Broadbent},\ and\ \citenamefont
  {Tapp}}]{Brassard2003multiparty}%
  \BibitemOpen
  \bibfield  {author} {\bibinfo {author} {\bibfnamefont {G.}~\bibnamefont
  {Brassard}}, \bibinfo {author} {\bibfnamefont {A.}~\bibnamefont
  {Broadbent}},\ and\ \bibinfo {author} {\bibfnamefont {A.}~\bibnamefont
  {Tapp}},\ }\bibfield  {title} {\bibinfo {title} {Multi-party
  pseudo-telepathy},\ }in\ \href {https://doi.org/10.1007/978-3-540-45078-8_1}
  {\emph {\bibinfo {booktitle} {Algorithms and Data Structures}}},\ \bibinfo
  {editor} {edited by\ \bibinfo {editor} {\bibfnamefont {F.}~\bibnamefont
  {Dehne}}, \bibinfo {editor} {\bibfnamefont {J.-R.}\ \bibnamefont {Sack}},\
  and\ \bibinfo {editor} {\bibfnamefont {M.}~\bibnamefont {Smid}}}\ (\bibinfo
  {publisher} {Springer Berlin Heidelberg},\ \bibinfo {address} {Berlin,
  Heidelberg},\ \bibinfo {year} {2003})\ pp.\ \bibinfo {pages}
  {1--11}\BibitemShut {NoStop}%
\bibitem [{\citenamefont {Brassard}\ \emph
  {et~al.}(2005{\natexlab{b}})\citenamefont {Brassard}, \citenamefont
  {Broadbent},\ and\ \citenamefont {Tapp}}]{brassard2005recasting}%
  \BibitemOpen
  \bibfield  {author} {\bibinfo {author} {\bibfnamefont {G.}~\bibnamefont
  {Brassard}}, \bibinfo {author} {\bibfnamefont {A.}~\bibnamefont
  {Broadbent}},\ and\ \bibinfo {author} {\bibfnamefont {A.}~\bibnamefont
  {Tapp}},\ }\bibfield  {title} {\bibinfo {title} {Recasting mermin's
  multi-player game into the framework of pseudo-telepathy},\ }\href
  {https://doi.org/10.26421/QIC5.7-2} {\bibfield  {journal} {\bibinfo
  {journal} {Quantum Information and Computation}\ }\textbf {\bibinfo {volume}
  {5}},\ \bibinfo {pages} {538} (\bibinfo {year}
  {2005}{\natexlab{b}})}\BibitemShut {NoStop}%
\bibitem [{\citenamefont {Anders}\ and\ \citenamefont
  {Browne}(2009)}]{Anders2009}%
  \BibitemOpen
  \bibfield  {author} {\bibinfo {author} {\bibfnamefont {J.}~\bibnamefont
  {Anders}}\ and\ \bibinfo {author} {\bibfnamefont {D.~E.}\ \bibnamefont
  {Browne}},\ }\bibfield  {title} {\bibinfo {title} {Computational power of
  correlations},\ }\href {https://doi.org/10.1103/PhysRevLett.102.050502}
  {\bibfield  {journal} {\bibinfo  {journal} {Phys. Rev. Lett.}\ }\textbf
  {\bibinfo {volume} {102}},\ \bibinfo {pages} {050502} (\bibinfo {year}
  {2009})}\BibitemShut {NoStop}%
\bibitem [{\citenamefont {Yao}(1977)}]{Yao1977}%
  \BibitemOpen
  \bibfield  {author} {\bibinfo {author} {\bibfnamefont {A.~C.-C.}\
  \bibnamefont {Yao}},\ }\bibfield  {title} {\bibinfo {title} {Probabilistic
  computations: Toward a unified measure of complexity},\ }in\ \href
  {https://doi.org/10.1109/SFCS.1977.24} {\emph {\bibinfo {booktitle} {18th
  Annual Symposium on Foundations of Computer Science (sfcs 1977)}}}\ (\bibinfo
  {year} {1977})\ pp.\ \bibinfo {pages} {222--227}\BibitemShut {NoStop}%
\bibitem [{\citenamefont {Kitaev}(2003)}]{Kitaev2003}%
  \BibitemOpen
  \bibfield  {author} {\bibinfo {author} {\bibfnamefont {A.}~\bibnamefont
  {Kitaev}},\ }\bibfield  {title} {\bibinfo {title} {Fault-tolerant quantum
  computation by anyons},\ }\href
  {https://doi.org/https://doi.org/10.1016/S0003-4916(02)00018-0} {\bibfield
  {journal} {\bibinfo  {journal} {Ann. Phys.}\ }\textbf {\bibinfo {volume}
  {303}},\ \bibinfo {pages} {2} (\bibinfo {year} {2003})}\BibitemShut {NoStop}%
\bibitem [{\citenamefont {Haah}(2016)}]{haah2016invariant}%
  \BibitemOpen
  \bibfield  {author} {\bibinfo {author} {\bibfnamefont {J.}~\bibnamefont
  {Haah}},\ }\bibfield  {title} {\bibinfo {title} {An invariant of
  topologically ordered states under local unitary transformations},\ }\href
  {https://doi.org/10.1007/s00220-016-2594-y} {\bibfield  {journal} {\bibinfo
  {journal} {Communications in Mathematical Physics}\ }\textbf {\bibinfo
  {volume} {342}},\ \bibinfo {pages} {771} (\bibinfo {year}
  {2016})}\BibitemShut {NoStop}%
\bibitem [{\citenamefont {Fradkin}(2017)}]{fradkin2017disorder}%
  \BibitemOpen
  \bibfield  {author} {\bibinfo {author} {\bibfnamefont {E.}~\bibnamefont
  {Fradkin}},\ }\bibfield  {title} {\bibinfo {title} {Disorder operators and
  their descendants},\ }\href {https://doi.org/10.1007/s10955-017-1737-7}
  {\bibfield  {journal} {\bibinfo  {journal} {Journal of Statistical Physics}\
  }\textbf {\bibinfo {volume} {167}},\ \bibinfo {pages} {427} (\bibinfo {year}
  {2017})}\BibitemShut {NoStop}%
\bibitem [{\citenamefont {Dua}\ \emph {et~al.}(2019)\citenamefont {Dua},
  \citenamefont {Kim}, \citenamefont {Cheng},\ and\ \citenamefont
  {Williamson}}]{Dua2019Sorting}%
  \BibitemOpen
  \bibfield  {author} {\bibinfo {author} {\bibfnamefont {A.}~\bibnamefont
  {Dua}}, \bibinfo {author} {\bibfnamefont {I.~H.}\ \bibnamefont {Kim}},
  \bibinfo {author} {\bibfnamefont {M.}~\bibnamefont {Cheng}},\ and\ \bibinfo
  {author} {\bibfnamefont {D.~J.}\ \bibnamefont {Williamson}},\ }\bibfield
  {title} {\bibinfo {title} {Sorting topological stabilizer models in three
  dimensions},\ }\href {https://doi.org/10.1103/PhysRevB.100.155137} {\bibfield
   {journal} {\bibinfo  {journal} {Phys. Rev. B}\ }\textbf {\bibinfo {volume}
  {100}},\ \bibinfo {pages} {155137} (\bibinfo {year} {2019})}\BibitemShut
  {NoStop}%
\bibitem [{\citenamefont {Nussinov}\ and\ \citenamefont
  {Ortiz}(2009)}]{Nussinov2007}%
  \BibitemOpen
  \bibfield  {author} {\bibinfo {author} {\bibfnamefont {Z.}~\bibnamefont
  {Nussinov}}\ and\ \bibinfo {author} {\bibfnamefont {G.}~\bibnamefont
  {Ortiz}},\ }\bibfield  {title} {\bibinfo {title} {A symmetry principle for
  topological quantum order},\ }\href
  {https://doi.org/doi.org/10.1016/j.aop.2008.11.002} {\bibfield  {journal}
  {\bibinfo  {journal} {Annals of Physics}\ }\textbf {\bibinfo {volume}
  {324}},\ \bibinfo {pages} {977} (\bibinfo {year} {2009})}\BibitemShut
  {NoStop}%
\bibitem [{\citenamefont {Gaiotto}\ \emph {et~al.}(2015)\citenamefont
  {Gaiotto}, \citenamefont {Kapustin}, \citenamefont {Seiberg},\ and\
  \citenamefont {Willett}}]{Gaiotto2015}%
  \BibitemOpen
  \bibfield  {author} {\bibinfo {author} {\bibfnamefont {D.}~\bibnamefont
  {Gaiotto}}, \bibinfo {author} {\bibfnamefont {A.}~\bibnamefont {Kapustin}},
  \bibinfo {author} {\bibfnamefont {N.}~\bibnamefont {Seiberg}},\ and\ \bibinfo
  {author} {\bibfnamefont {B.}~\bibnamefont {Willett}},\ }\bibfield  {title}
  {\bibinfo {title} {Generalized global symmetries},\ }\href
  {https://doi.org/10.1007/JHEP02(2015)172} {\bibfield  {journal} {\bibinfo
  {journal} {J. High Energ. Phys.}\ }\textbf {\bibinfo {volume} {2015}}\bibinfo
   {number} { (172)}}\BibitemShut {NoStop}%
\bibitem [{\citenamefont {McGreevy}(2023)}]{McGreevy2022}%
  \BibitemOpen
\bibfield  {number} {  }\bibfield  {author} {\bibinfo {author} {\bibfnamefont
  {J.}~\bibnamefont {McGreevy}},\ }\bibfield  {title} {\bibinfo {title}
  {Generalized symmetries in condensed matter},\ }\href
  {https://doi.org/10.1146/annurev-conmatphys-040721-021029} {\bibfield
  {journal} {\bibinfo  {journal} {Annual Review of Condensed Matter Physics}\
  }\textbf {\bibinfo {volume} {14}},\ \bibinfo {pages} {57} (\bibinfo {year}
  {2023})}\BibitemShut {NoStop}%
\bibitem [{\citenamefont {Fredenhagen}\ and\ \citenamefont
  {Marcu}(1983)}]{Fredenhagen1983}%
  \BibitemOpen
  \bibfield  {author} {\bibinfo {author} {\bibfnamefont {K.}~\bibnamefont
  {Fredenhagen}}\ and\ \bibinfo {author} {\bibfnamefont {M.}~\bibnamefont
  {Marcu}},\ }\bibfield  {title} {\bibinfo {title} {{Charged states in
  $\mathbb{Z}_2$ gauge theories}},\ }\href
  {https://doi.org/10.1007/BF01206315/METRICS} {\bibfield  {journal} {\bibinfo
  {journal} {Communications in Mathematical Physics}\ }\textbf {\bibinfo
  {volume} {92}},\ \bibinfo {pages} {81} (\bibinfo {year} {1983})}\BibitemShut
  {NoStop}%
\bibitem [{\citenamefont {Hamma}\ \emph {et~al.}(2005)\citenamefont {Hamma},
  \citenamefont {Zanardi},\ and\ \citenamefont {Wen}}]{Hamma3DTC}%
  \BibitemOpen
  \bibfield  {author} {\bibinfo {author} {\bibfnamefont {A.}~\bibnamefont
  {Hamma}}, \bibinfo {author} {\bibfnamefont {P.}~\bibnamefont {Zanardi}},\
  and\ \bibinfo {author} {\bibfnamefont {X.-G.}\ \bibnamefont {Wen}},\
  }\bibfield  {title} {\bibinfo {title} {String and membrane condensation on
  three-dimensional lattices},\ }\href
  {https://doi.org/10.1103/PhysRevB.72.035307} {\bibfield  {journal} {\bibinfo
  {journal} {Phys. Rev. B}\ }\textbf {\bibinfo {volume} {72}},\ \bibinfo
  {pages} {035307} (\bibinfo {year} {2005})}\BibitemShut {NoStop}%
\bibitem [{Note1()}]{Note1}%
  \BibitemOpen
  \bibinfo {note} {We will often omit the isomorphism $*$ between a cell
  complex $X$ and its dual $X^*$ for brevity. Expressions such as $f \in
  \partial R^*$ should be understood as $f \in *\partial R^*$.}\BibitemShut
  {Stop}%
\bibitem [{\citenamefont {Nandkishore}\ and\ \citenamefont
  {Hermele}(2019)}]{Nandkishore2019fractons}%
  \BibitemOpen
  \bibfield  {author} {\bibinfo {author} {\bibfnamefont {R.~M.}\ \bibnamefont
  {Nandkishore}}\ and\ \bibinfo {author} {\bibfnamefont {M.}~\bibnamefont
  {Hermele}},\ }\bibfield  {title} {\bibinfo {title} {Fractons},\ }\href
  {https://doi.org/https://doi.org/10.1146/annurev-conmatphys-031218-013604}
  {\bibfield  {journal} {\bibinfo  {journal} {Annual Review of Condensed Matter
  Physics}\ }\textbf {\bibinfo {volume} {10}},\ \bibinfo {pages} {295}
  (\bibinfo {year} {2019})}\BibitemShut {NoStop}%
\bibitem [{\citenamefont {Castelnovo}\ \emph {et~al.}(2010)\citenamefont
  {Castelnovo}, \citenamefont {Chamon},\ and\ \citenamefont
  {Sherrington}}]{CastelnovoXcube}%
  \BibitemOpen
  \bibfield  {author} {\bibinfo {author} {\bibfnamefont {C.}~\bibnamefont
  {Castelnovo}}, \bibinfo {author} {\bibfnamefont {C.}~\bibnamefont {Chamon}},\
  and\ \bibinfo {author} {\bibfnamefont {D.}~\bibnamefont {Sherrington}},\
  }\bibfield  {title} {\bibinfo {title} {Quantum mechanical and information
  theoretic view on classical glass transitions},\ }\href
  {https://doi.org/10.1103/PhysRevB.81.184303} {\bibfield  {journal} {\bibinfo
  {journal} {Phys. Rev. B}\ }\textbf {\bibinfo {volume} {81}},\ \bibinfo
  {pages} {184303} (\bibinfo {year} {2010})}\BibitemShut {NoStop}%
\bibitem [{\citenamefont {Vijay}\ \emph {et~al.}(2016)\citenamefont {Vijay},
  \citenamefont {Haah},\ and\ \citenamefont {Fu}}]{Vijay2016}%
  \BibitemOpen
  \bibfield  {author} {\bibinfo {author} {\bibfnamefont {S.}~\bibnamefont
  {Vijay}}, \bibinfo {author} {\bibfnamefont {J.}~\bibnamefont {Haah}},\ and\
  \bibinfo {author} {\bibfnamefont {L.}~\bibnamefont {Fu}},\ }\bibfield
  {title} {\bibinfo {title} {Fracton topological order, generalized lattice
  gauge theory, and duality},\ }\href
  {https://doi.org/10.1103/PhysRevB.94.235157} {\bibfield  {journal} {\bibinfo
  {journal} {Phys. Rev. B}\ }\textbf {\bibinfo {volume} {94}},\ \bibinfo
  {pages} {235157} (\bibinfo {year} {2016})}\BibitemShut {NoStop}%
\bibitem [{\citenamefont {Ma}\ \emph {et~al.}(2017)\citenamefont {Ma},
  \citenamefont {Lake}, \citenamefont {Chen},\ and\ \citenamefont
  {Hermele}}]{Ma2017Coupled}%
  \BibitemOpen
  \bibfield  {author} {\bibinfo {author} {\bibfnamefont {H.}~\bibnamefont
  {Ma}}, \bibinfo {author} {\bibfnamefont {E.}~\bibnamefont {Lake}}, \bibinfo
  {author} {\bibfnamefont {X.}~\bibnamefont {Chen}},\ and\ \bibinfo {author}
  {\bibfnamefont {M.}~\bibnamefont {Hermele}},\ }\bibfield  {title} {\bibinfo
  {title} {Fracton topological order via coupled layers},\ }\href
  {https://doi.org/10.1103/PhysRevB.95.245126} {\bibfield  {journal} {\bibinfo
  {journal} {Phys. Rev. B}\ }\textbf {\bibinfo {volume} {95}},\ \bibinfo
  {pages} {245126} (\bibinfo {year} {2017})}\BibitemShut {NoStop}%
\bibitem [{Note2()}]{Note2}%
  \BibitemOpen
  \bibinfo {note} {We consider graphs $G$ where $G$ and its geometric dual
  $G^*$ are \protect \emph {loopless}, i.e., they contain no self-loops. A
  self-loop corresponds to a trivial, disentangled qubit, since either $X_i$ or
  $Z_i$ belongs to the stabilizer group for some $i$ corresponding to the
  self-loop.}\BibitemShut {Stop}%
\bibitem [{\citenamefont {Dennis}\ \emph {et~al.}(2002)\citenamefont {Dennis},
  \citenamefont {Kitaev}, \citenamefont {Landahl},\ and\ \citenamefont
  {Preskill}}]{Dennis2002}%
  \BibitemOpen
  \bibfield  {author} {\bibinfo {author} {\bibfnamefont {E.}~\bibnamefont
  {Dennis}}, \bibinfo {author} {\bibfnamefont {A.}~\bibnamefont {Kitaev}},
  \bibinfo {author} {\bibfnamefont {A.}~\bibnamefont {Landahl}},\ and\ \bibinfo
  {author} {\bibfnamefont {J.}~\bibnamefont {Preskill}},\ }\bibfield  {title}
  {\bibinfo {title} {{Topological quantum memory}},\ }\href
  {https://doi.org/10.1063/1.1499754} {\bibfield  {journal} {\bibinfo
  {journal} {Journal of Mathematical Physics}\ }\textbf {\bibinfo {volume}
  {43}},\ \bibinfo {pages} {4452} (\bibinfo {year} {2002})}\BibitemShut
  {NoStop}%
\bibitem [{Note3()}]{Note3}%
  \BibitemOpen
  \bibinfo {note} {Unlike Ref.~\cite {BBSgame}, the strategy presented herein
  does not require the application of a nonlocal unitary operator, but gives
  rise to the same $p_\protect \text {q}(\mathinner {|{\psi }\rangle })$ given
  some generic resource state $\mathinner {|{\psi }\rangle }$.}\BibitemShut
  {Stop}%
\bibitem [{\citenamefont {Freedman}\ and\ \citenamefont
  {Meyer}(2001)}]{freedman2001projective}%
  \BibitemOpen
  \bibfield  {author} {\bibinfo {author} {\bibfnamefont {M.~H.}\ \bibnamefont
  {Freedman}}\ and\ \bibinfo {author} {\bibfnamefont {D.~A.}\ \bibnamefont
  {Meyer}},\ }\bibfield  {title} {\bibinfo {title} {Projective plane and planar
  quantum codes},\ }\href {https://doi.org/10.1007/s102080010013} {\bibfield
  {journal} {\bibinfo  {journal} {Foundations of Computational Mathematics}\
  }\textbf {\bibinfo {volume} {1}},\ \bibinfo {pages} {325} (\bibinfo {year}
  {2001})}\BibitemShut {NoStop}%
\bibitem [{\citenamefont {Bombin}\ and\ \citenamefont
  {Martin-Delgado}(2007)}]{Bombin2006}%
  \BibitemOpen
  \bibfield  {author} {\bibinfo {author} {\bibfnamefont {H.}~\bibnamefont
  {Bombin}}\ and\ \bibinfo {author} {\bibfnamefont {M.~A.}\ \bibnamefont
  {Martin-Delgado}},\ }\bibfield  {title} {\bibinfo {title} {{Homological error
  correction: Classical and quantum codes}},\ }\href
  {https://doi.org/10.1063/1.2731356} {\bibfield  {journal} {\bibinfo
  {journal} {Journal of Mathematical Physics}\ }\textbf {\bibinfo {volume}
  {48}},\ \bibinfo {pages} {052105} (\bibinfo {year} {2007})}\BibitemShut
  {NoStop}%
\bibitem [{\citenamefont {Breuckmann}(2017)}]{BreuckmannThesis}%
  \BibitemOpen
  \bibfield  {author} {\bibinfo {author} {\bibfnamefont {N.~P.}\ \bibnamefont
  {Breuckmann}},\ }\emph {\bibinfo {title} {{H}omological quantum codes beyond
  the toric code}},\ \href {https://doi.org/10.18154/RWTH-2018-01100} {\bibinfo
  {type} {Dissertation}},\ \bibinfo  {school} {RWTH Aachen University},
  \bibinfo {address} {Aachen} (\bibinfo {year} {2017})\BibitemShut {NoStop}%
\bibitem [{Note4()}]{Note4}%
  \BibitemOpen
  \bibinfo {note} {The requirement that the bits are associated to a linearly
  independent set of stabilizer generators can straightforwardly be relaxed. If
  additional stabilizer generators are included, the map from $\protect \vec
  {x}$ to bits $\protect \vec {a}$ and $\protect \vec {b}$ is surjective but
  not injective.}\BibitemShut {Stop}%
\bibitem [{\citenamefont {Levin}\ and\ \citenamefont
  {Wen}(2005)}]{LevinWen2005}%
  \BibitemOpen
  \bibfield  {author} {\bibinfo {author} {\bibfnamefont {M.~A.}\ \bibnamefont
  {Levin}}\ and\ \bibinfo {author} {\bibfnamefont {X.-G.}\ \bibnamefont
  {Wen}},\ }\bibfield  {title} {\bibinfo {title} {String-net condensation: A
  physical mechanism for topological phases},\ }\href
  {https://doi.org/10.1103/PhysRevB.71.045110} {\bibfield  {journal} {\bibinfo
  {journal} {Phys. Rev. B}\ }\textbf {\bibinfo {volume} {71}},\ \bibinfo
  {pages} {045110} (\bibinfo {year} {2005})}\BibitemShut {NoStop}%
\bibitem [{\citenamefont {Aravind}(2002)}]{aravind2002bell}%
  \BibitemOpen
  \bibfield  {author} {\bibinfo {author} {\bibfnamefont {P.~K.}\ \bibnamefont
  {Aravind}},\ }\bibfield  {title} {\bibinfo {title} {Bell’s theorem without
  inequalities and only two distant observers},\ }\href
  {https://doi.org/10.1023/A:1021272729475} {\bibfield  {journal} {\bibinfo
  {journal} {Foundations of Physics Letters}\ }\textbf {\bibinfo {volume}
  {15}},\ \bibinfo {pages} {397} (\bibinfo {year} {2002})}\BibitemShut
  {NoStop}%
\bibitem [{\citenamefont {Aravind}(2003)}]{aravind2003simple}%
  \BibitemOpen
  \bibfield  {author} {\bibinfo {author} {\bibfnamefont {P.~K.}\ \bibnamefont
  {Aravind}},\ }\href@noop {} {\bibinfo {title} {A simple demonstration of
  bell's theorem involving two observers and no probabilities or inequalities}}
  (\bibinfo {year} {2003}),\ \Eprint {https://arxiv.org/abs/quant-ph/0206070}
  {arXiv:quant-ph/0206070 [quant-ph]} \BibitemShut {NoStop}%
\bibitem [{\citenamefont {Mermin}(1990{\natexlab{c}})}]{Mermin1990simple}%
  \BibitemOpen
  \bibfield  {author} {\bibinfo {author} {\bibfnamefont {N.~D.}\ \bibnamefont
  {Mermin}},\ }\bibfield  {title} {\bibinfo {title} {Simple unified form for
  the major no-hidden-variables theorems},\ }\href
  {https://doi.org/10.1103/PhysRevLett.65.3373} {\bibfield  {journal} {\bibinfo
   {journal} {Phys. Rev. Lett.}\ }\textbf {\bibinfo {volume} {65}},\ \bibinfo
  {pages} {3373} (\bibinfo {year} {1990}{\natexlab{c}})}\BibitemShut {NoStop}%
\bibitem [{\citenamefont {Ellison}\ \emph {et~al.}(2022)\citenamefont
  {Ellison}, \citenamefont {Chen}, \citenamefont {Dua}, \citenamefont
  {Shirley}, \citenamefont {Tantivasadakarn},\ and\ \citenamefont
  {Williamson}}]{Ellison2022}%
  \BibitemOpen
  \bibfield  {author} {\bibinfo {author} {\bibfnamefont {T.~D.}\ \bibnamefont
  {Ellison}}, \bibinfo {author} {\bibfnamefont {Y.-A.}\ \bibnamefont {Chen}},
  \bibinfo {author} {\bibfnamefont {A.}~\bibnamefont {Dua}}, \bibinfo {author}
  {\bibfnamefont {W.}~\bibnamefont {Shirley}}, \bibinfo {author} {\bibfnamefont
  {N.}~\bibnamefont {Tantivasadakarn}},\ and\ \bibinfo {author} {\bibfnamefont
  {D.~J.}\ \bibnamefont {Williamson}},\ }\bibfield  {title} {\bibinfo {title}
  {Pauli stabilizer models of twisted quantum doubles},\ }\href
  {https://doi.org/10.1103/PRXQuantum.3.010353} {\bibfield  {journal} {\bibinfo
   {journal} {PRX Quantum}\ }\textbf {\bibinfo {volume} {3}},\ \bibinfo {pages}
  {010353} (\bibinfo {year} {2022})}\BibitemShut {NoStop}%
\bibitem [{\citenamefont {Kawagoe}\ and\ \citenamefont
  {Levin}(2020)}]{Kawagoe2020microscopic}%
  \BibitemOpen
  \bibfield  {author} {\bibinfo {author} {\bibfnamefont {K.}~\bibnamefont
  {Kawagoe}}\ and\ \bibinfo {author} {\bibfnamefont {M.}~\bibnamefont
  {Levin}},\ }\bibfield  {title} {\bibinfo {title} {Microscopic definitions of
  anyon data},\ }\href {https://doi.org/10.1103/PhysRevB.101.115113} {\bibfield
   {journal} {\bibinfo  {journal} {Phys. Rev. B}\ }\textbf {\bibinfo {volume}
  {101}},\ \bibinfo {pages} {115113} (\bibinfo {year} {2020})}\BibitemShut
  {NoStop}%
\bibitem [{\citenamefont {Laversanne-Finot}\ \emph {et~al.}(2017)\citenamefont
  {Laversanne-Finot}, \citenamefont {Ketterer}, \citenamefont {Barros},
  \citenamefont {Walborn}, \citenamefont {Coudreau}, \citenamefont {Keller},\
  and\ \citenamefont {Milman}}]{LaversanneFinot_2017}%
  \BibitemOpen
  \bibfield  {author} {\bibinfo {author} {\bibfnamefont {A.}~\bibnamefont
  {Laversanne-Finot}}, \bibinfo {author} {\bibfnamefont {A.}~\bibnamefont
  {Ketterer}}, \bibinfo {author} {\bibfnamefont {M.~R.}\ \bibnamefont
  {Barros}}, \bibinfo {author} {\bibfnamefont {S.~P.}\ \bibnamefont {Walborn}},
  \bibinfo {author} {\bibfnamefont {T.}~\bibnamefont {Coudreau}}, \bibinfo
  {author} {\bibfnamefont {A.}~\bibnamefont {Keller}},\ and\ \bibinfo {author}
  {\bibfnamefont {P.}~\bibnamefont {Milman}},\ }\bibfield  {title} {\bibinfo
  {title} {General conditions for maximal violation of non-contextuality in
  discrete and continuous variables},\ }\href
  {https://doi.org/10.1088/1751-8121/aa6016} {\bibfield  {journal} {\bibinfo
  {journal} {Journal of Physics A: Mathematical and Theoretical}\ }\textbf
  {\bibinfo {volume} {50}},\ \bibinfo {pages} {155304} (\bibinfo {year}
  {2017})}\BibitemShut {NoStop}%
\bibitem [{Note5()}]{Note5}%
  \BibitemOpen
  \bibinfo {note} {Measurement of the unitary operators $U_i = \DOTSI \sumop
  \slimits@ _j \omega ^{s_j} \mathinner {|{j}\rangle }\mathinner {\langle
  {j}|}$ with $s_j \in \protect \mathbb {Z}_d$ may be thought of as measuring
  the Hermitian operator $\DOTSI \sumop \slimits@ _j s_j \mathinner
  {|{j}\rangle }\mathinner {\langle {j}|}$. Alternatively, phase estimation can
  be performed, which is exact in the present setting~\cite
  {okay2017topological}.}\BibitemShut {Stop}%
\bibitem [{\citenamefont {Devakul}\ and\ \citenamefont
  {Williamson}(2021)}]{Devakul2020b}%
  \BibitemOpen
  \bibfield  {author} {\bibinfo {author} {\bibfnamefont {T.}~\bibnamefont
  {Devakul}}\ and\ \bibinfo {author} {\bibfnamefont {D.~J.}\ \bibnamefont
  {Williamson}},\ }\bibfield  {title} {\bibinfo {title} {Fractalizing quantum
  codes},\ }\href {https://doi.org/10.22331/q-2021-04-22-438} {\bibfield
  {journal} {\bibinfo  {journal} {{Quantum}}\ }\textbf {\bibinfo {volume}
  {5}},\ \bibinfo {pages} {438} (\bibinfo {year} {2021})}\BibitemShut {NoStop}%
\bibitem [{\citenamefont {Song}\ \emph {et~al.}(2024)\citenamefont {Song},
  \citenamefont {Tantivasadakarn}, \citenamefont {Shirley},\ and\ \citenamefont
  {Hermele}}]{SongFractonSelf}%
  \BibitemOpen
  \bibfield  {author} {\bibinfo {author} {\bibfnamefont {H.}~\bibnamefont
  {Song}}, \bibinfo {author} {\bibfnamefont {N.}~\bibnamefont
  {Tantivasadakarn}}, \bibinfo {author} {\bibfnamefont {W.}~\bibnamefont
  {Shirley}},\ and\ \bibinfo {author} {\bibfnamefont {M.}~\bibnamefont
  {Hermele}},\ }\bibfield  {title} {\bibinfo {title} {Fracton
  self-statistics},\ }\href {https://doi.org/10.1103/PhysRevLett.132.016604}
  {\bibfield  {journal} {\bibinfo  {journal} {Phys. Rev. Lett.}\ }\textbf
  {\bibinfo {volume} {132}},\ \bibinfo {pages} {016604} (\bibinfo {year}
  {2024})}\BibitemShut {NoStop}%
\bibitem [{\citenamefont {Gidney}(2022)}]{Gidney2022stability}%
  \BibitemOpen
  \bibfield  {author} {\bibinfo {author} {\bibfnamefont {C.}~\bibnamefont
  {Gidney}},\ }\bibfield  {title} {\bibinfo {title} {Stability {E}xperiments:
  {T}he {O}verlooked {D}ual of {M}emory {E}xperiments},\ }\href
  {https://doi.org/10.22331/q-2022-08-24-786} {\bibfield  {journal} {\bibinfo
  {journal} {{Quantum}}\ }\textbf {\bibinfo {volume} {6}},\ \bibinfo {pages}
  {786} (\bibinfo {year} {2022})}\BibitemShut {NoStop}%
\bibitem [{\citenamefont {Okay}\ \emph {et~al.}(2017)\citenamefont {Okay},
  \citenamefont {Roberts}, \citenamefont {Bartlett},\ and\ \citenamefont
  {Raussendorf}}]{okay2017topological}%
  \BibitemOpen
  \bibfield  {author} {\bibinfo {author} {\bibfnamefont {C.}~\bibnamefont
  {Okay}}, \bibinfo {author} {\bibfnamefont {S.}~\bibnamefont {Roberts}},
  \bibinfo {author} {\bibfnamefont {S.~D.}\ \bibnamefont {Bartlett}},\ and\
  \bibinfo {author} {\bibfnamefont {R.}~\bibnamefont {Raussendorf}},\
  }\bibfield  {title} {\bibinfo {title} {Topological proofs of contextuality in
  qunatum mechanics},\ }\href {https://doi.org/10.26421/QIC17.13-14-5}
  {\bibfield  {journal} {\bibinfo  {journal} {Quantum Information \&
  Computation}\ }\textbf {\bibinfo {volume} {17}},\ \bibinfo {pages} {1135}
  (\bibinfo {year} {2017})}\BibitemShut {NoStop}%
\bibitem [{\citenamefont {Delcamp}\ and\ \citenamefont
  {Tiwari}(2019)}]{delcamp2019_2form}%
  \BibitemOpen
  \bibfield  {author} {\bibinfo {author} {\bibfnamefont {C.}~\bibnamefont
  {Delcamp}}\ and\ \bibinfo {author} {\bibfnamefont {A.}~\bibnamefont
  {Tiwari}},\ }\bibfield  {title} {\bibinfo {title} {On 2-form gauge models of
  topological phases},\ }\href {https://doi.org/10.1007/JHEP05(2019)064}
  {\bibfield  {journal} {\bibinfo  {journal} {Journal of High Energy Physics}\
  }\textbf {\bibinfo {volume} {2019}},\ \bibinfo {pages} {1} (\bibinfo {year}
  {2019})}\BibitemShut {NoStop}%
\end{thebibliography}%

\end{document}